%% file: note2739.tex
\RequirePackage[switch]{lineno_babar_revtex}
\documentclass[twocolumn,showpacs,aps,floatfix,prd,nofootinbib]{revtex4}
\catcode`\@=11\relax
\def\@makecol{%
 \setbox\@outputbox\vbox{%
  \boxmaxdepth\@maxdepth
 \protected@write\@auxout{}{%
 \string\@LN@col{\@ifnum{\pagegrid@cur=\@ne}{1}{2}}
      }%
  \@tempdima\dp\@cclv
  \unvbox\@cclv
  \vskip-\@tempdima
 }%
 \xdef\@freelist{\@freelist\@midlist}\global\let\@midlist\@empty
 \@combinefloats
 \@combineinserts\@outputbox\footins
  \set@adj@colht\dimen@
  \count@\vbadness
  \vbadness\@M
  \setbox\@outputbox\vbox to\dimen@{%
   \@texttop
   \dimen@\dp\@outputbox
   \unvbox\@outputbox
   \vskip-\dimen@
   \@textbottom
  }%
  \vbadness\count@
 \global\maxdepth\@maxdepth
}%
\def\balance@two#1#2{%
\outputdebug@sw{{\tracingall\scrollmode\showbox#1\showbox#2}}{}%
 \setbox\@ne\vbox{%
  \@ifvoid#1{}{%
   \unvcopy#1\recover@footins
   \@ifvoid#2{}{\marry@baselines}%
  }%
  \@ifvoid#2{}{%
   \unvcopy#2\recover@footins
  }%
 }%
 \dimen@\ht\@ne\divide\dimen@\tw@
 \dimen@i\dimen@
 \vbadness\@M
 \vfuzz\maxdimen
 \loopwhile{%
  \dimen@i=.5\dimen@i
  \outputdebug@sw{\saythe\dimen@\saythe\dimen@i\saythe\dimen@ii}{}%
  \setbox\z@\copy\@ne\setbox\tw@\vsplit\z@ to\dimen@
  \setbox\z@ \vbox{%
 \protected@write\@auxout{}{%
 \string\@LN@col{\@ifnum{\pagegrid@cur=\@ne}{1}{2}}
      }%
   \unvcopy\z@
   \setbox\z@\vbox{\unvbox\z@ \setbox\z@\lastbox\aftergroup\vskip\aftergroup-\expandafter}\the\dp\z@\relax
  }%
  \setbox\tw@\vbox{%
   \unvcopy\tw@
   \setbox\z@\vbox{\unvbox\tw@\setbox\z@\lastbox\aftergroup\vskip\aftergroup-\expandafter}\the\dp\z@\relax
  }%
  \dimen@ii\ht\tw@\advance\dimen@ii-\ht\z@
  \@ifdim{\dimen@i>.5\p@}{%
   \advance\dimen@\@ifdim{\dimen@ii<\z@}{}{-}\dimen@i
   \true@sw
  }{%
   \@ifdim{\dimen@ii<\z@}{%
    \advance\dimen@\tw@\dimen@i
    \true@sw
   }{%
    \false@sw
   }%
  }%
 }%
 \outputdebug@sw{\saythe\dimen@\saythe\dimen@i\saythe\dimen@ii}{}%
\@ifdim{\ht\z@=\z@}{%
\@ifdim{\ht\tw@=\z@}{%
\true@sw
}{%
\false@sw
}%
}{%
\true@sw
}%
{%
}{%
\ltxgrid@info{Unsatifactorily balanced columns: giving up}%
\setbox\tw@\box#1%
\setbox\z@ \box#2%
}%
 \setbox\tw@\vbox{\unvbox\tw@\vskip\z@skip}%
 \setbox\z@ \vbox{\unvbox\z@ \vskip\z@skip}%
 \set@colroom
\dimen@\ht\z@\@ifdim{\dimen@<\ht\tw@}{\dimen@\ht\tw@}{}%
\@ifdim{\dimen@>\@colroom}{\dimen@\@colroom}{}%
 \outputdebug@sw{\saythe{\ht\z@}\saythe{\ht\tw@}\saythe\@colroom\saythe\dimen@}{}%
\setbox#1\vbox to\dimen@{\unvbox\tw@\unskip\raggedcolumn@skip}%
\setbox#2\vbox to\dimen@{\unvbox\z@ \unskip\raggedcolumn@skip}%
\outputdebug@sw{{\tracingall\scrollmode\showbox#1\showbox#2}}{}%
}%
\catcode`\@=12\relax

\usepackage{graphicx}
\usepackage{dcolumn}
\usepackage{array, longtable}
\usepackage{epsfig}
\usepackage{amsmath}
\usepackage{colordvi}
\usepackage{hhline}
\usepackage{color}

\newcommand{\BaBarYear}{21}
\newcommand{\BaBarNumber}{005}
\newcommand{\SLACPubNumber}{17619}

 \newcommand{\BaBarType}      {PUB}  

\input babarsym

\def\Ecm          {\ensuremath {E_{\rm c.m.}}\xspace}

\def\mgg  {\ensuremath {m(\gamma\gamma)}\xspace}

\long\def\inst#1{\par\nobreak\kern 4pt\nobreak
    {\it #1}\par\vskip 10pt plus 3pt minus 3pt}

\begin{document}

\begin{flushleft}
\babar-\BaBarType-\BaBarYear/\BaBarNumber \\
SLAC-PUB-\SLACPubNumber \\
\end{flushleft}


\title{\large \bf
\boldmath
Study of the reactions  $\epem\to\pi^+\pi^-\pi^0\pi^0\pi^0\pi^0$ and
$\pi^+\pi^-\pi^0\pi^0\pi^0\eta$  at
 center-of-mass energies from threshold to 4.5 GeV using initial-state
radiation
} 

\input authors_sep2021_frozen


\begin{abstract}
We study the processes $\epem\to
\pipi\ppz\ppz\gamma$ and $\pipi\ppz\piz\eta\gamma$ in which an energetic
photon is radiated from the initial state. 
The data were collected with the \babar~ detector at the SLAC National
Accelerator Laboratory.
 About 7300 and 870 events, respectively, are
selected from a data sample corresponding to an integrated
luminosity of 469~\invfb.
The invariant mass of the hadronic final state defines the effective \epem
center-of-mass energy. The center-of-mass energies range from threshold to 4.5~\gev.
From the mass spectra, 
the first ever measurements of
the $\epem\to\pipi\ppz\ppz$ and  the $\epem\to\pipi\ppz\piz\eta$  cross sections are  performed.
The contributions from  $\omega\ppz\piz$, $\eta\pipi\piz$, $\omega\eta$, and other
intermediate  states are presented. 
We observe the $J/\psi$ and $\psi(2S)$ in most of these final states and
measure the
corresponding branching fractions, many of them for the first time.
\end{abstract}

\pacs{13.66.Bc, 14.40.Cs, 13.25.Gv, 13.25.Jx, 13.20.Jf}

\vfill
\maketitle

\setcounter{footnote}{0}


\section{Introduction}
\label{sec:Introduction}

Many precision Standard Model (SM) predictions require 
the hadronic vacuum polarization (HVP) terms  to be taken into account.
At a relatively large
momentum transfer, these terms are measured by studying the inclusive
hadron production in  \epem annihilation and are relatively well
calculated by perturbative quantum chromodynamics.
However, in the energy region from the hadronic threshold to about 2 GeV,
the inclusive hadronic cross section cannot be measured or calculated 
reliably, and a sum of exclusive states must be used. It is particularly
important for the calculation of the muon anomalous
magnetic moment ($g_\mu-2$), which is most sensitive to the low-energy
region. Despite  large data sets of  \epem cross sections, accumulated in the past years, and the
studies performed~\cite{dehz,theoryg2}, there still is a discrepancy
between the SM calculation  and the experimental
value. With the latest result of the ($g_\mu-2$)
experiment at Fermilab~\cite{fermilab}, this discrepancy increased to 4.2 sigma.

Electron-positron annihilation events with initial-state radiation
(ISR) can be used to study processes over a wide range of energies
below the nominal \epem center-of-mass (c.m.) energy (\Ecm),
as proposed in Ref.~\cite{baier}.
The possibility of exploiting ISR to make precise measurements of
low-energy cross sections at high-luminosity $\phi$ and $B$ factories
is discussed in Refs.~\cite{arbus, kuehn, ivanch},
and motivates the studies described in this paper.
Not all accessible states have yet been measured; thus
new measurements will improve the reliability of the HVP calculation.
In addition, studies of ISR events at $B$ factories 
are interesting in their own right, because they provide
information on resonance spectroscopy for masses up to the
charmonium region. 
                                
Studies of  hadron ($h$) production in the ISR processes
$\epem \to h\gamma$  
using data from the \babar\ experiment at SLAC
have been previously reported
~\cite{Druzhinin1,isr2pi,isr2k,isr2p,isr4pi,isr2k2pi,isr6pi,isr3pi,isr5pi,isr2pi2pi0,isr2pi3pi0,isrkkpi,isrkskl,isretapipi,isr4pi3pi0}.
Initial-state radiation events with detection of the ISR photon are characterized by good
reconstruction efficiency and by well understood kinematics,
demonstrated in the references given above.
The \babar~ detector performance
         (tracking, particle identification, \piz, \KS, and
 \KL reconstruction) is well suited to the study of  ISR processes.
 
This paper reports on analyses of the $\pipi4\piz$ and
$\pipi3\piz\eta$ final states produced in
conjunction with an energetic photon, assumed to result from ISR.
While \babar\ data cover effective c.m.\@ energies up to
10.58  \gev, 
this analysis is restricted to 
energies below 4.5 \gev because of backgrounds from $\Upsilon(4S)$
decays. 

There are no previous measurements of the $\epem\to \pipi4\piz$ 
and $\epem\to \pipi3\piz\eta$ cross sections. The six-pion cross
sections have a sizable value below 2~GeV~\cite{isr6pi}
and the two-charged plus four-neutral pion processes
 are currently included in the HPV
calculation by assuming  isospin relations~\cite{dehz}.  The direct
measurement of this channel can reduce the calculation uncertainty.
It is also important to extract the contribution of the intermediate
resonances, because  the 
total cross section calculation depends on
their decay rate to the measured final states.
Below, we present the 
measurements of $\epem\to\omega\ppz\piz$,   $\epem\to\eta\pipi\piz$, and
$\epem\to\omega\eta$ cross sections, with $\eta\to\ppz\piz$, that
contribute to the $\epem\to\pipi4\piz$ final state.

A clear $J/\psi$ signal is observed for both the $\pipi4\piz$ and
$\pipi3\piz\eta$ channels, and the
corresponding $J/\psi$ branching fractions are measured.

\section{\boldmath The \babar\ detector and data set}
\label{sec:babar}

The data used in this analysis were collected with the \babar\ detector at
the \pep2\ asymmetric-energy \epem\ storage ring. 
The total integrated luminosity used is 468.6~\invfb~\cite{lumi}, 
which includes data collected at the $\Upsilon(4S)$
resonance (424.7~\invfb) and at a c.m.\ energy 40~\mev below this
resonance (43.9~\invfb). 

The \babar\ detector is described in detail elsewhere~\cite{babar}. 
Charged particles are reconstructed using the \babar\ tracking system,
which is comprised of the silicon vertex tracker (SVT) and the drift
chamber (DCH), both located 
inside a 1.5 T solenoid.
Separation of pions and kaons is accomplished by means of the detector of
internally reflected Cherenkov light (DIRC) and energy-loss measurements in
the SVT and DCH. 
Photons and \KL mesons are detected in the electromagnetic calorimeter (EMC).  
Muon identification is provided by the instrumented flux return (IFR).

To evaluate the detector acceptance and efficiency, 
we have developed a special package of Monte Carlo (MC) simulation programs for
radiative processes based on 
the approach of K\"uhn  and Czy\.z~\cite{kuehn2}.  
Multiple collinear soft-photon emission from the initial \epem state 
is implemented with the structure function technique~\cite{kuraev,strfun}, 
while additional photon radiation from final-state particles is
simulated using the PHOTOS package~\cite{PHOTOS}.  
The precision of the radiative simulation is such that it contributes less than 1\% to
the uncertainty in the measured hadronic cross sections.

\begin{figure}[tbh]
\begin{center}
\includegraphics[width=0.95\linewidth]{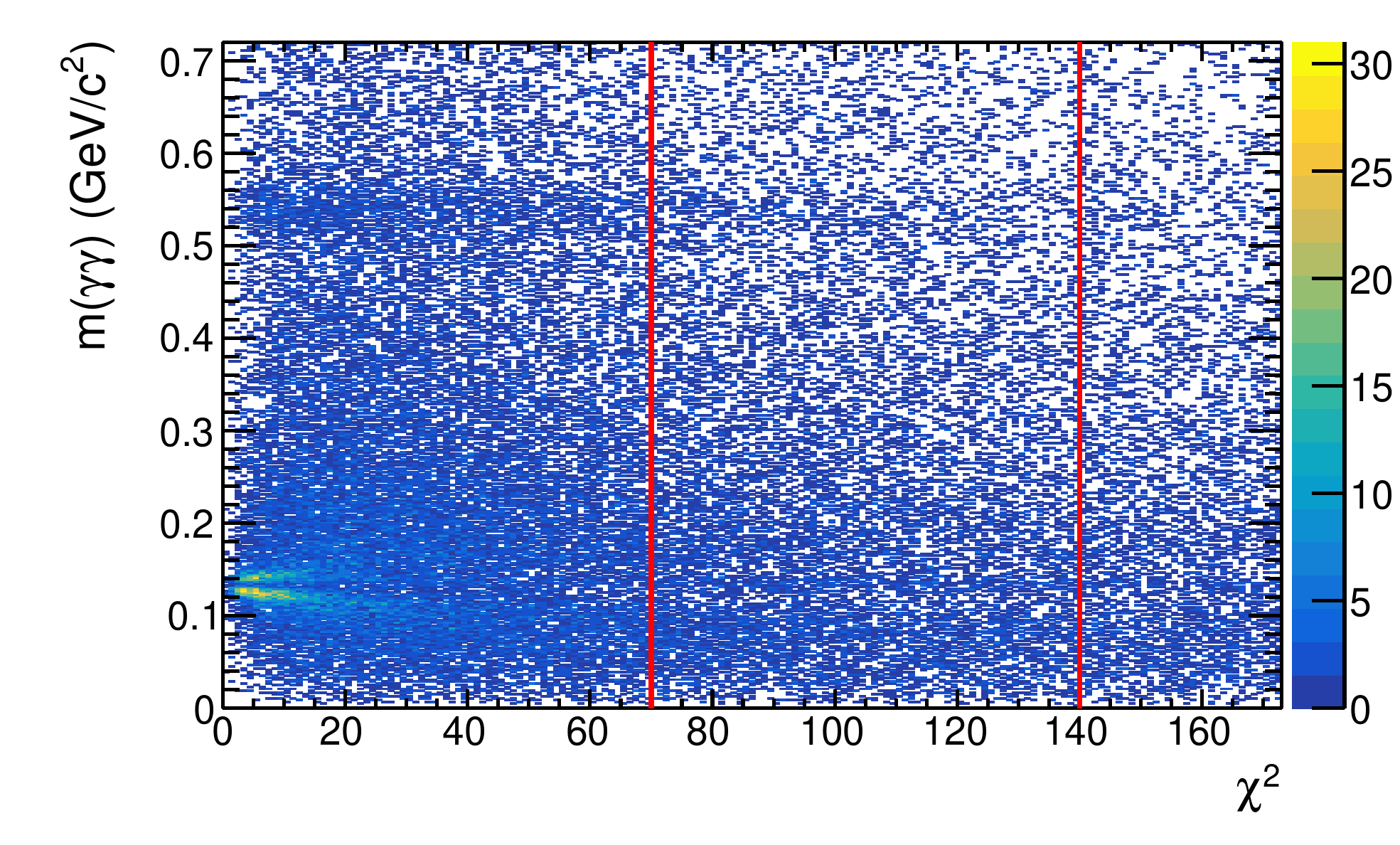}
\put(-40,100){\makebox(0,0)[lb]{\textcolor{black}{\bf (a)}}}\\
\vspace{-0.2cm}
\includegraphics[width=0.95\linewidth]{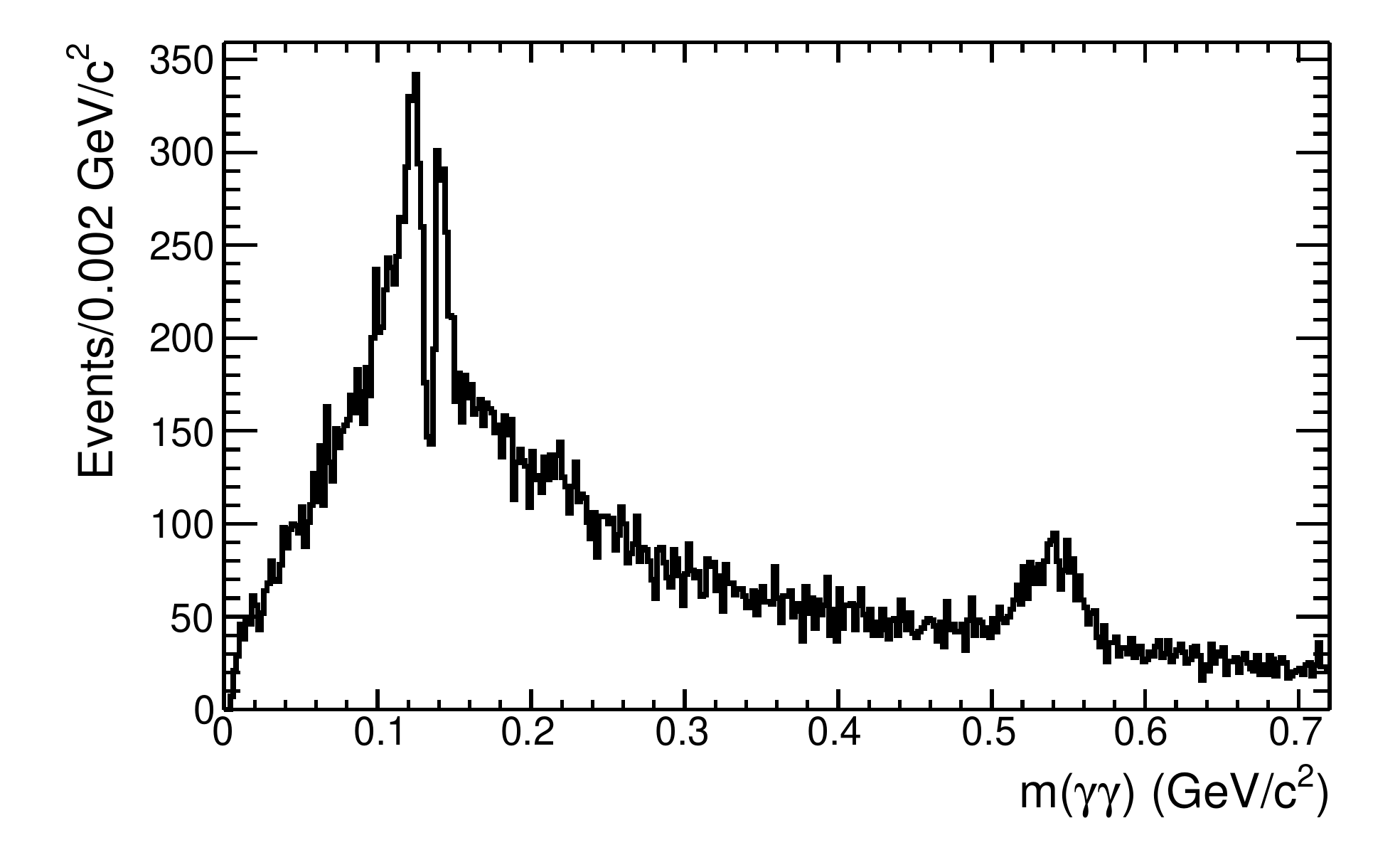}
\put(-50,100){\makebox(0,0)[lb]{\bf(b)}}
\vspace{-0.5cm}
\caption{(a) The invariant mass $\mgg$ of the fourth photon pair  vs
  $\chisq_{2\pi3\piz\gamma\gamma}$.
(b) The $\mgg$ distribution for 
$\chisq_{2\pi3\piz\gamma\gamma} <70$ with additional selection
criteria applied as described in the text.
The double-peak structure near the $\piz$ mass is produced by the
   reconstruction procedure, as explained in the text.
}
\label{2pi4pi0_chi2_all}
\end{center}
\end{figure}
We simulate $\epem\to \pipi\ppz\ppz\gamma$ events assuming production
through the $\omega(782)\eta$ intermediate channel,
with decay of  the $\omega$ to three pions and
decay of the $\eta$ to all  its measured decay modes~\cite{PDG}.

A sample of 460k simulated events is generated  for the
signal reaction  and
processed through the detector response simulation, based on the GEANT4 package~\cite{GEANT4}. These
events are reconstructed using
the same software chain as the data. 
Most of the experimental events contain additional soft photons due to machine background
or interactions in the detector material. Variations in the detector
and background conditions are included in the simulation.

For the purpose of background estimation,  large samples of events from the
main relevant ISR processes ($5\pi\gamma$, $\rho\eta\gamma$, 
$\pipi\ppz\gamma$, etc.) are simulated.  
To evaluate the background from the relevant
 non-ISR processes, namely $\epem\to\qqbar$ $(q=u, d, s)$ and
 $\epem\to\tau^+\tau^-$, 
 simulated samples with integrated luminosities about
that of the data are generated 
using the \textsc{jetset}~\cite{jetset}
and \textsc{koralb}~\cite{koralb} programs,
respectively.
The cross sections for the above processes
are  known with an accuracy slightly better than
10\%, which is sufficient for the present purposes.

\begin{figure}[t]
\begin{center}
\includegraphics[width=0.95\linewidth]{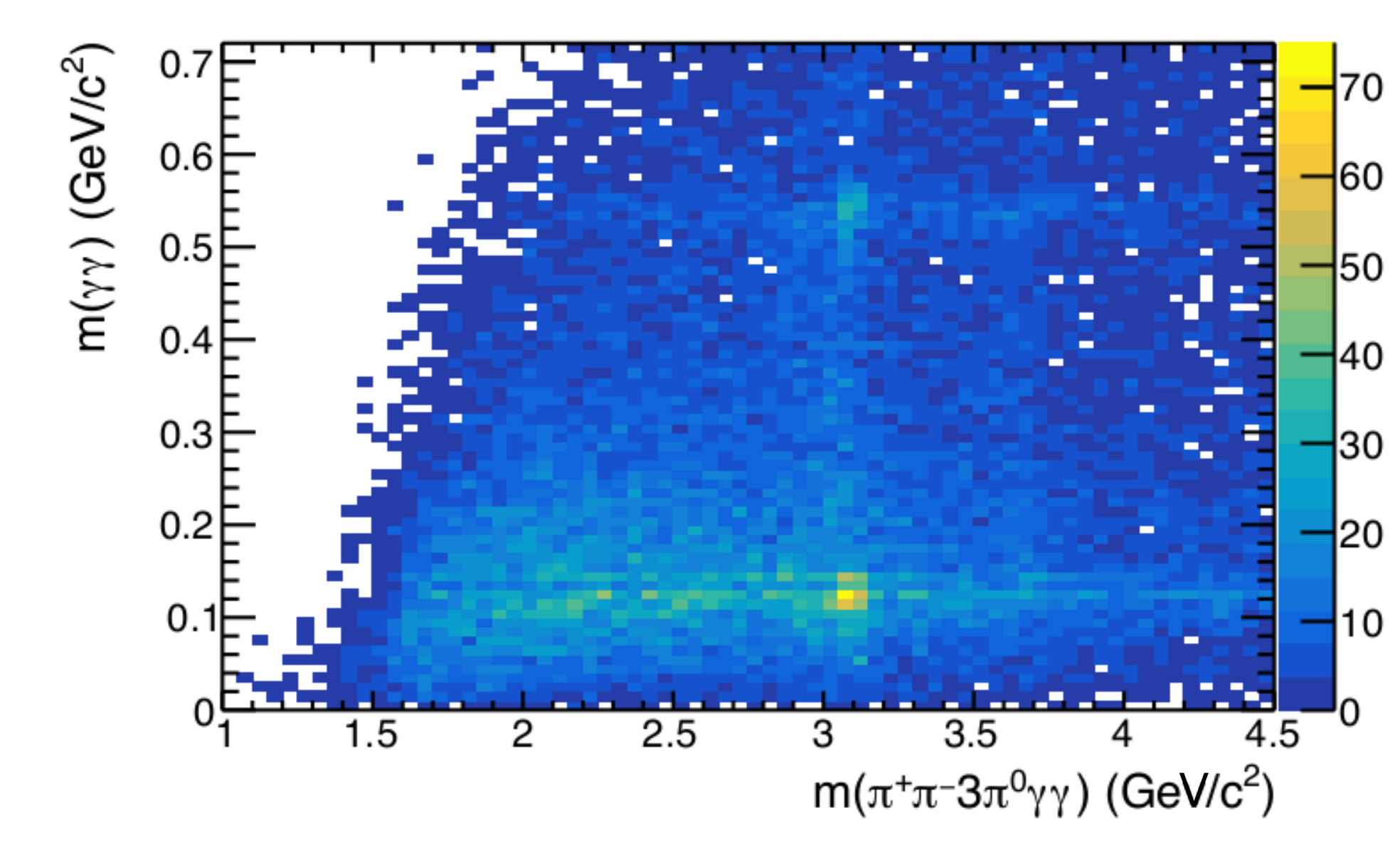}
\put(-185,100){\makebox(0,0)[lb]{\bf(a)}}\\
\vspace{-0.2cm}
\includegraphics[width=0.95\linewidth]{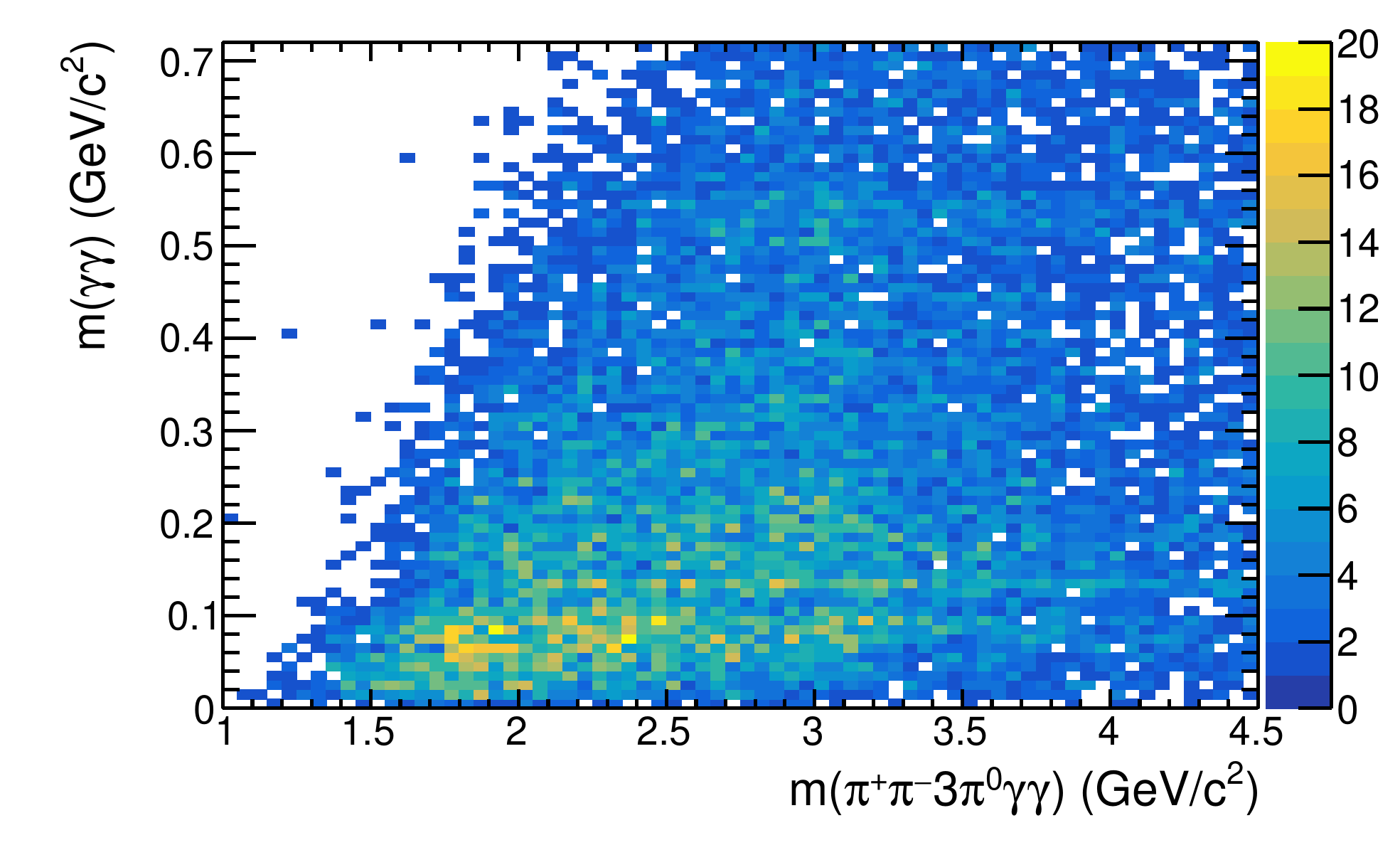}
\put(-185,100){\makebox(0,0)[lb]{\bf(b)}}
\vspace{-0.3cm}
\caption{(a) The fourth-photon-pair invariant mass $\mgg$ vs
  $m(2\pi3\piz\gamma\gamma)$ for (a) $\chisq_{2\pi3\piz\gamma\gamma}<
  70$ and
(b) $70< \chisq_{2\pi3\piz\gamma\gamma}< 140$.
}
\label{2pi4pi0_mass_all}
\end{center}
\end{figure}

\section{\boldmath Event Selection and Kinematic Fit}
\label{sec:Fits}

A relatively clean sample of 
$\pipi4\piz\gamma$ and $\pipi3\piz\eta\gamma$  events is selected
by requiring that there be two tracks reconstructed in the DCH,
SVT, or both, and nine or more photons (sometimes up to 20), with an energy above
0.02~\gev in the EMC. We assume the photon with the highest energy to be  the ISR
photon, and we require its c.m. energy to be larger than 3~\gev.
 
We allow  either  two or  three tracks in an event, 
with exactly one opposite-sign pair that extrapolates
within 0.25 cm of the beam axis and 3.0 cm 
of the nominal collision point along that axis.
The reason a third track is allowed is to capture a
relatively small fraction of signal events that contain a
background track.
The two tracks that satisfy the extrapolation criteria are
fit to a vertex, which is used as the point of origin
in the calculation of the photon directions.

We subject each candidate event to a set of constrained 
kinematic fits and use the fit results,
along with charged-particle identification,
to select the final states of interest and evaluate backgrounds
from other processes.
 The kinematic fits make use of the
four-momenta and covariance matrices of the initial $e^+$, $e^-$, and the
set of selected tracks and photons.
The fitted three-momenta of each track and photon are then used 
in further calculations.

Excluding the photon with the highest c.m.\  energy,
which is assumed to arise from ISR, we consider all independent sets of
eight other photons, and   combine them into four pairs.
For each set of eight photons, we test all
possible independent combinations of four photon pairs.
 We consider those combinations in which the di-photon mass
 of at least three pairs lies within $\pm$35~\mevcc ($\pm 3\sigma$ of
 the resolution) of the
\piz mass $m_{\piz}$~\cite{PDG}.
The selected combinations are subjected to a fit
 in which the di-photon masses of the three pairs with
$|m(\gamma\gamma)-m_{\pi^0}|<35~\mevcc$
 are constrained to $m_{\piz}$.
For  the signal hypothesis
$\epem\to\pipi3\piz\gamma\gamma\gamma_{ISR}$ with the constraints
due to four-momentum conservation, there are thus seven
 constraints (7C) in the fit.  The photons in the
 remaining (``fourth'') pair are treated as being independent.
If all four photon pairs in the combination
 satisfy $|m(\gamma\gamma)-m_{\pi^0}|<35~\mevcc$,
 we rotate the combinations,
  allowing each of the four di-photon pairs in turn
 to be the fourth pair, i.e., the pair without the
 $m_{\piz}$ constraint.
The combination with the smallest \chisq is retained, along with the obtained
$\chisq_{2\pi3\piz\gamma\gamma}$ ($\chisq_{7C}$) value and the fitted three-momenta of each
track and photon. 

 The above procedure allows us not only to search 
for events with  $\piz\to\gamma\gamma$  in the fourth photon pair, but
also for  events with $\eta\to\gamma\gamma$. 

Each retained event is also subjected to a 7C fit under the
$\epem\to\pipi3\piz\gamma_{ISR}$ background hypothesis, and the smallest
 $\chisq_{2\pi3\piz}$ value from all photon combinations is retained.  
The $\pipi3\piz$ process has a comparable
cross section to the $\pipi4\piz$ signal process and can
contribute to the background when two or more background photons are present.

\section{Additional selection criteria}

The results of the 7C fit     
to events with two tracks and at least nine photon candidates
are used to perform the final selection of the six-pion and the five-pion plus eta sample. 
We require the tracks to lie within the fiducial
region of the DCH (0.45-2.40 radians) and to be
inconsistent with being a kaon or muon.
The photon candidates are required to lie within the
fiducial region of the EMC (0.35-2.40 radians) and to have
an energy larger than 0.035 GeV.
A requirement that there be no charged
tracks within 1 radian of the ISR photon reduces the $\tau^+\tau^-$ background
to a negligible level. 
A requirement that any extra photons in an event each
have an energy below 0.7 GeV slightly reduces the multi-photon
background. 

Figure~\ref{2pi4pi0_chi2_all}(a) shows the invariant mass $\mgg$ of the fourth
photon pair vs $\chisq_{2\pi3\piz\gamma\gamma}$. Clear $\piz$
and $\eta$ peaks are visible at small  \chisq values.
We require $\chisq_{2\pi3\piz\gamma\gamma}< 70$ to select the signal  events,
and apply $\chi_{2\pi 3\piz}^2 >30$ condition if these events also
satisfy  the $2\pi 3\piz$ background hypothesis. 
This requirement  reduces the contamination due to    
 $2\pi 3\piz$ events from ~30\% to about 1-2\% while reducing
the signal efficiency by only 5\%.

Figure~\ref{2pi4pi0_chi2_all}(b)
shows the $\mgg$ distribution  after the
above requirements have been applied. 
The dip in this distribution at the $\piz$ mass value is a
consequence of the kinematic fit constraint of the best
three photon pairs to the $\piz$ mass. Also, because of this constraint, the fourth photon pair is
sometimes formed from photon candidates that are less
well measured.

Figure~\ref{2pi4pi0_mass_all} shows the $\mgg$ 
distribution vs the invariant mass $m(2\pi 3\piz\gamma\gamma)$ 
 for events (a) in the signal region
$\chisq_{2\pi3\piz\gamma\gamma}< 70$ and (b) in a control
region defined by $70 < \chisq_{2\pi3\piz\gamma\gamma}< 140$. Events from the
$\epem\to\pipi4\piz$ and $\pipi3\piz\eta$ processes are
clearly seen in the signal region, as well as $J/\psi$ decays to these
final states.
No significant structures are seen in the control region,
and we use these events to evaluate background.

Our strategy to extract the signals for the $\epem\to\pipi\ppz\ppz$
and $\pipi\ppz\piz\eta$ processes 
is to perform a fit to the $\piz$ and $\eta$ yields
in intervals of 0.05~\gevcc in the distribution of
the  invariant mass $m(\pipi3\piz\gamma\gamma)$. 

\begin{figure}[tbh]
\begin{center}
\includegraphics[width=1.0\linewidth]{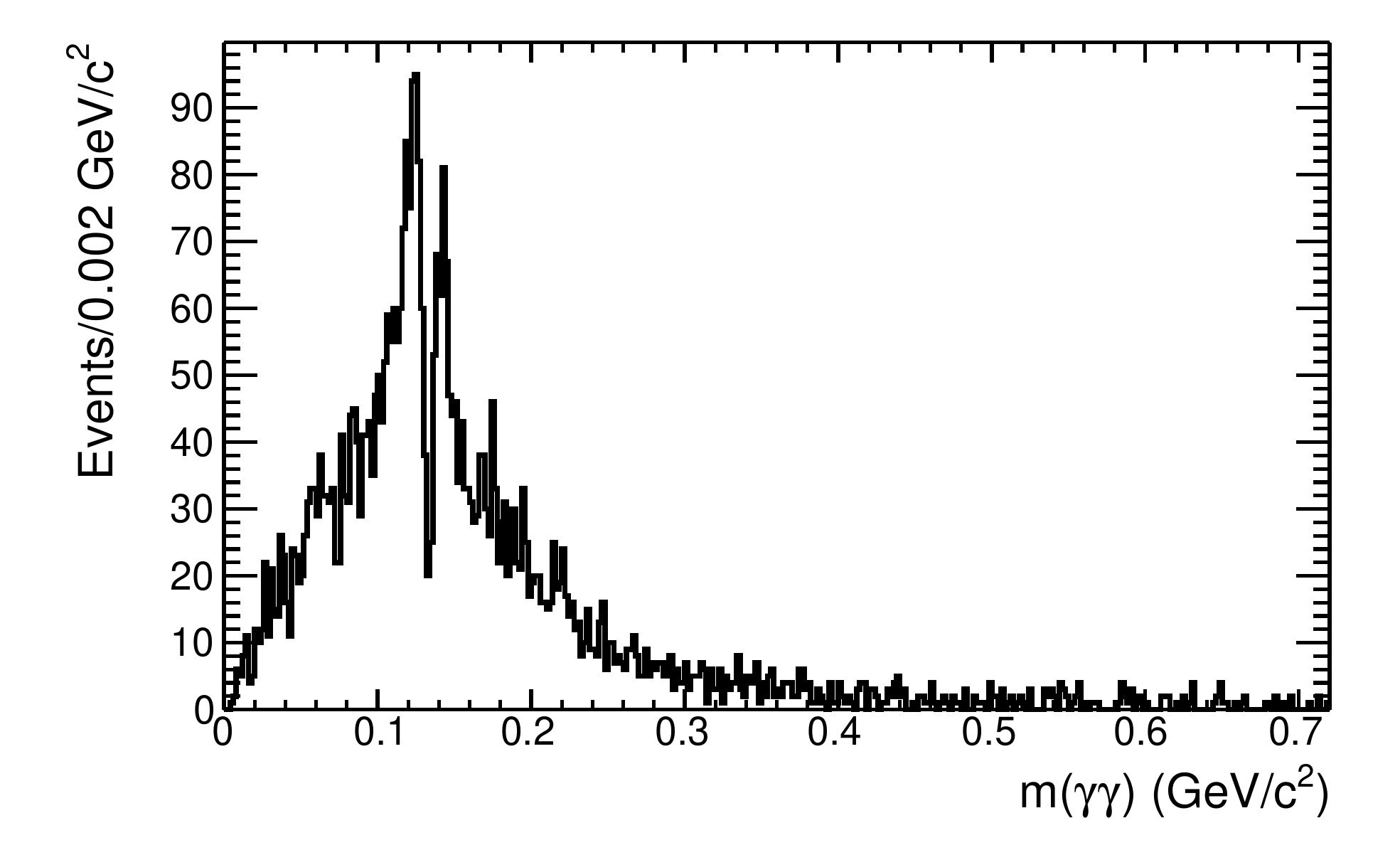}
\put(-50,110){\makebox(0,0)[lb]{\bf(a)}}\\
\vspace{-0.2cm}
\includegraphics[width=0.95\linewidth]{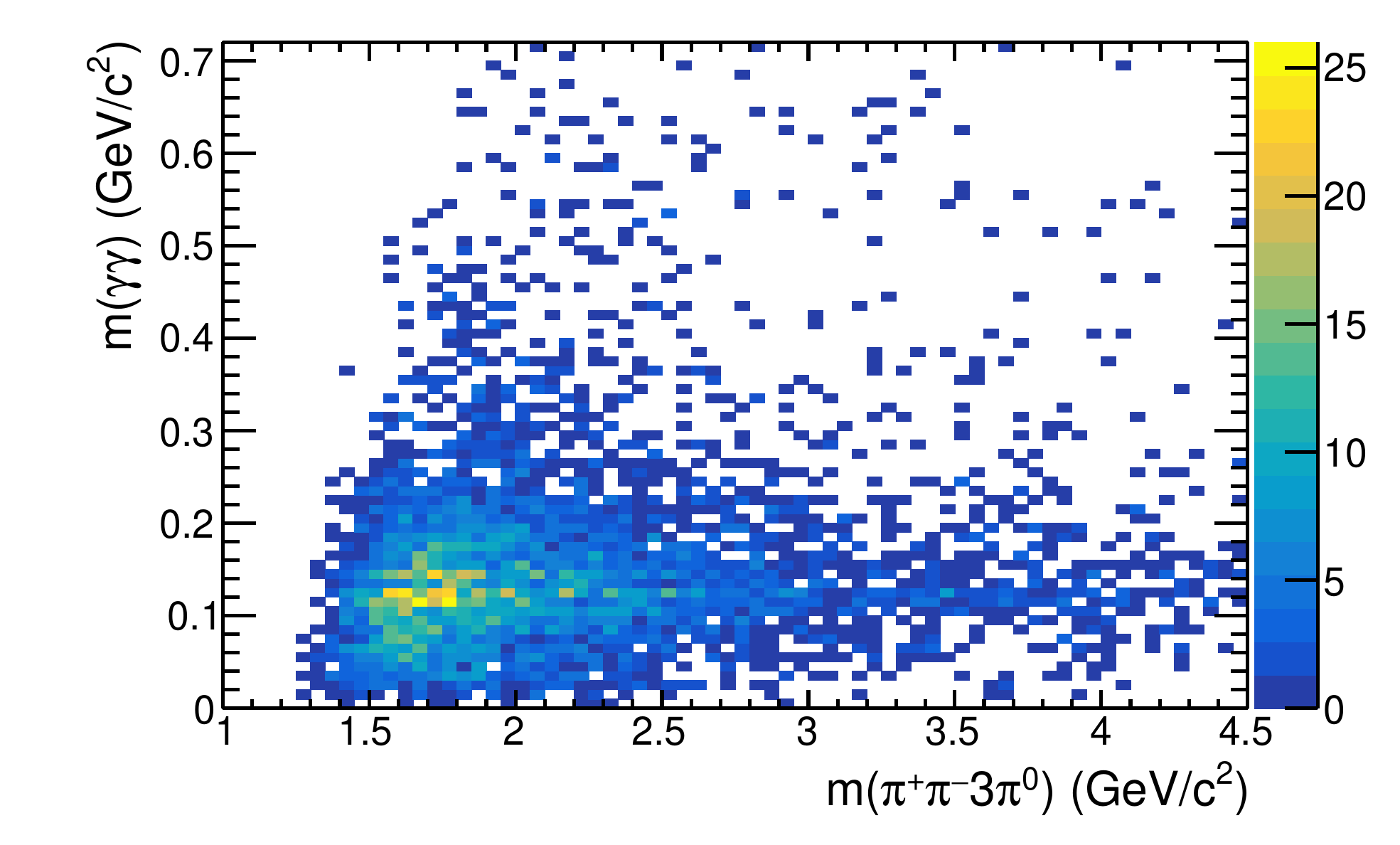}
\put(-50,110){\makebox(0,0)[lb]{\bf(b)}}
\vspace{-0.5cm}
\caption{
The  MC-simulated distribution for $\epem\to\omega\eta$ events of (a)
the fourth-photon-pair invariant mass $\mgg$, and (b) $\mgg$ vs $m(\pipi3\piz\gamma\gamma)$.
}
\label{mgg_eta2pi}
\end{center}
\end{figure}

\section{\boldmath Detection efficiency}\label{sec:efficiency}
\subsection{Number of signal events in simulation}
As mentioned in Sec.~\ref{sec:babar}, the model used in the MC simulation
assumes that the six-pion final state arises primarily through 
$\omega\eta$ production, with  $\omega$
decays to three pions and $\eta$ decays to 3$\piz$. As shown below,
events with $\eta$ and $\omega$ dominate in the observed cross sections.

The selection procedure applied to the data is also applied to the
MC-simulated events. Figure~\ref{mgg_eta2pi} shows (a)
the $\mgg$ distribution for the \chisq signal region and (b) the distribution
of
$\mgg$ vs $m(2\pi3\piz\gamma\gamma)$ for the simulated $\omega\eta$ events.
The $\piz$ signal shape is not Gaussian due to the procedure explained in
the previous section. It also includes a combinatoric
background arising from the combination of background photons, included in the
simulation, with the photons from
the signal reactions.
\begin{figure}[tbh]
\begin{center}
\includegraphics[width=0.88\linewidth]{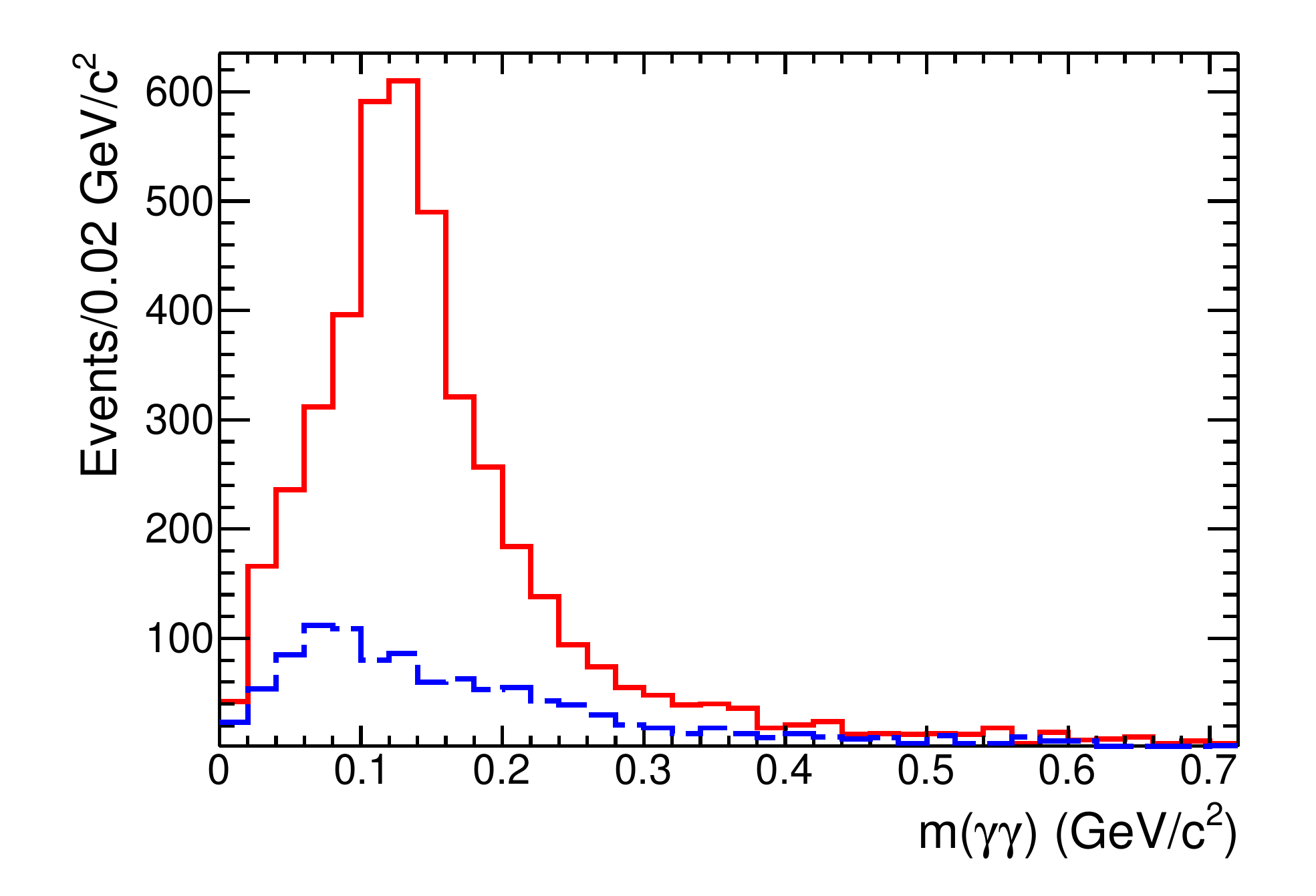}
\put(-50,100){\makebox(0,0)[lb]{\bf(a)}}\\
\vspace{-0.2cm}
\includegraphics[width=0.88\linewidth]{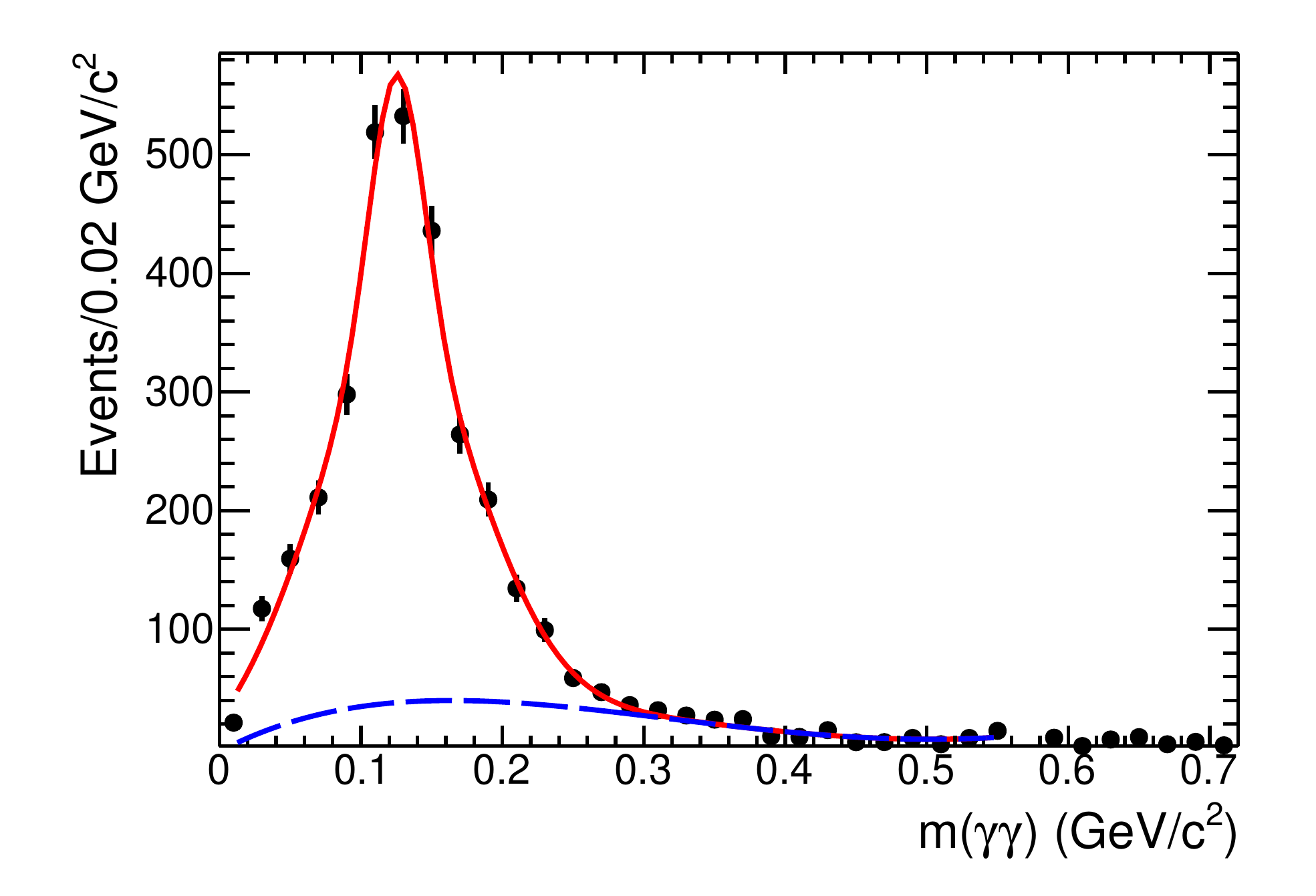}
\put(-50,100){\makebox(0,0)[lb]{\bf(b)}}
\vspace{-0.5cm}
\caption{
The MC-simulated $\mgg$ distribution for
(a) $\epem\to\omega\eta$ in the signal \chisq region (solid histogram) and
control region (dashed), and (b) background-subtracted $\mgg$
distribution. The fit function is described in the text.
The dashed curve shows the remaining background contribution.
}
\label{mgg_eta2pi_fit}
\end{center}
\end{figure}
\begin{figure*}[tbh]
\begin{center}
\includegraphics[width=0.34\linewidth]{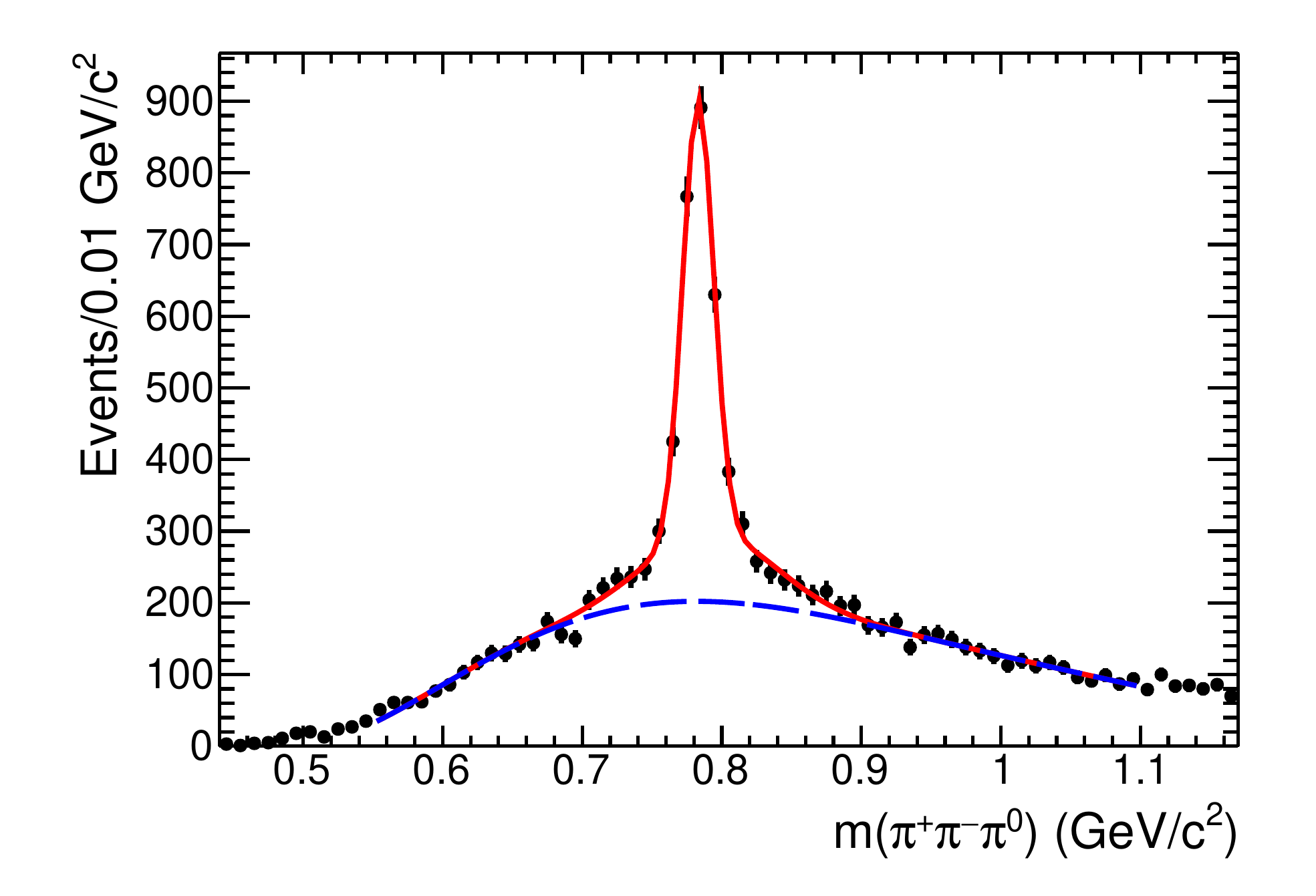}
\put(-50,90){\makebox(0,0)[lb]{\bf(a)}}
\includegraphics[width=0.34\linewidth]{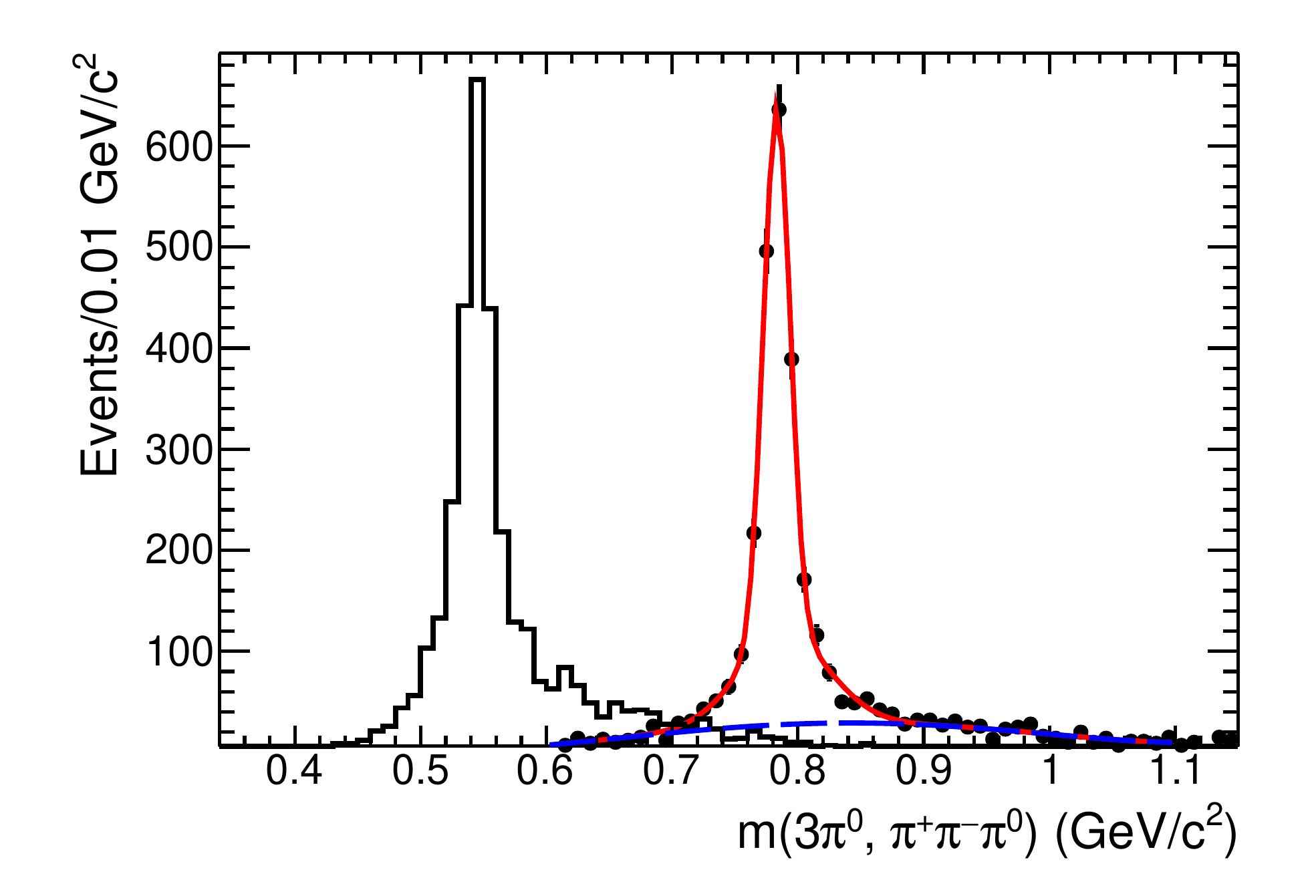}
\put(-50,90){\makebox(0,0)[lb]{\bf(b)}}
\includegraphics[width=0.34\linewidth]{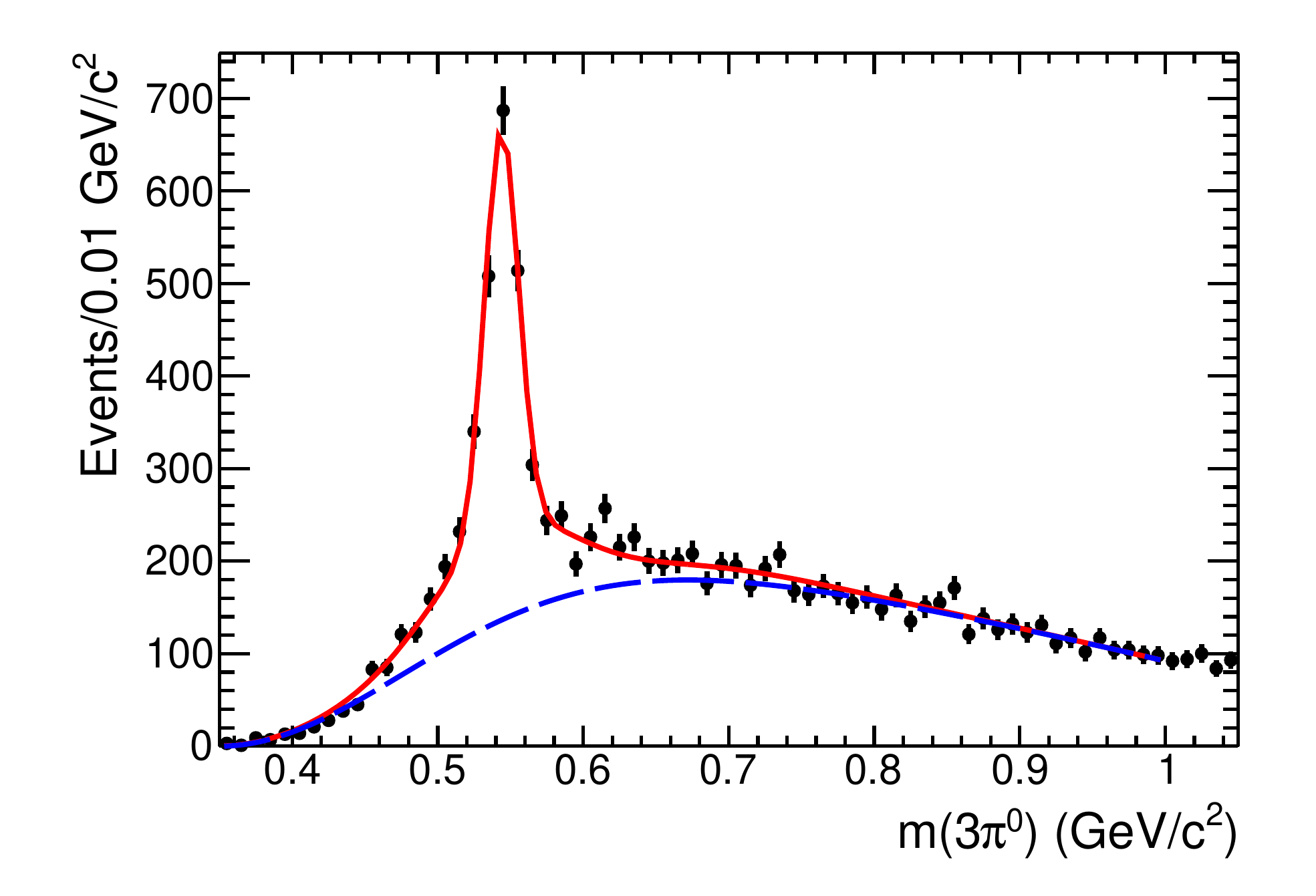}
\put(-50,90){\makebox(0,0)[lb]{\bf(c)}}
\vspace{-0.5cm}
\caption{The MC-simulated $\epem\to\omega\eta$ events.
(a) The $\pipi\piz$ invariant mass (four entries/event) with BW fit
function (solid curve). The dashed curve shows the combinatoric background.
(b) The $\ppz\piz$ invariant mass combination closest to the $\eta$
mass (histogram), and the remaining $\pipi\piz$ invariant mass
distribution (dots).
The solid curve shows the $\omega$ signal fit and the dashed curve shows
the remaining background.
(c) The $\ppz\piz$ invariant mass for all selected MC-simulated
events (four entries/event) with the fit functions used to determine the $\eta$
signal.
}
\label{m3pi_omega_eta_mc}
\end{center}
\end{figure*}
This background is subtracted as illustrated in Fig.~\ref{mgg_eta2pi_fit},
which shows the simulated $\mgg$ distribution from Fig.~\ref{mgg_eta2pi}(a) with
a bin width of 0.02~\gevcc.
The solid histogram in Fig.~\ref{mgg_eta2pi_fit}(a) corresponds to the
two-photon mass distribution obtained from the \chisq signal region. The dashed
histogram is obtained instead from the control region and represents a
combinatoric background distribution, which is subtracted assuming a scale
factor that is varied to estimate the uncertainty in its contribution.
The signal yield is then extracted by fitting the \piz\ peak of this
distribution with a sum of three Gaussian functions for the signal plus a
second-order polynomial function to account for a residual combinatoric background.
If a scale factor 1.5  is used, the background level becomes
negligible, and we can determine and fix parameters for the signal function.
If then  we change the scale factor to 1.0 or to 0.0 in the fit, the obtained
signal yield does not change by more than 3\%.
The result, for a scale factor of 1.0, is shown by the points in
Fig.\,\ref{mgg_eta2pi_fit}(b).
The  fit  is shown by the smooth solid curve,
while the dashed curve shows the contribution of the remaining combinatoric
background.
The fitted signal yields $2639\pm66$ events.
We apply a similar fit procedure in each
0.05~\gevcc interval of the $m(\pipi3\piz\gamma\gamma)$ invariant mass
distribution.

Alternatively, for $\omega\eta$ events,
the $\omega$ mass peak can be used.
Figure~\ref{m3pi_omega_eta_mc}(a) shows the
$\pipi\piz$ invariant mass (four entries per event) for selected
MC-simulated events.
A Breit-Wigner (BW) function,
convolved with a Gaussian distribution to account
for the detector resolution, is used to describe the $\omega$ signal.
A second-order polynomial is used to describe the background. We
obtain $2699\pm75$ events in total.
We also obtain the number of events by fitting
$m(\pipi\piz)$ in 0.05~\gevcc intervals of the
$m(\pipi3\piz\gamma\gamma)$ invariant mass.

Because in our simulation the $\omega$ and $\eta$ mesons are produced in
correlation, we can significantly reduce combinatorial background by selecting
only one (from four) combination, in which the $3\piz$ invariant mass is
closest to the $\eta$ mass.
The distribution of $m(3\piz)$ for this combination is shown in
Fig.~\ref{m3pi_omega_eta_mc}(b) by the histogram. In the remaining $\pipi\piz$
combination the $\omega$ signal, shown by dots in
Fig.~\ref{m3pi_omega_eta_mc}(b), has much lower background, and the fit yields
$2796\pm76$ events in total. The $\pipi3\piz\gamma\gamma$ mass distribution is
obtained by similar fitting
in each 0.05~\gevcc interval.

Similarly, as an alternative for
the $\omega\eta$ events, we determine the number of
events by fitting the
$\eta$ signal from the $\eta\to\ppz\piz$ decay: the simulated
distribution is shown in Fig.~\ref{m3pi_omega_eta_mc}(c) (four
entries per event). The fit functions are the sum of three
Gaussian functions and a polynomial for the combinatoric background.
This fit yields $2569\pm79$ events in total. The $\pipi3\piz\gamma\gamma$
mass distribution is also obtained 
in each 0.05~\gevcc interval.

\subsection{Efficiency evaluation}
\begin{figure}[tbh]
\begin{center}
\includegraphics[width=0.95\linewidth]{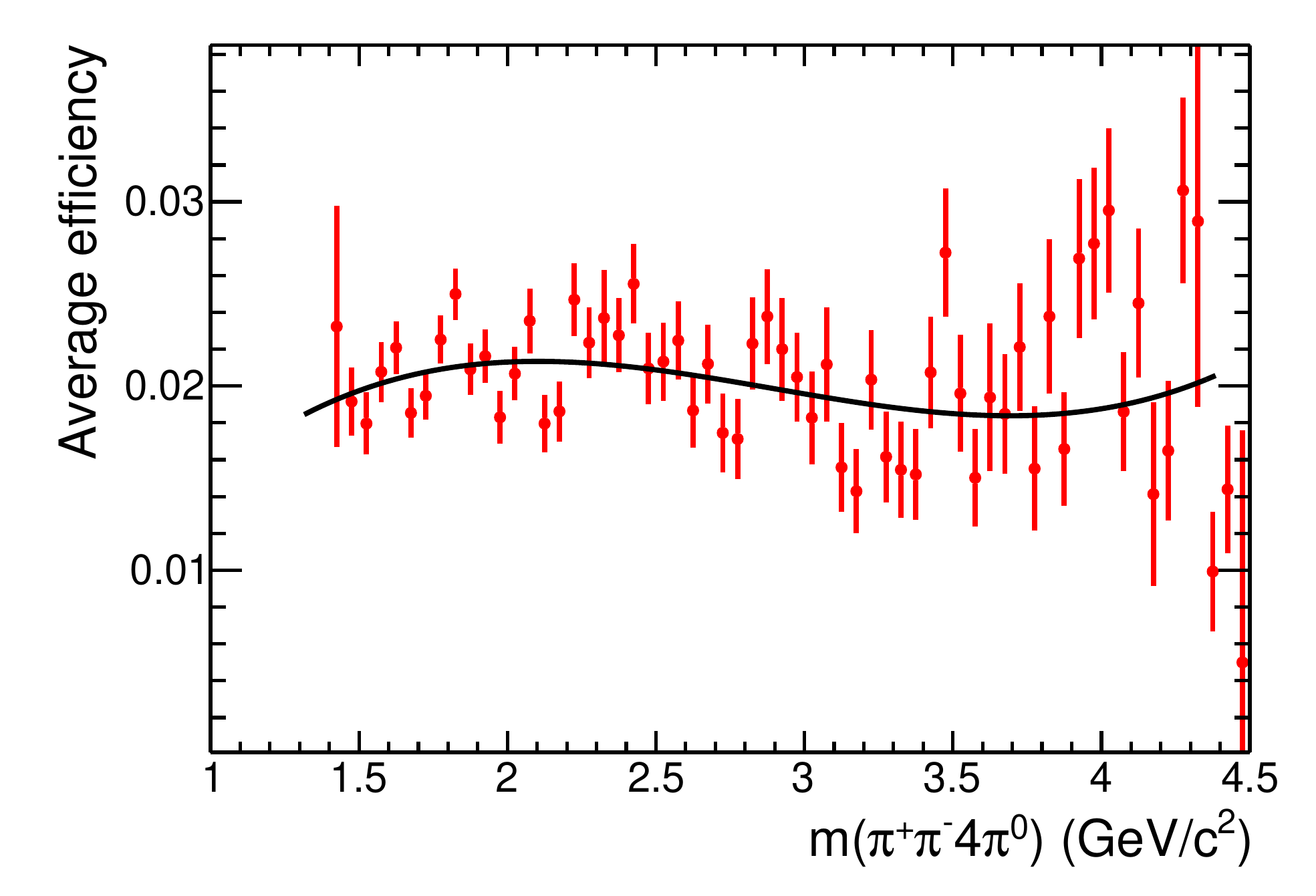}
\vspace{-0.5cm}
\caption{ The energy-dependent reconstruction efficiency for 
  the  $\epem\to\pipi\ppz\ppz$ events.
The curve shows the fit result, which is used in the cross
section calculation.
}
\label{mc_acc}
\end{center}
\end{figure}

The mass-dependent  detection
efficiency is obtained by dividing the number of fitted MC
events in each 0.05~\gevcc mass interval by the number generated in          
the same interval.
By comparing the results of the four different
methods, 
we conclude that the total efficiency does not change by more than 
5\% because of variations of the functions used to extract the number
of events or the use of different background subtraction procedures. This
value is taken as an estimate of the systematic uncertainty in the efficiency
associated with the simulation model used and with the fit procedure.
We average the four efficiencies in each 0.05~\gevcc mass interval and
fit the result with a third-order polynomial function, shown in
Fig.~\ref{mc_acc}. Although the signal simulation accounts for all
$\eta$ decay modes, the efficiency calculation
considers only the $\eta\to\ppz\piz$ decay mode.
From Fig.~\ref{mc_acc} it is seen that the reconstruction
efficiency is about 2\%, roughly independent of mass.
The result of this fit is used for the cross section calculation. 

This efficiency estimate takes into account the geometrical acceptance of the detector 
for the final-state photons and the charged pions, the inefficiency of 
the  detector subsystems, and  the event loss due to additional
soft-photon  emission from the initial and final states.
Corrections to the efficiency that account for data-MC differences are discussed below.

\begin{figure}[tbh]
\begin{center}
\includegraphics[width=0.9\linewidth]{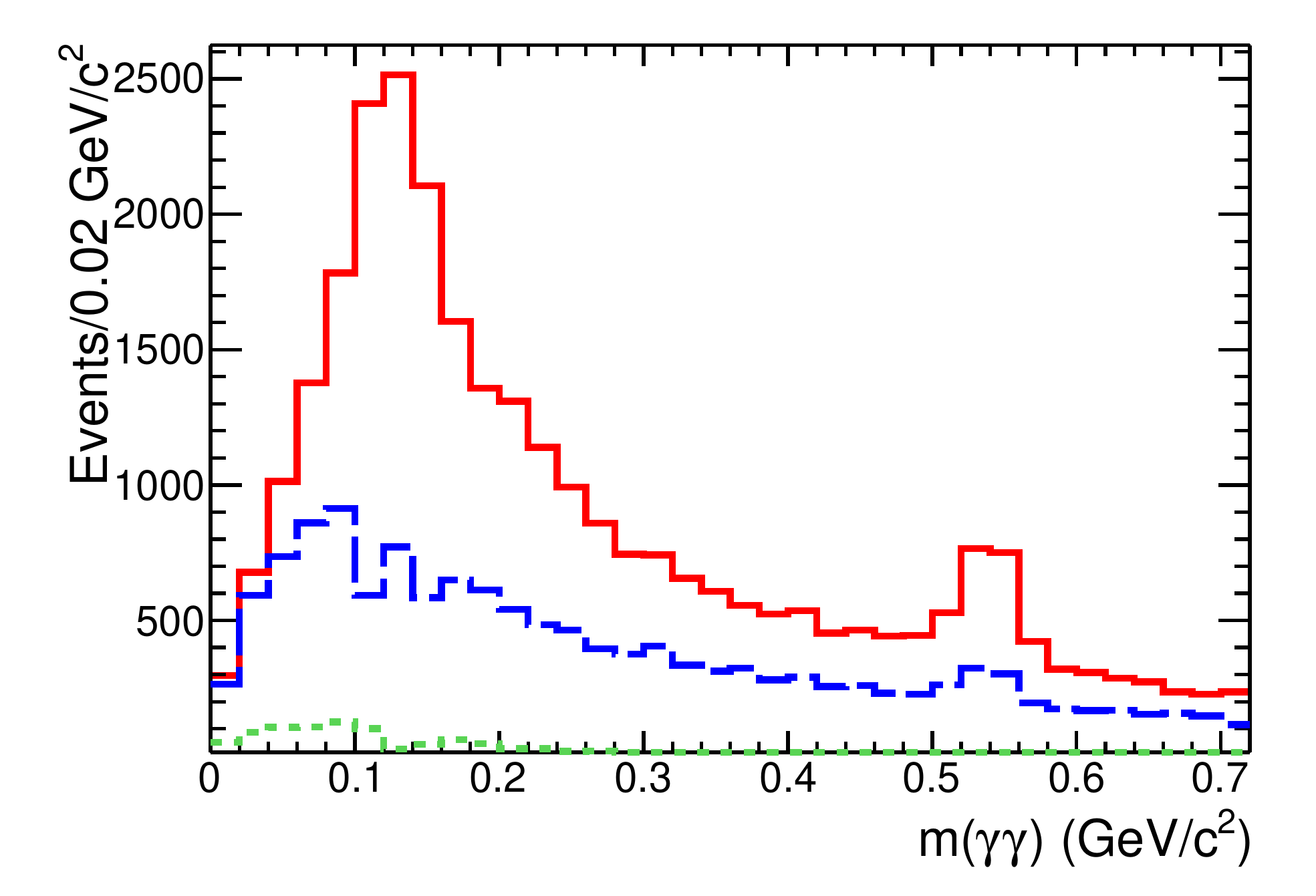}
\put(-50,90){\makebox(0,0)[lb]{\bf(a)}}\\
\vspace{-0.2cm}
\includegraphics[width=0.9\linewidth]{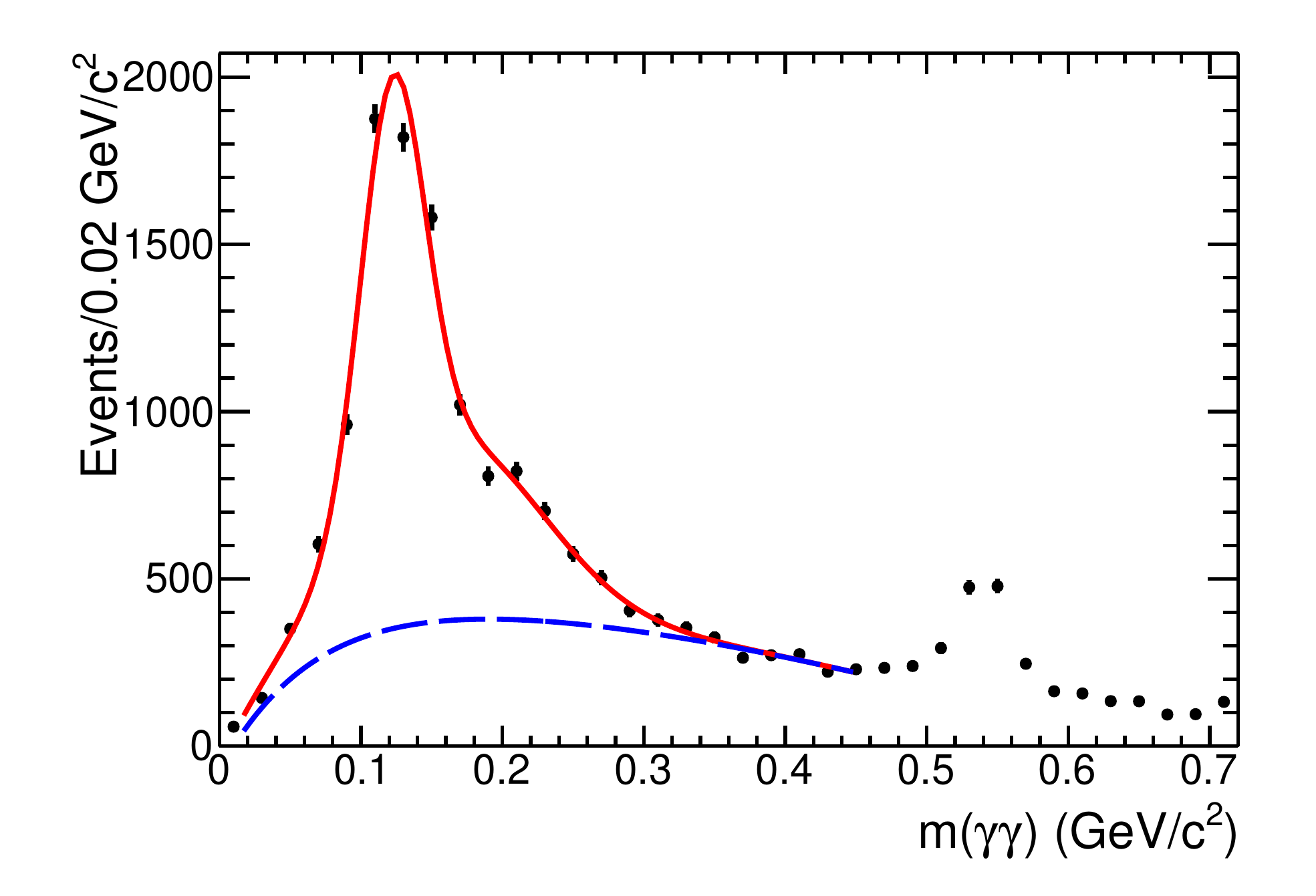}
\put(-50,90){\makebox(0,0)[lb]{\bf(b)}}
\vspace{-0.5cm}
\caption{
(a) The fourth-photon-pair invariant mass $\mgg$ for data in
the signal (solid) and \chisq control (dashed) regions. 
The dotted histogram shows the estimated remaining background in the signal region from
$\epem\to\pipi3\piz$.
(b) The $\mgg$ invariant mass for data after background
subtraction. The curves are the fit results as described in the text.
}
\label{2pi2pi0_bkg}
\end{center}
\end{figure}

\section{The {\boldmath $\pipi4\piz$} final state}
\subsection{\boldmath Number of $\pipi4\piz$ events}\label{sec:signal}

The solid histogram in Fig.~\ref{2pi2pi0_bkg}(a) shows the $\mgg$
data of Fig.~\ref{2pi4pi0_chi2_all}(b) binned in mass intervals of 0.02~\gevcc.
The dashed histogram shows the distribution of data from the 
\chisq control region. The dotted
histogram is the estimated
remaining background from the $\epem\to\pipi3\piz$ process. 
No evidence for a peaking background is seen below 0.45~\gevcc in either
of the two background distributions.
We subtract the background evaluated using the \chisq control region with the  scale factor 1.0.
The resulting $\mgg$ distribution is shown in Fig.~\ref{2pi2pi0_bkg}(b).

\begin{figure}[tbh]
\begin{center}
  \includegraphics[width=0.95\linewidth]{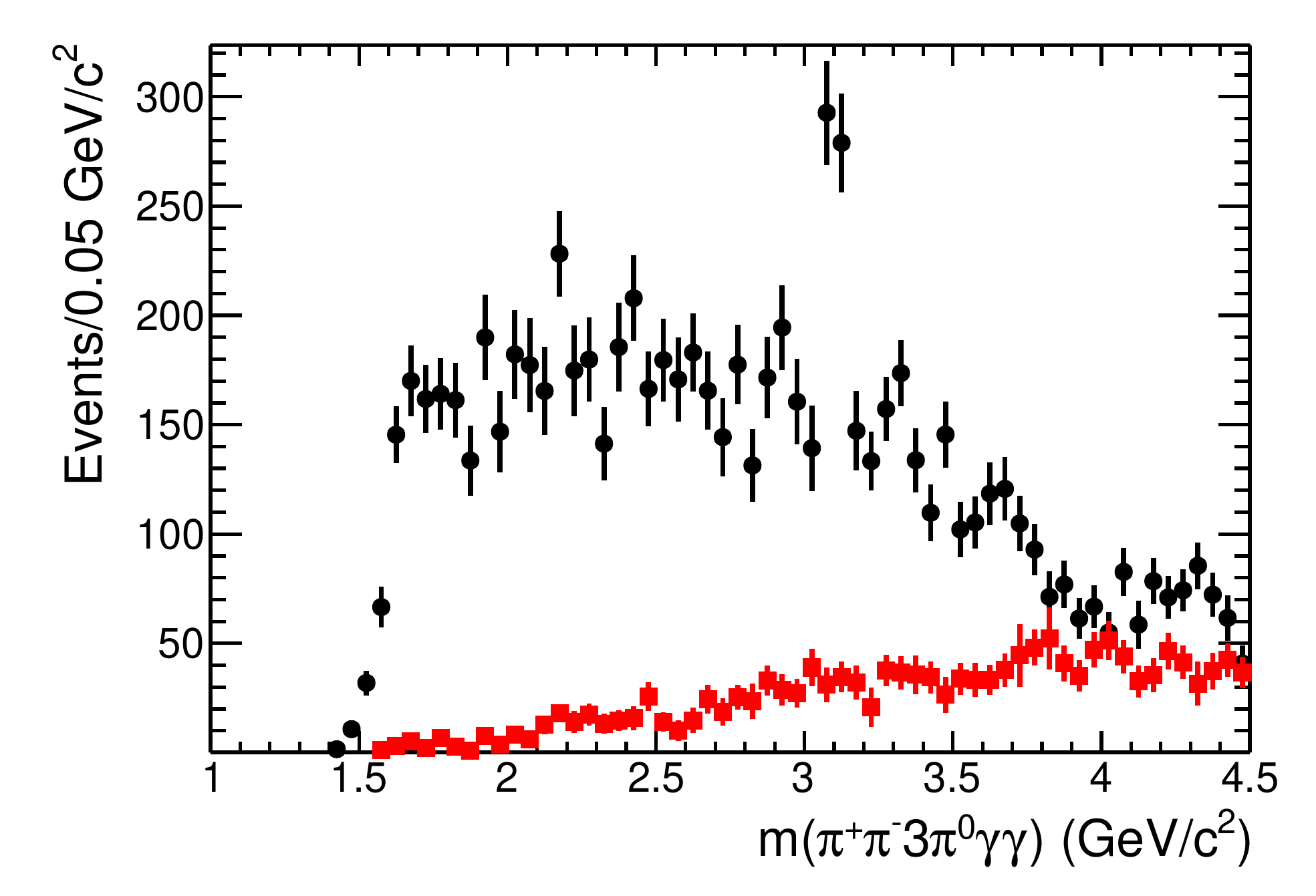}
\vspace{-0.5cm}
\caption{ The invariant mass distribution of 
  $\pipi4\piz$ events (black curcles), obtained from the fit to the $\piz$ mass peak.
The contribution from non-ISR $uds$ background is shown by red squares.
}
\label{nev_2pi4pi0_data}
\end{center}
\end{figure}

We fit the data of Fig.~\ref{2pi2pi0_bkg}(b) with a combination of a
signal function, taken from  a fit to simulated data, and a
background function, taken to be a third-order polynomial.
The fit is performed in the $\mgg$ mass range from
0.0 to 0.45~\gevcc.
The result of the fit is shown by the solid and
dashed curves in Fig.~\ref{2pi2pi0_bkg}(b). In total $7306\pm164$
events are obtained.
Note that this number includes a relatively small
peaking background component, due to $\qqbar$ events,
which is discussed in Sect.~\ref{sec:udsbkg}.
The same fit is applied to the corresponding \mgg
distribution in each 0.05~\gevcc interval in the
$\pipi3\piz\gamma\gamma$  invariant mass.
The resulting number of $\pipi4\piz$ event candidates
as a function of $m(\pipi4\piz)$, including the
peaking $\qqbar$ background, is shown by the data
points in Fig.~\ref{nev_2pi4pi0_data}. 

\begin{figure}[tbh]
\begin{center}
\includegraphics[width=0.95\linewidth]{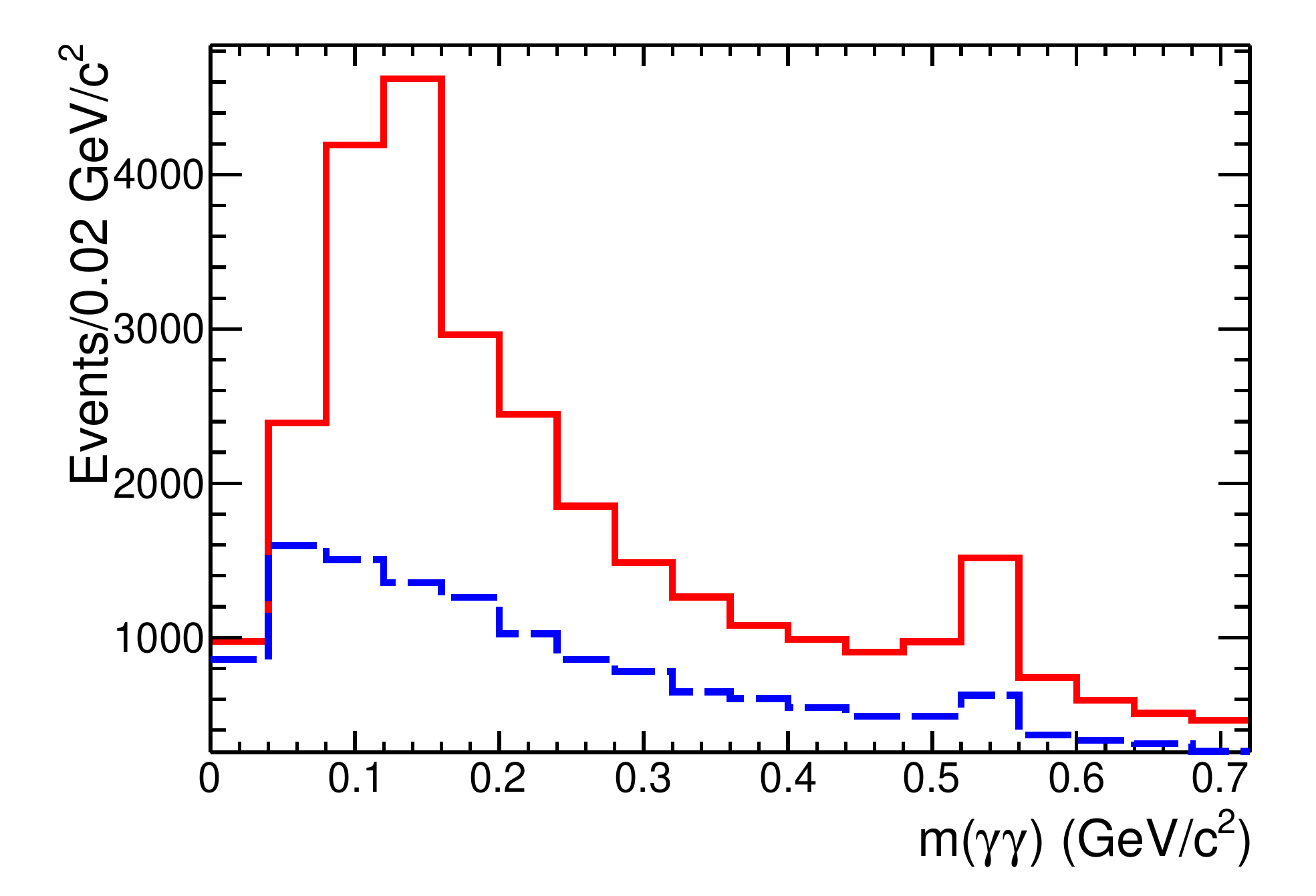}
\vspace{-0.5cm}
\caption{The fourth-photon-pair invariant mass for the $uds$ simulation
 for the signal region
$\chisq_{2\pi3\piz\gamma\gamma} <70$ (solid histogram), and
the control region $70<\chisq_{2\pi3\piz\gamma\gamma} <140$
(dashed histogram). 
}
\label{udsbkg}
\end{center}
\end{figure}
\subsection{Peaking background}\label{sec:udsbkg}
The major background producing a $\piz$ peak
following application of the
selection criteria of Sect. IV.A is
from non-ISR \qqbar events, the most important
channel being $\epem\to\pipi\ppz\ppz\piz$
in which one of the
neutral pions decays asymmetrically, 
yielding a high energy photon
that mimics an ISR photon.
We apply all our selection criteria and fit procedures to
the non-ISR light quark
$\qqbar$ ($uds$) simulation.
Figure~\ref{udsbkg} shows the  fourth-photon-pair invariant
mass for  $\chisq_{2\pi3\piz\gamma\gamma} <70$ and 
$70<\chisq_{2\pi3\piz\gamma\gamma} <140$: clear signals from $\piz$ and 
$\eta$ are seen.

To normalize the $uds$ simulation, we form the diphoton invariant
mass distribution of the ISR candidate with each of the other
photons in the event. 
  A $\piz$ peak is observed, with approximately the same
number of events in data and simulation, leading
to a normalization factor of $1.0\pm0.1$.
The resulting $uds$ background is shown
in Fig.~\ref{nev_2pi4pi0_data}: the $uds$ background is negligible
below 2~\gevcc, but accounts for more than half 
the total spectrum for around 4~\gevcc and above. We subtract this
background for the cross section calculation.

\begin{figure}[tbh]
\begin{center}
\includegraphics[width=1.05\linewidth]{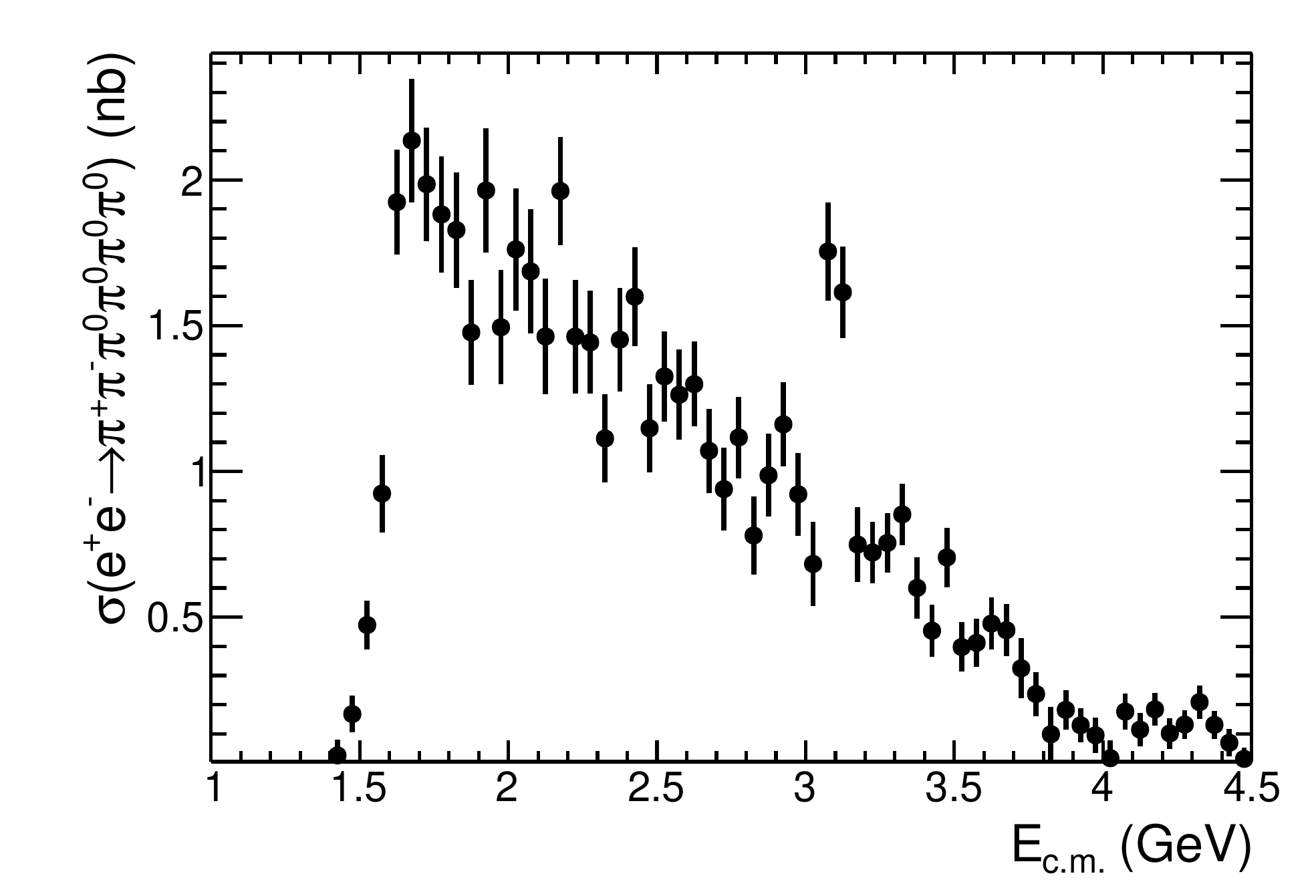}
\vspace{-0.5cm}
\caption{
The  measured $\epem\to\pipi\ppz\ppz$ cross section.
The uncertainties are statistical only.
}
\label{2pi4pi0_ee_babar}
\end{center}
\end{figure} 

\input{xs2pi4pi0_table}

\begin{figure*}[tbh]
\begin{center}
\includegraphics[width=0.315\linewidth]{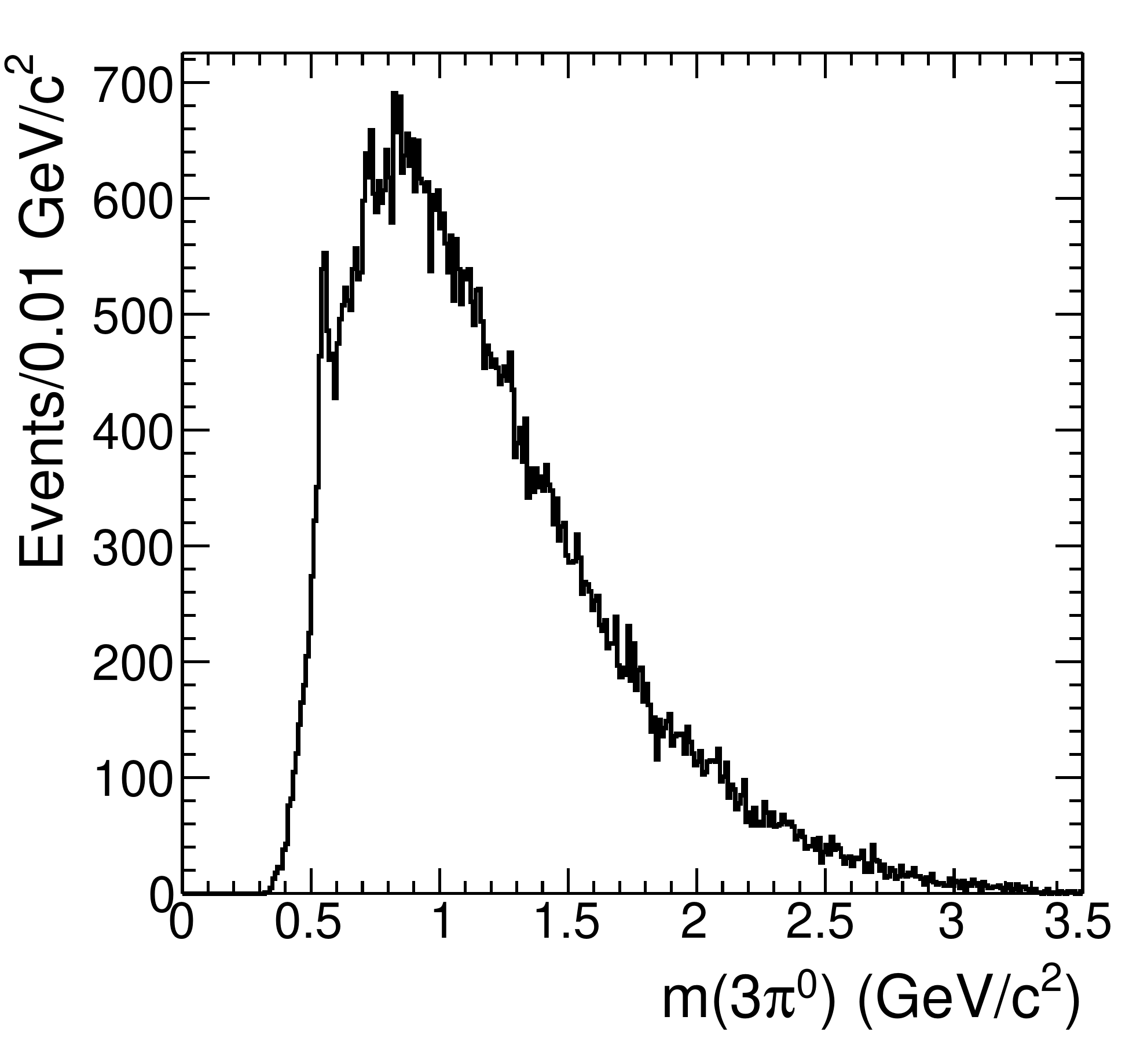}
\put(-50,120){\makebox(0,0)[lb]{\bf(a)}}
\includegraphics[width=0.315\linewidth]{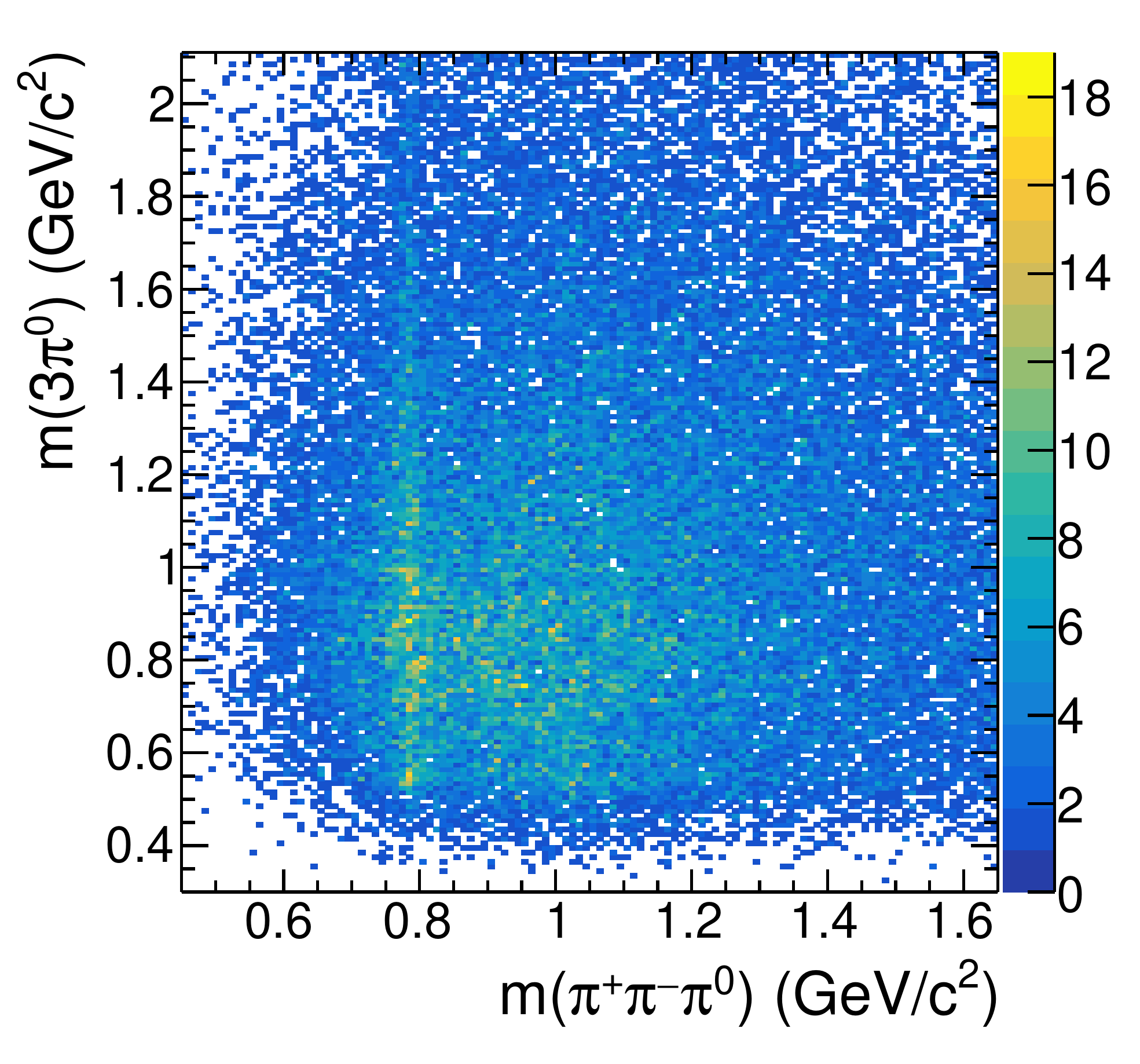}
\put(-50,120){\makebox(0,0)[lb]{\bf(b)}}
\includegraphics[width=0.315\linewidth]{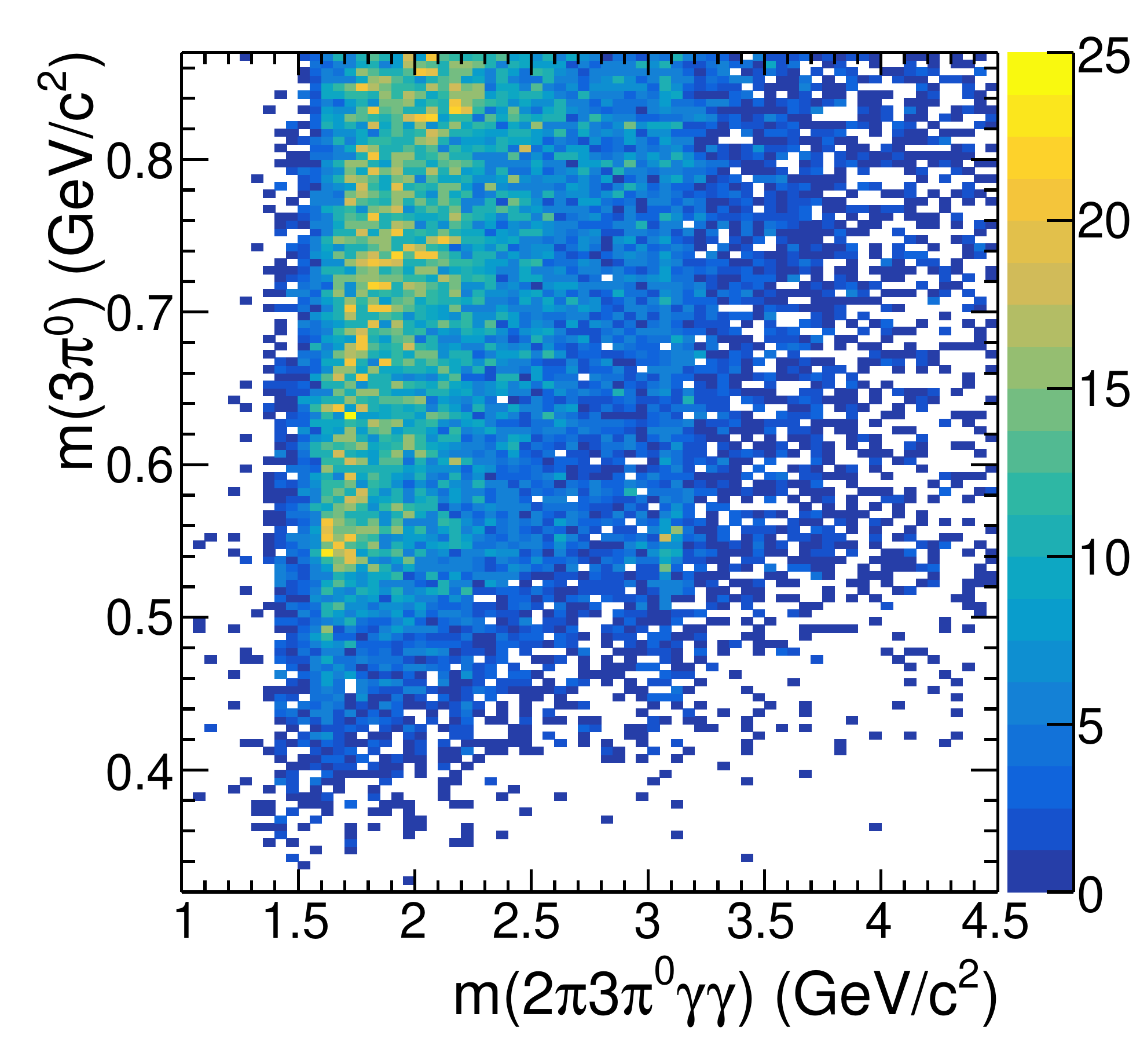}
\put(-135,120){\makebox(0,0)[lb]{\bf(c)}}
\caption{
(a) The $\ppz\piz$  invariant mass (four combinations per event).
(b) The $\ppz\piz$ vs the $\pipi\piz$ invariant mass.
(c)  The $\ppz\piz$  invariant mass vs the six-pion invariant mass.
}
\label{3pi0vs7pi}  
\end{center}
%
\begin{center}
\includegraphics[width=0.315\linewidth]{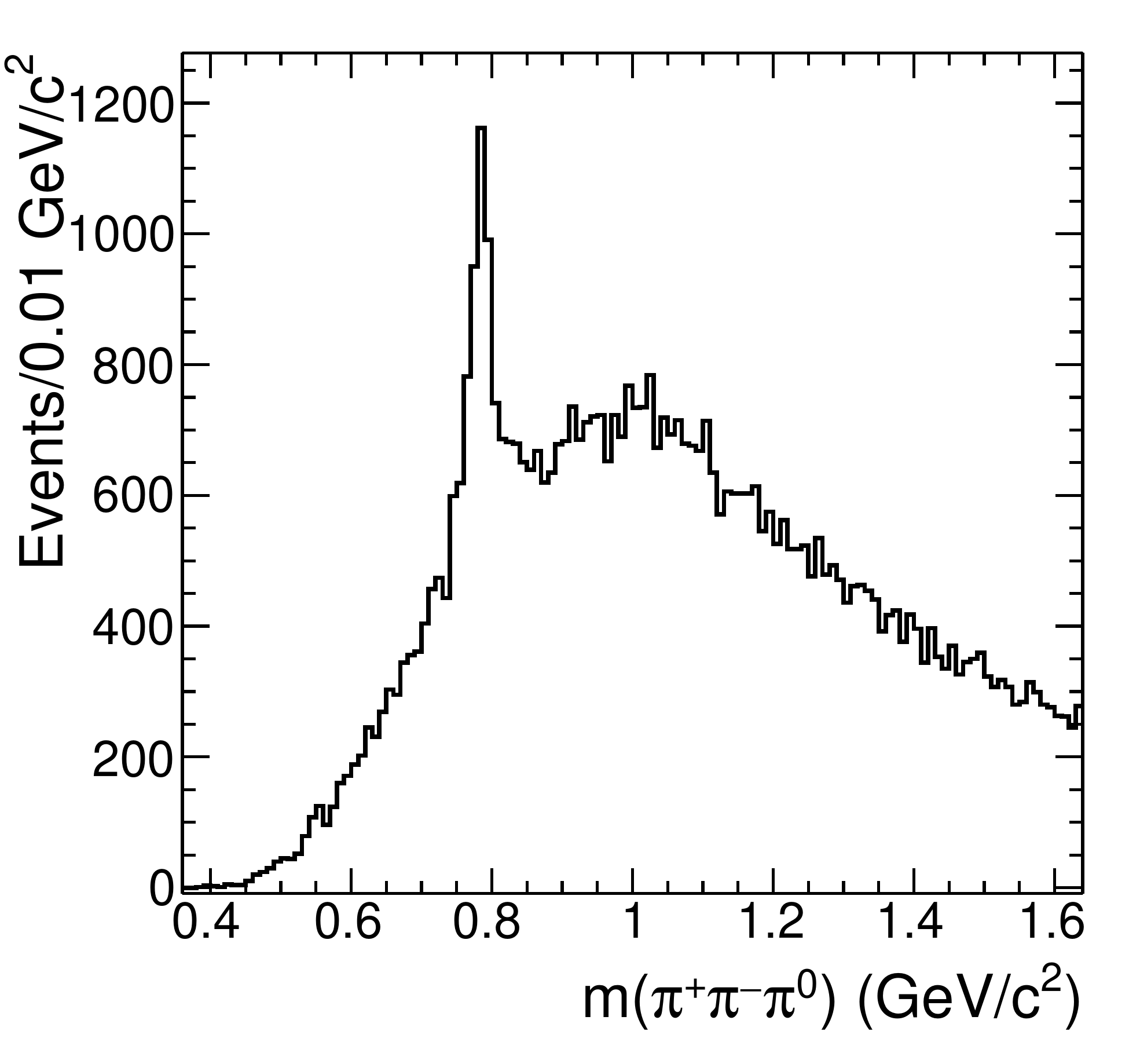}
\put(-50,120){\makebox(0,0)[lb]{\bf(a)}}
\includegraphics[width=0.315\linewidth]{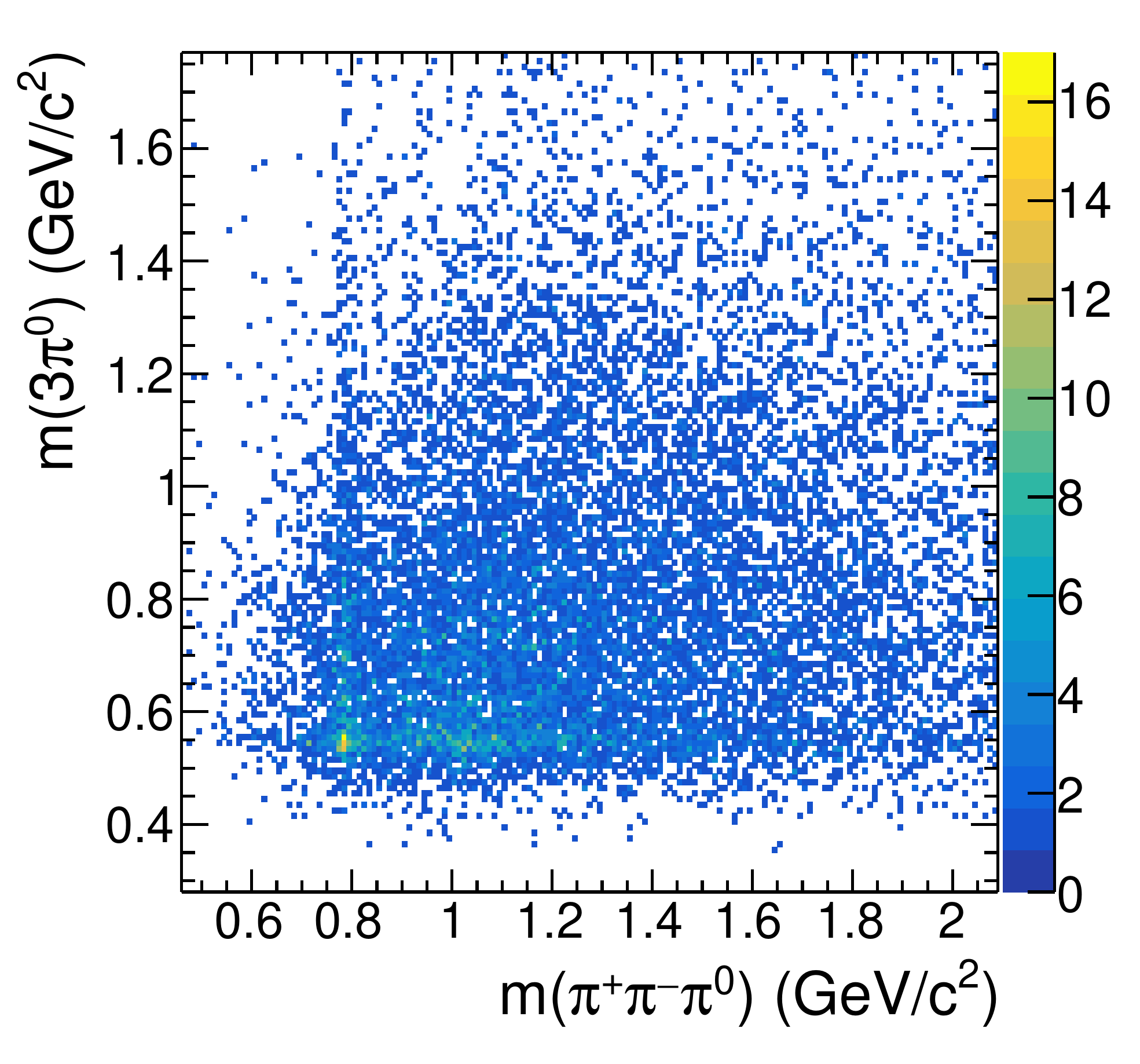}
\put(-130,120){\makebox(0,0)[lb]{\bf(b)}}
\includegraphics[width=0.315\linewidth]{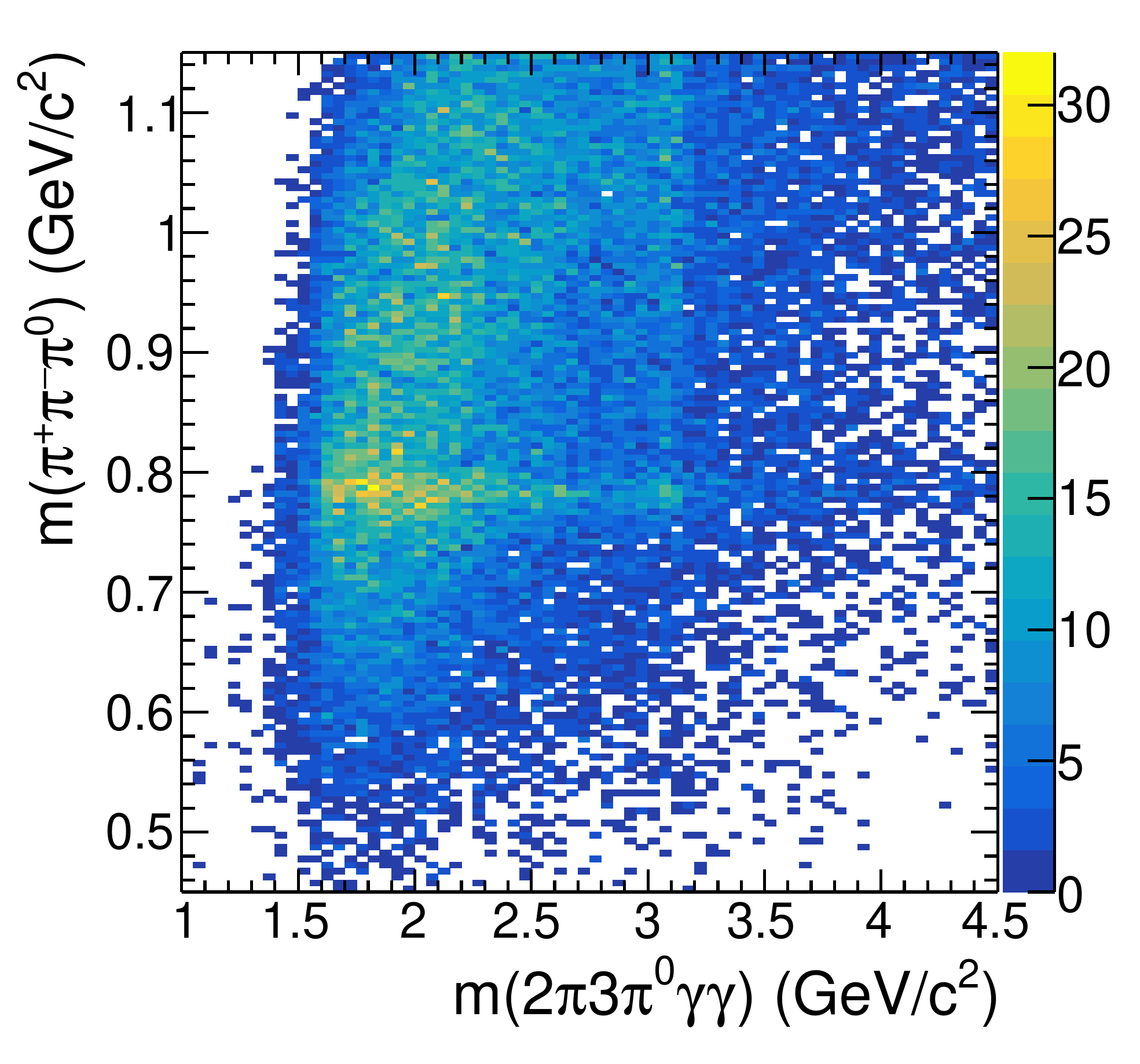}
\put(-135,120){\makebox(0,0)[lb]{\bf(c)}}
\caption{
(a) The $\pipi\piz$  invariant mass (four combinations per event).
(b) The same as Fig.~\ref{3pi0vs7pi}(b) but the $\ppz\piz$ invariant mass closest
 to the $\eta$ mass is selected (one entry per event).
(c)  The $\pipi\piz$  invariant mass vs the six-pion invariant mass.
}
\label{3pivs7pi}  
\end{center}
%
\begin{center}
\includegraphics[width=0.315\linewidth]{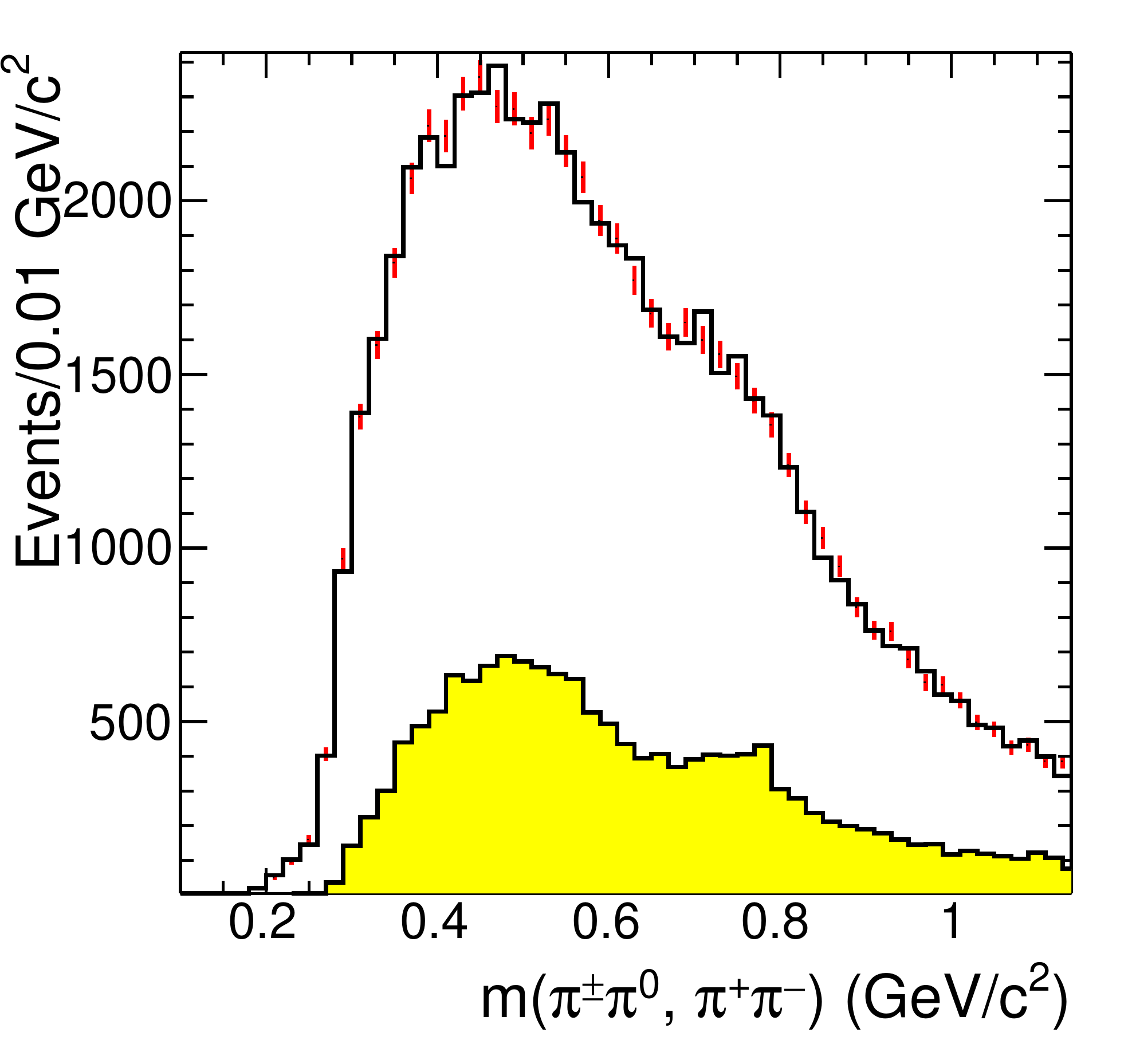}
\put(-50,120){\makebox(0,0)[lb]{\bf(a)}}
\includegraphics[width=0.315\linewidth]{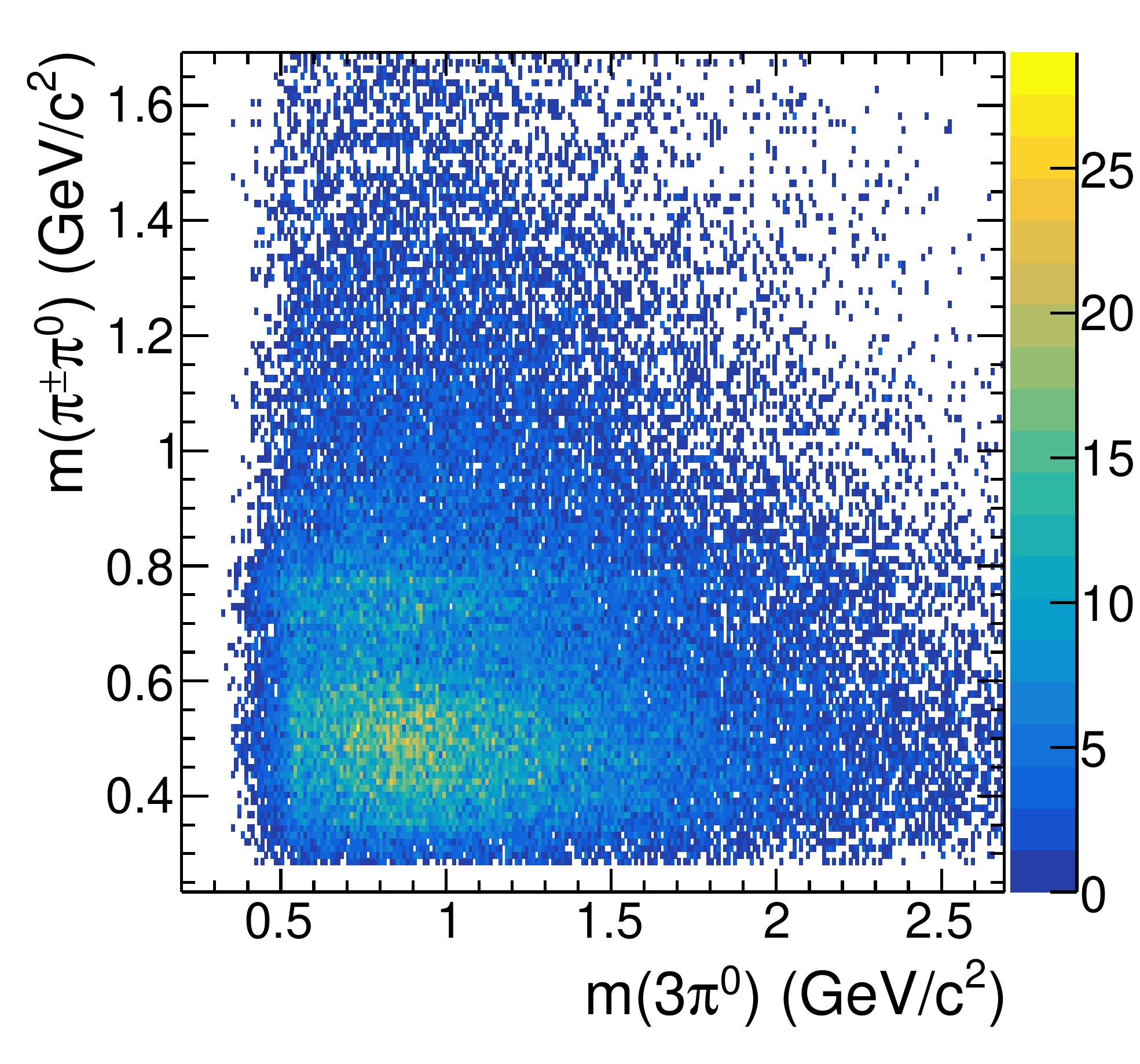}
\put(-35,120){\makebox(0,0)[lb]{\bf(b)}}
\includegraphics[width=0.315\linewidth]{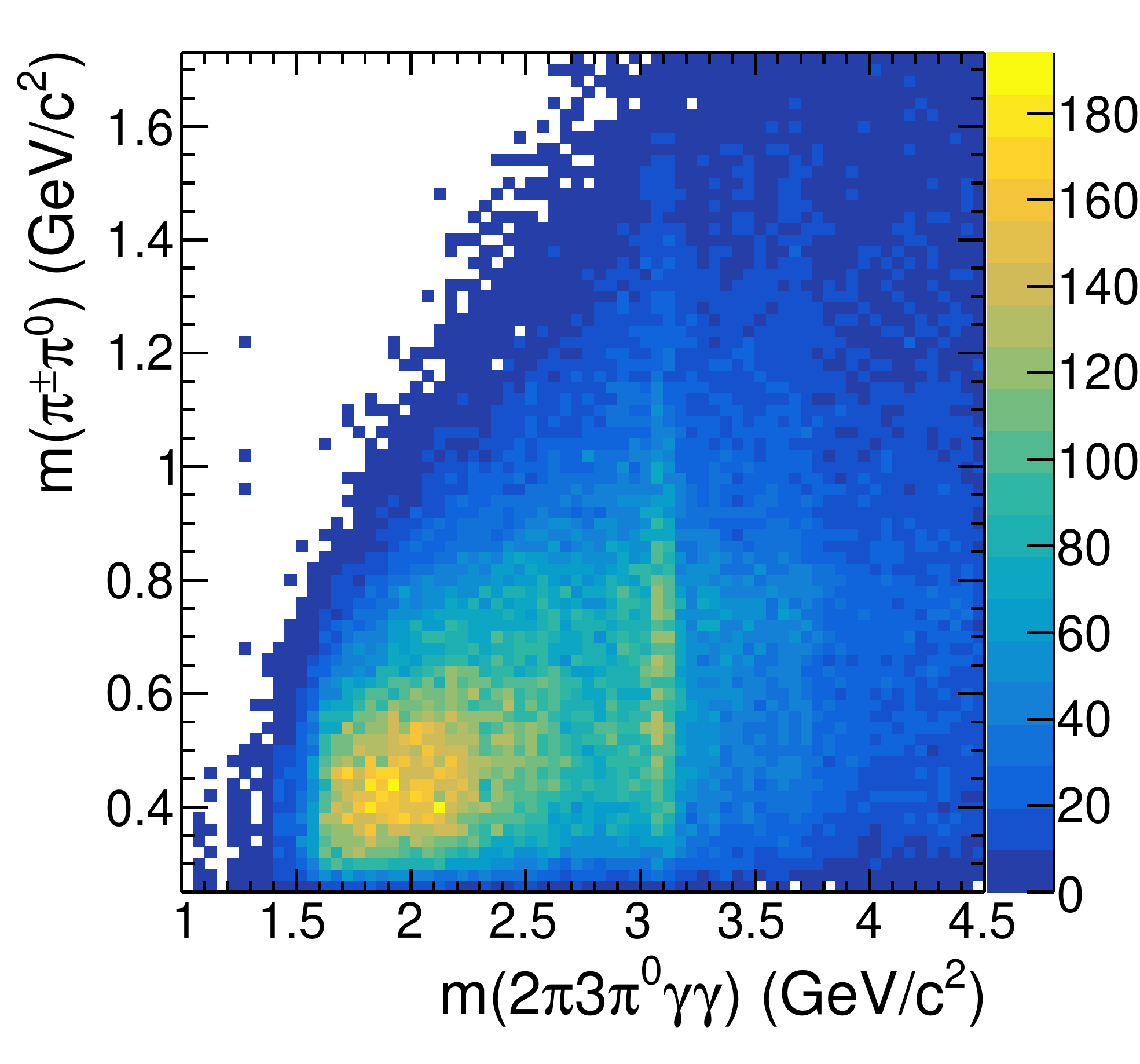}
\put(-130,120){\makebox(0,0)[lb]{\bf(c)}}
\caption{
(a) The $\pipi$ (shaded), $\pi^+\piz$ (solid), and $\pi^-\piz$ (points with errors)  invariant masses
(four combinations per event for $\rho^+$ and for $\rho^-$ ). 
(b) The $\pi^{\pm}\piz$ vs the $3\piz$ invariant mass (eight combinations per event).
(c)  The $\pi^{\pm}\piz$  invariant mass vs the six-pion invariant mass.
}
\label{pipi0vs7pi}  
\end{center}
\end{figure*}

\subsection{\boldmath Cross section for $\epem\to \pipi\ppz\ppz$}
\label{2pi4pi0}

The $\epem\to\pipi\ppz\ppz$ Born cross section is  determined from
\begin{equation}
  \sigma(2\pi4\piz)(\Ecm)
  = \frac{dN_{6\pi\gamma}(\Ecm)}
         {d{\cal L}(E_{\rm c.m.})\epsilon_{6\pi}^{\rm corr}
          \epsilon_{6\pi}^{\rm MC}(E_{\rm c.m.})(1+\delta_R)}\ ,
\label{xseq}
\end{equation}
where $\Ecm$ is the invariant mass of
the six-pion system; $dN_{6\pi\gamma}$ is the background-subtracted number of selected
six-pion events in the interval $dE_{\rm c.m.}$,  and
$\epsilon_{6\pi}^{\rm MC}(E_{\rm c.m.})$ is the corresponding detection
efficiency from simulation. The factor $\epsilon_{6\pi}^{\rm corr}$ 
accounts for the difference between data and
simulation in the tracking
(1.0$\pm$1.0\%/per track)~\cite{isr4pi} and $\piz$
(3.0$\pm$1.0\% per pion)~\cite{isr2pi2pi0} reconstruction efficiencies.  
The ISR differential luminosity, $d{\cal L}$, is calculated using the 
total integrated \babar~ luminosity of 469 fb$^{-1}$~\cite{isr3pi}.
The initial- and final-state soft-photon emission is accounted for
by the radiative correction factor $(1+\delta_R)$, which is
close to unity within a percent level for our selection criteria.
The cross section results contain the effect of
 vacuum polarization because this effect is not accounted for in
the luminosity calculation.

Our results for the $\epem\to\pipi\ppz\ppz$ cross section
are shown in Fig.~\ref{2pi4pi0_ee_babar}.  The cross section exhibits a
structure around 1.7~\gev with  a peak value of about 2~nb, 
followed by a monotonic decrease toward higher energies. 
Because we present our data in bins of width 0.050~\gevcc, compatible   
with the experimental resolution, we do not apply an unfolding procedure to the data.
Numerical values for the cross section are presented in Table~\ref{2pi4pi0_tab}.
The $J/\psi$ region is discussed later.

\subsection{\boldmath Summary of the systematic studies}
\label{sec:Systematics}
The systematic 
uncertainties, presented in the previous sections, are summarized in
Table~\ref{error_tab},  along 
with the corrections that are applied to the measurements.
\begin{table}[tbh]
\caption{
Summary of the systematic uncertainties in the $\epem\to
\pipi\ppz\ppz$ cross section measurement. The total uncertainly is computed assuming no correlations.
}
\label{error_tab}
\begin{tabular}{l c c} 
\hline
Source & Correction & Uncertainty\\
\hline
Luminosity  &  --  &  $1\%$ \\
  MC-data difference in\\
  ISR photon efficiency & +1.5\%  & $1\%$\\ 
\chisq cut uncertainty & -- & $3\%$\\
Fit and background subtraction & -- &  $10\%$ \\
MC-data difference in track losses & $+2\%$ & $1\%$ \\
MC-data difference in $\pi^0$ losses & $+12\%$ & $4\%$ \\
Radiative corrections accuracy & -- & $1\%$ \\
Efficiency from MC \\(model-fit-dependent) & -- & $5\%$  \\
\hline
Total    &  $+15.5\%$   & $12.4\%$  \\
\end{tabular}
\end{table}

The three corrections applied to the cross sections sum
up to 15.5\%. The systematic uncertainties are estimated as
 12.4\%.
The largest systematic uncertainty arises from the fitting
and background subtraction procedures.
This is estimated by varying the background levels and the parameters of the functions used.

\subsection{Overview of the intermediate structures}
The $\epem\to\pipi\ppz\ppz$ process has a rich internal
substructure. 
To study this substructure, we impose the restriction   $\mgg <
0.35$~\gevcc, eliminating the region populated
by $\epem\to\pipi\ppz\piz\eta$, but with some level of the background remaining.  We then assume that
the $m(\pipi3\piz\gamma\gamma)$ invariant mass can be taken to
represent $m(\pipi4\piz)$.

Figure~\ref{3pi0vs7pi}(a) shows the distribution of the $\ppz\piz$
invariant mass (four entries per event).  The distribution is seen to exhibit a  
prominent $\eta$ peak, which is due to the
$\epem\to\eta\pipi\piz$ reaction.
Figure~\ref{3pi0vs7pi}(b) presents a scatter plot of the $\ppz\piz$
vs the $\pipi\piz$ invariant mass.
From this plot, the $\omega\eta$ intermediate state is seen.
Figure~\ref{3pi0vs7pi}(c)  presents a scatter plot of the $3\piz$
invariant mass versus $m(\pipi3\piz\gamma\gamma)$.

The distribution of the $\pipi\piz$ invariant mass (four entries per event)
is shown in Fig.~\ref{3pivs7pi}(a).  A prominent $\omega$ peak from
$\epem\to\omega3\piz$ is seen.
The scatter plot in Fig.~\ref{3pivs7pi}(b) shows  $\ppz\piz$  vs
the $\pipi\piz$ invariant mass for events from Fig.~\ref{3pi0vs7pi}(b)
when only the $3\piz$
combination with the invariant mass closest to the nominal $\eta$ mass is kept. Correlated
$\eta$ and $\omega$ production is seen. 
A scatter plot of the $\pipi\piz$ vs the $\pipi3\piz\gamma\gamma$
 mass is shown in Fig.~\ref{3pivs7pi}(c).
A clear signal for a $J/\psi$ peak is also observed.

Figure~\ref{pipi0vs7pi}(a) shows the
$\pi^+\piz$(solid) and $\pi^-\piz$(points) invariant masses (four entries per
event). Prominent $\rho(770)^{\pm}$ peaks, corresponding to  
$\epem\to\rho^{\pm}\pi^{\mp}3\piz$ (or $\rho^{\pm}\rho^{\mp}2\piz$),
are visible. The shaded histogram shows the presence of the $\rho^{0}$ signal.
The scatter plot in Fig.~\ref{pipi0vs7pi}(b) shows the $\pi^{\pm}\piz$  vs
the $3\piz$ invariant mass.  An indication of the $\rho^{\pm}\pi^{\mp}\eta$ (or $\rho^{0}\eta\piz$ - not shown)
intermediate state is visible. Figure~\ref{pipi0vs7pi}(c) shows
the $\pi^{\pm}\piz$ invariant mass  vs the six-pion invariant mass: a clear signal for
the $J/\psi$ and an indication for the $\psi(2S)$  are seen. 

\begin{figure}[tbh]
  \begin{center}
    \vspace{-0.3cm}
  \includegraphics[width=0.95\linewidth]{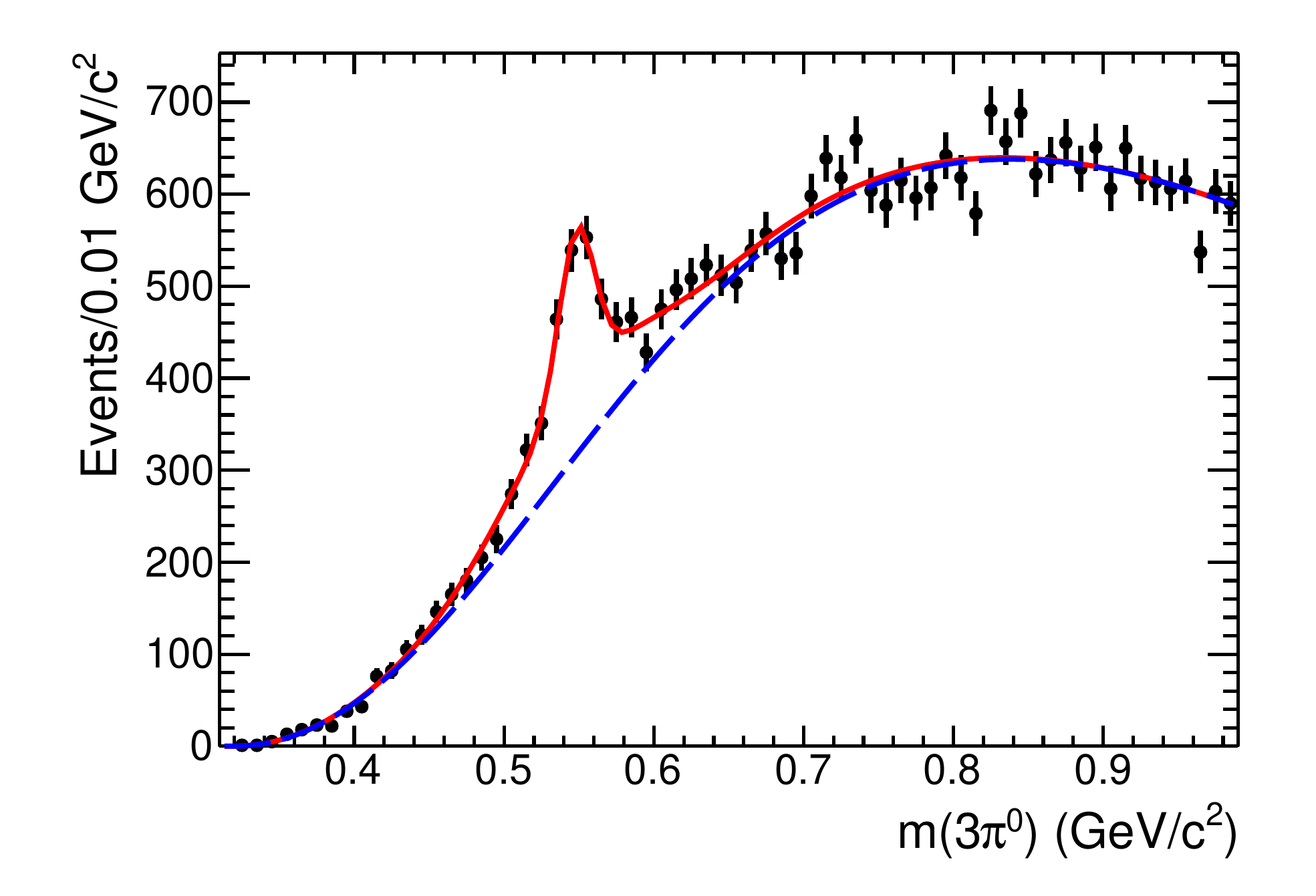}
   \vspace{-0.5cm}
\caption{The $3\piz$ invariant mass for data. 
  The curves show the fit functions.  The solid curve shows 
the $\eta$ peak (based on MC simulation) plus the non-$\eta$ continuum
background (dashed).
}
\label{3pi0slices}
\end{center}
%
%
\begin{center}
  \vspace{-0.2cm}
  \includegraphics[width=0.95\linewidth]{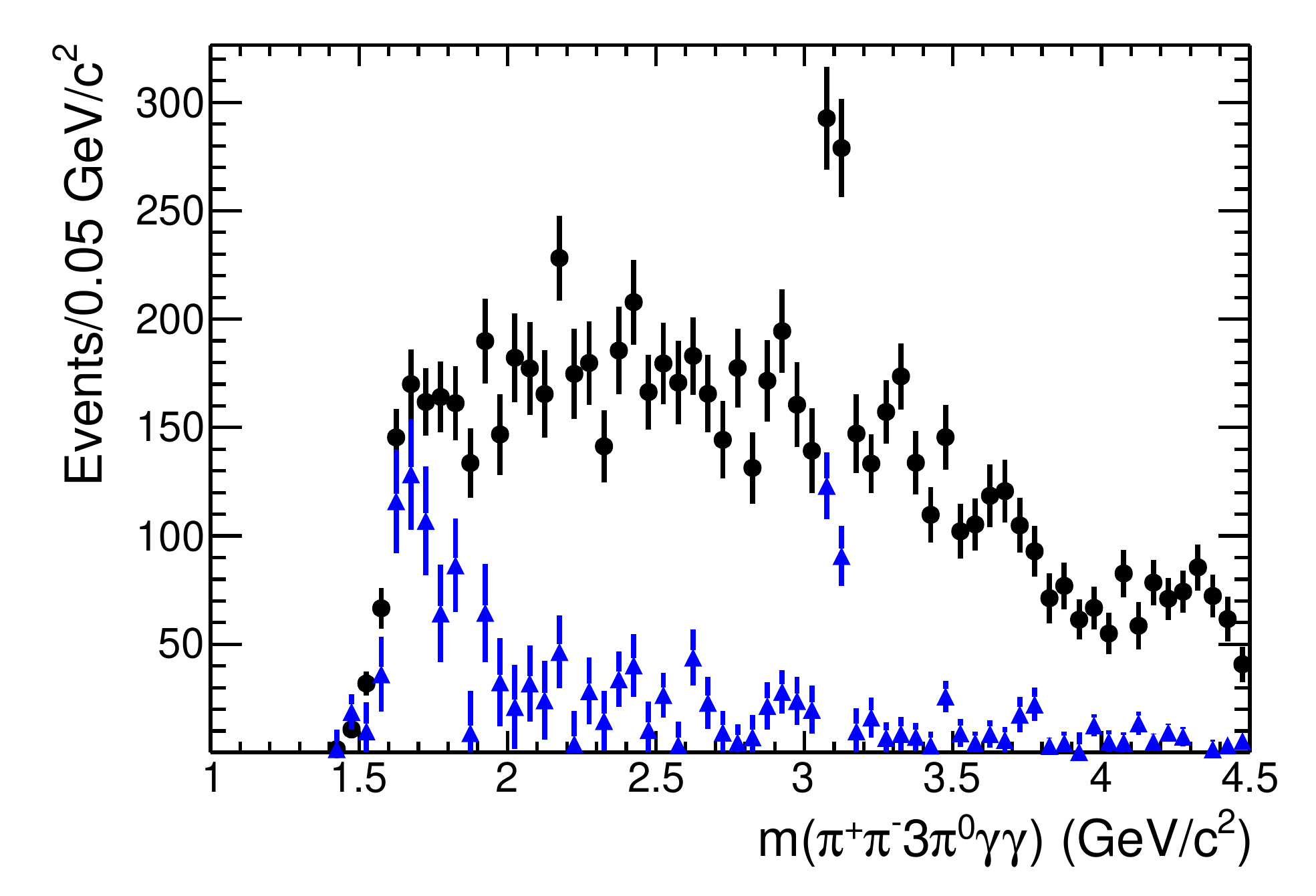}
   \vspace{-0.5cm}
\caption{ The $m(\pipi4\piz)$ invariant mass dependence of the selected data events
for $\epem\to\eta\pipi\piz, \eta\to 3\piz$ (triangles) in comparison with
all six-pion events (dots).
}
\label{neveta2pi}
\end{center}
\end{figure}
\begin{figure}[tbh]
\begin{center}
  \includegraphics[width=1.0\linewidth]{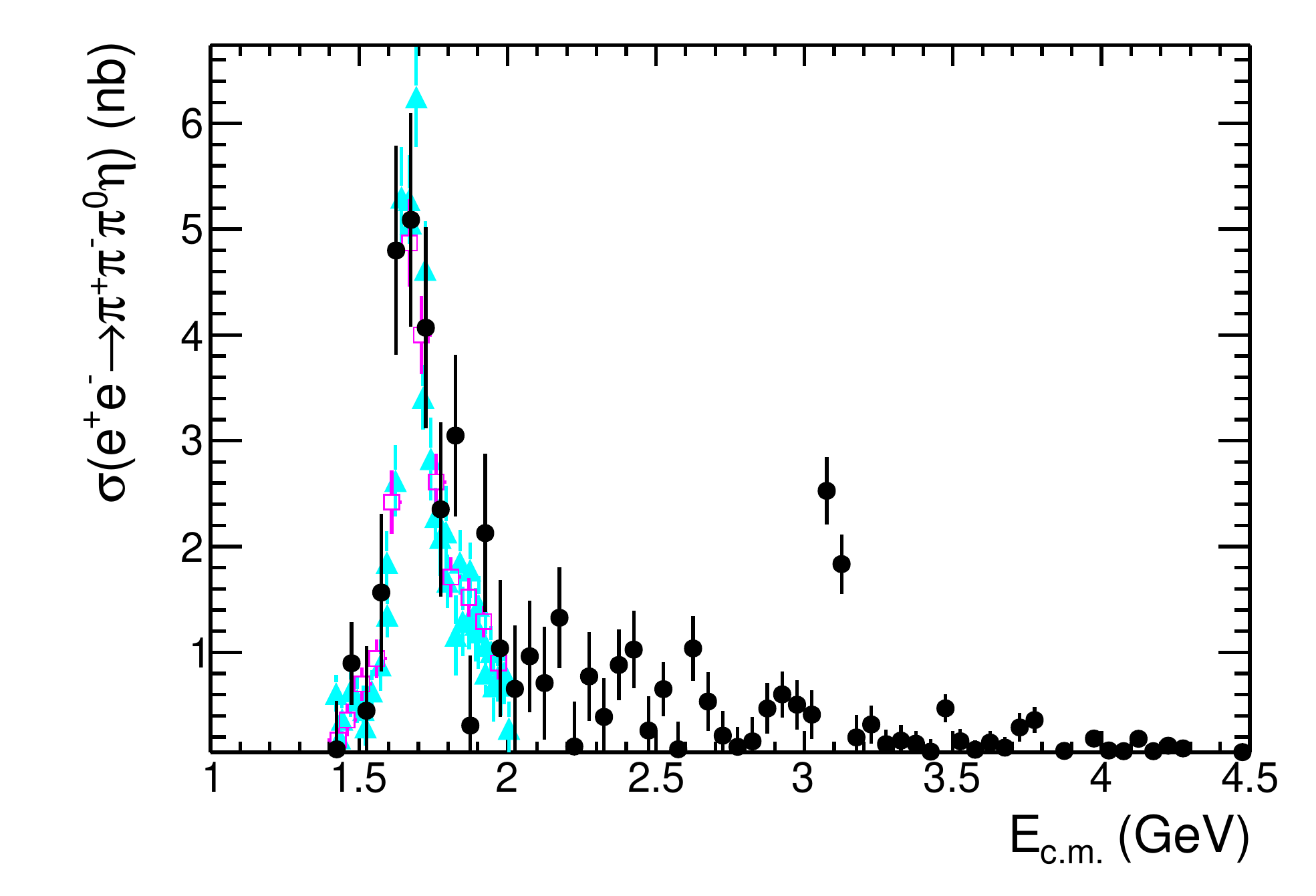}
  \put(-140,82){
  \includegraphics[width=0.5\linewidth,height=0.3\linewidth]{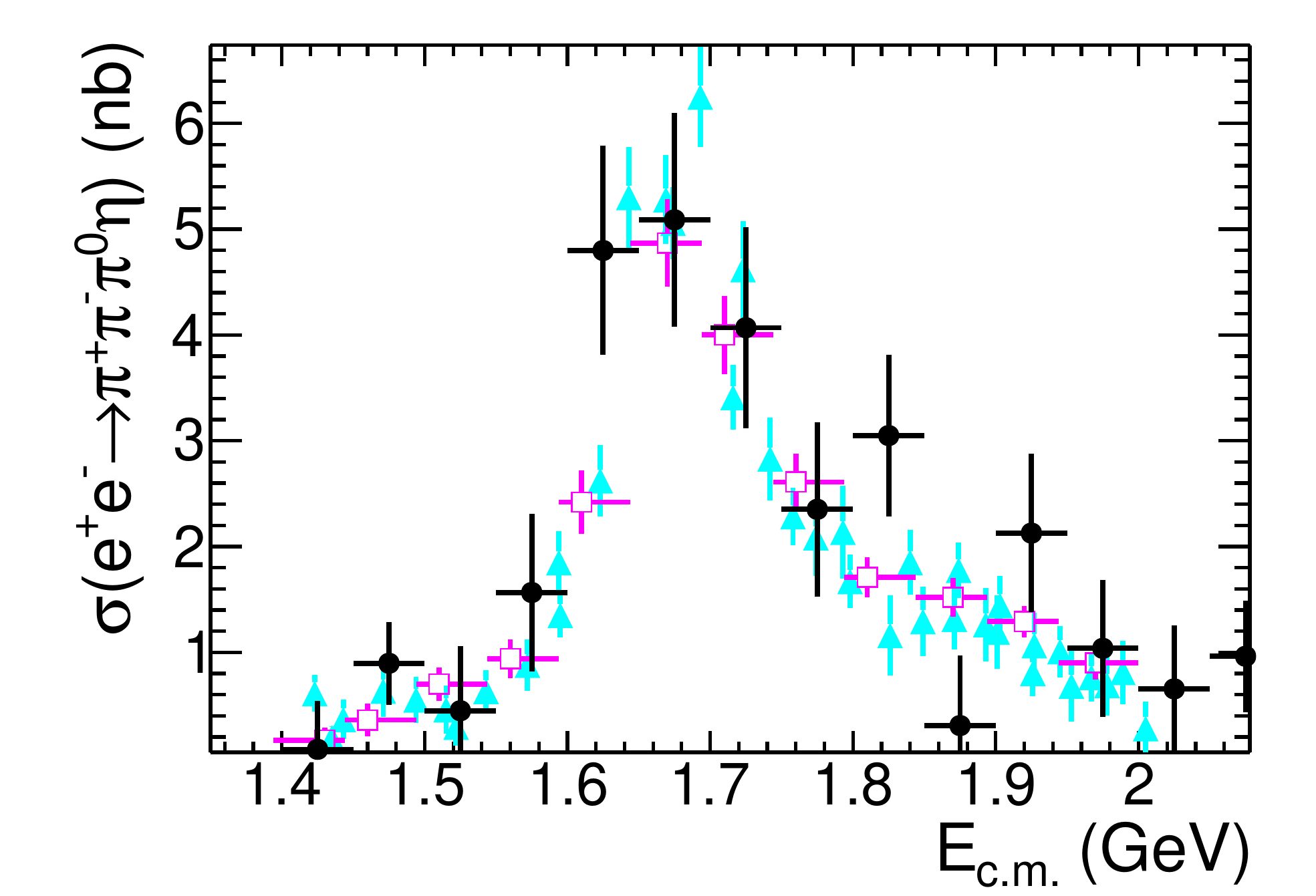}}
  \vspace{-0.5cm}
\caption{Comparison of the current results (dots) for
the $\epem\to\pipi\piz\eta$ cross section with those
from the SND experiment with $\eta\to\gamma\gamma$,  shown by squares~\cite{sndeta3pi} and
with those from the CMD-3 experiment, also based on
$\eta\to\gamma\gamma$, shown by triangles~\cite{cmdeta3pi}. 
The insert shows an expanded view of the resonant region.
}
\label{xs_2pieta}
\end{center}
\end{figure}
\input{xs3pieta_table}

\subsection{\bf\boldmath The $\eta\pipi\piz$ intermediate state}
\label{Sec:etapipi}
To determine the contribution of the $\eta\pipi\piz$
 intermediate state, we fit the events of Fig.~\ref{3pi0vs7pi}(a)
using a triple-Gaussian function to describe the
signal peak, as in Fig.~\ref{m3pi_omega_eta_mc}(c), and a polynomial to
describe the background.
The result of the fit is shown in Fig.~\ref{3pi0slices}.
We obtain $1539\pm89$ $\eta\pipi\piz$ events. 
The number of $\eta\pipi\piz$ events as a function of the six-pion invariant mass
is determined by performing an analogous
fit to the  events in Fig.~\ref{3pi0vs7pi}(c) in each 0.05~\gevcc interval of $m(\pipi4\piz)$.  
The resulting distribution is shown in Fig.~\ref{neveta2pi} by
triangles in comparison with all $\pipi4\piz$ events (dots).
\input{xsomegaeta_table}

Using Eq.~(\ref{xseq}), we determine the cross section for the
$\epem\to\eta\pipi\piz$ process.
The results, which account for the
$\eta\to 3\piz$ branching fractions of 0.327, are reported in Fig.~\ref{xs_2pieta}
and Table~\ref{3pieta_table}.
Systematic uncertainties in this measurement are the same as those listed in Table~\ref{error_tab}.
Figure~\ref{xs_2pieta} shows our measurement
in comparison to the SND result~\cite{sndeta3pi}  and to
those from the CMD-3 experiment~\cite{cmdeta3pi}.
These previous results are based on a different
$\eta$ decay mode from that considered here.
The insert shows an expanded view for the c.m.\ energies below 2~GeV,
where the resonance, interpreted as the $\omega(1650)$, dominates. 
The results of the three experiments are seen to agree within the
uncertainties.

\begin{figure}[tbh]
  \begin{center}
\vspace{-0.3cm}
\includegraphics[width=0.49\linewidth]{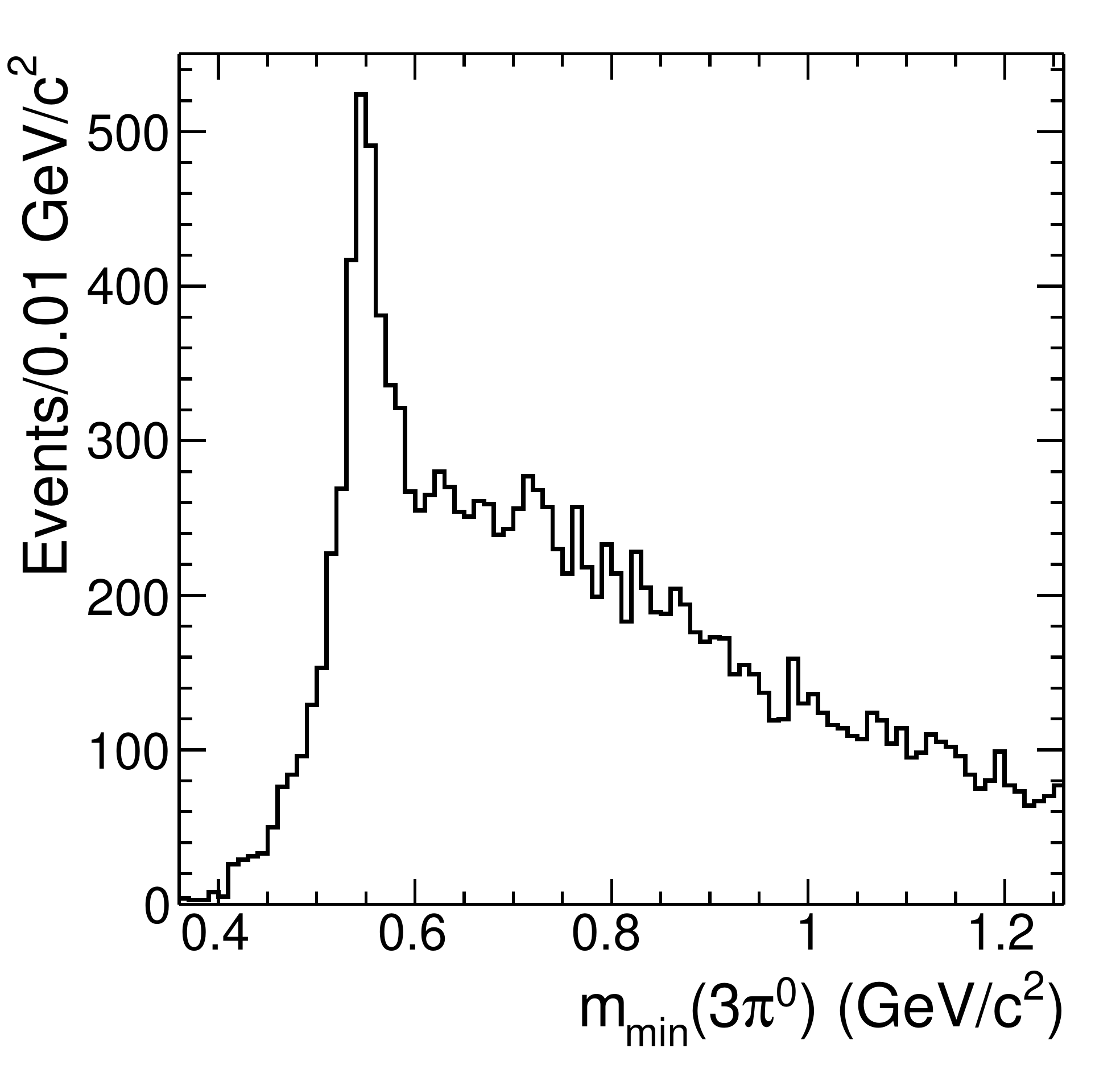}
\put(-50,90){\makebox(0,0)[lb]{\bf(a)}}
\includegraphics[width=0.49\linewidth]{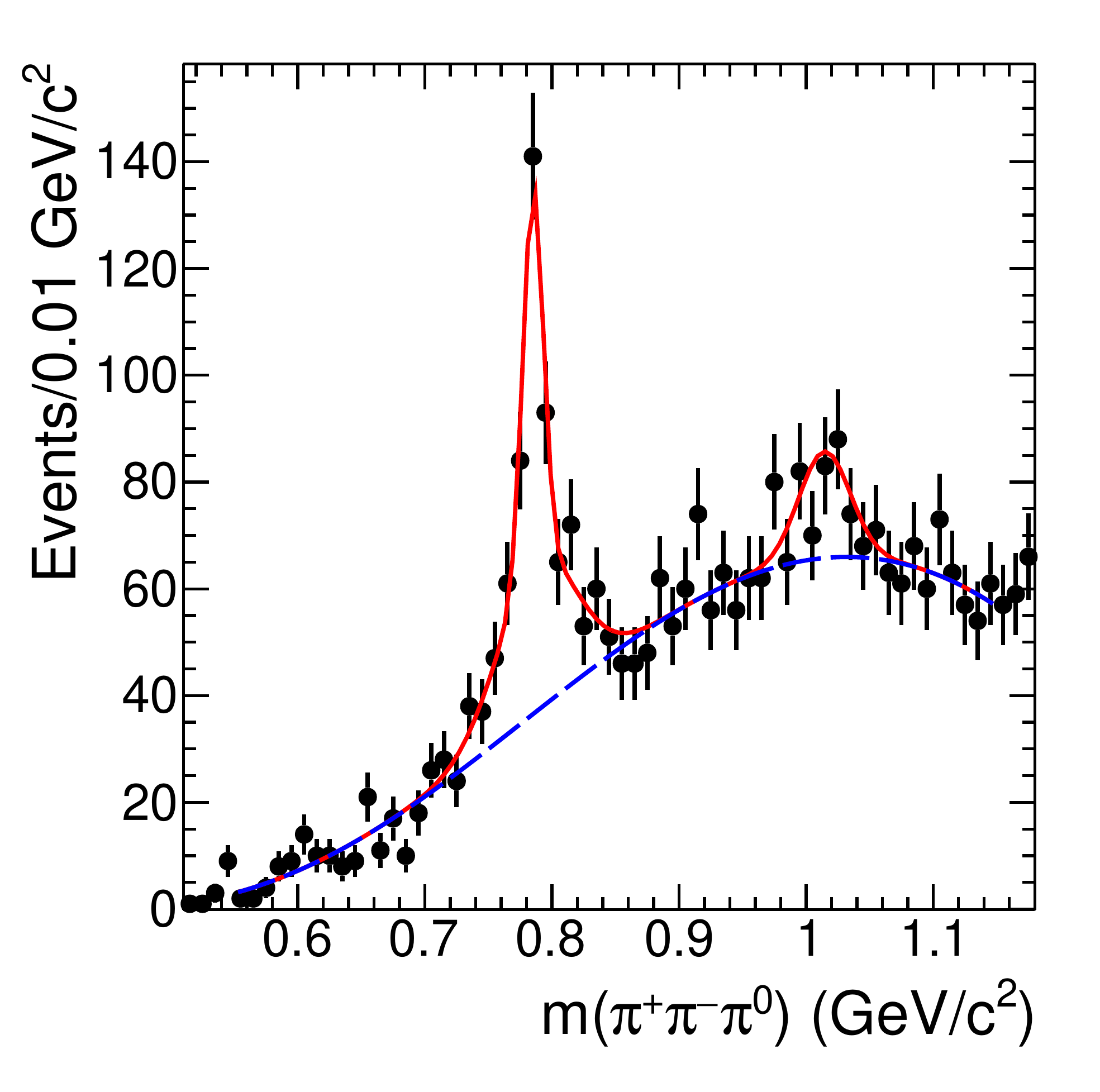}
\put(-50,90){\makebox(0,0)[lb]{\bf(b)}}
   \vspace{-0.5cm}
   \caption{(a) The $3\piz$ invariant mass closest to the $\eta$ mass.
     (b) The $\pipi\piz$ invariant mass for events with $m_\text{min}(3\piz) <$0.7~\gevcc.
  The curves show the fit functions.  The solid curve shows 
the $\omega$ and $\phi$ peak (based on MC simulation fit) plus the continuum
background (dashed).
}
\label{3pi0omega} 
\end{center}
\end{figure}
\begin{figure}[tbh]
\begin{center}
  \includegraphics[width=1.0\linewidth]{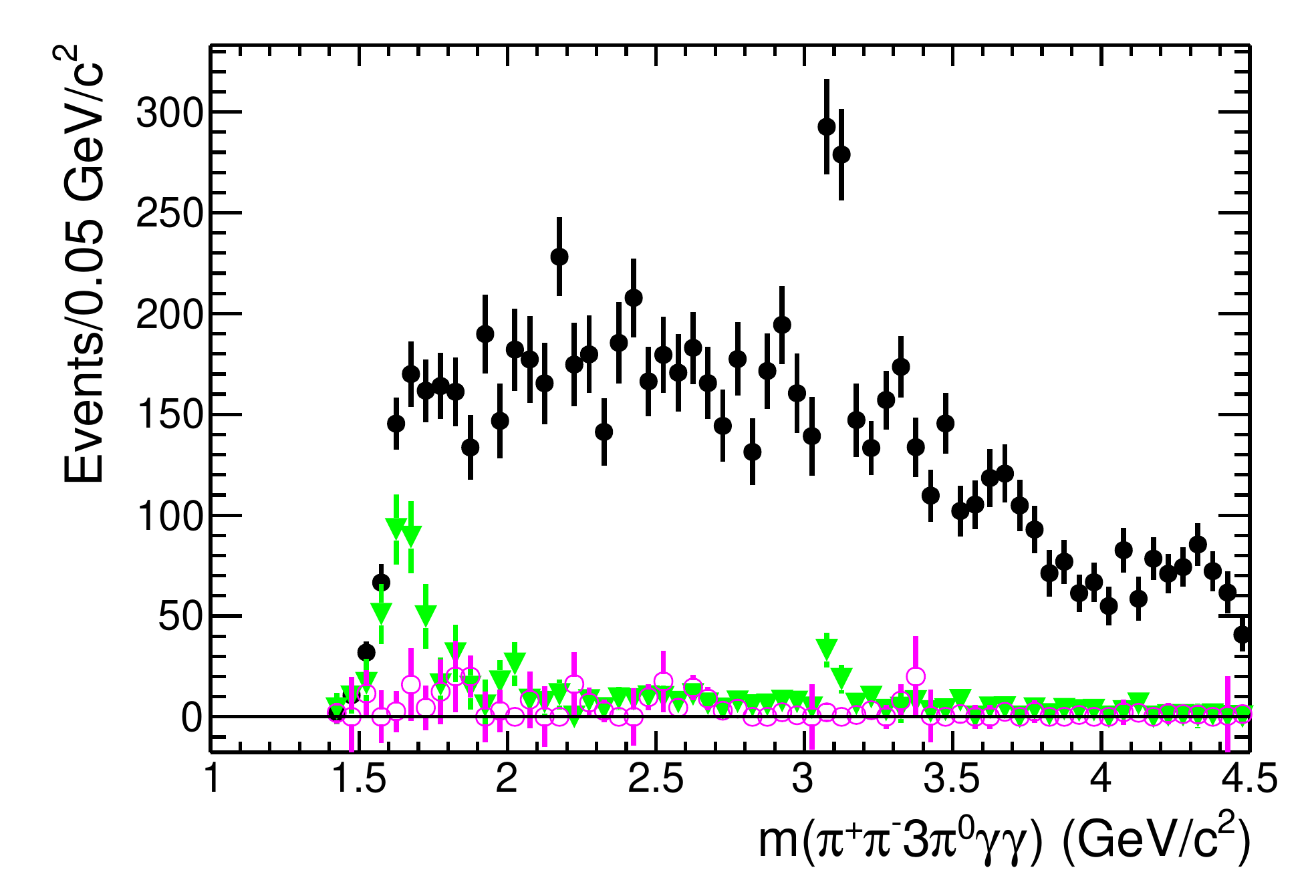}
   \vspace{-0.8cm}
\caption{ The $m(\pipi4\piz)$ invariant mass dependence of the selected data events
for $\epem\to\eta\omega, \eta\to 3\piz$ (triangles) and
$\epem\to\eta\phi, \eta\to 3\piz$ (open circles) in comparison with
all six-pion events (dots).
}
\label{nevetaomega}
\end{center}
\end{figure}
\begin{figure}[tbh]
\begin{center}
  \includegraphics[width=1.0\linewidth]{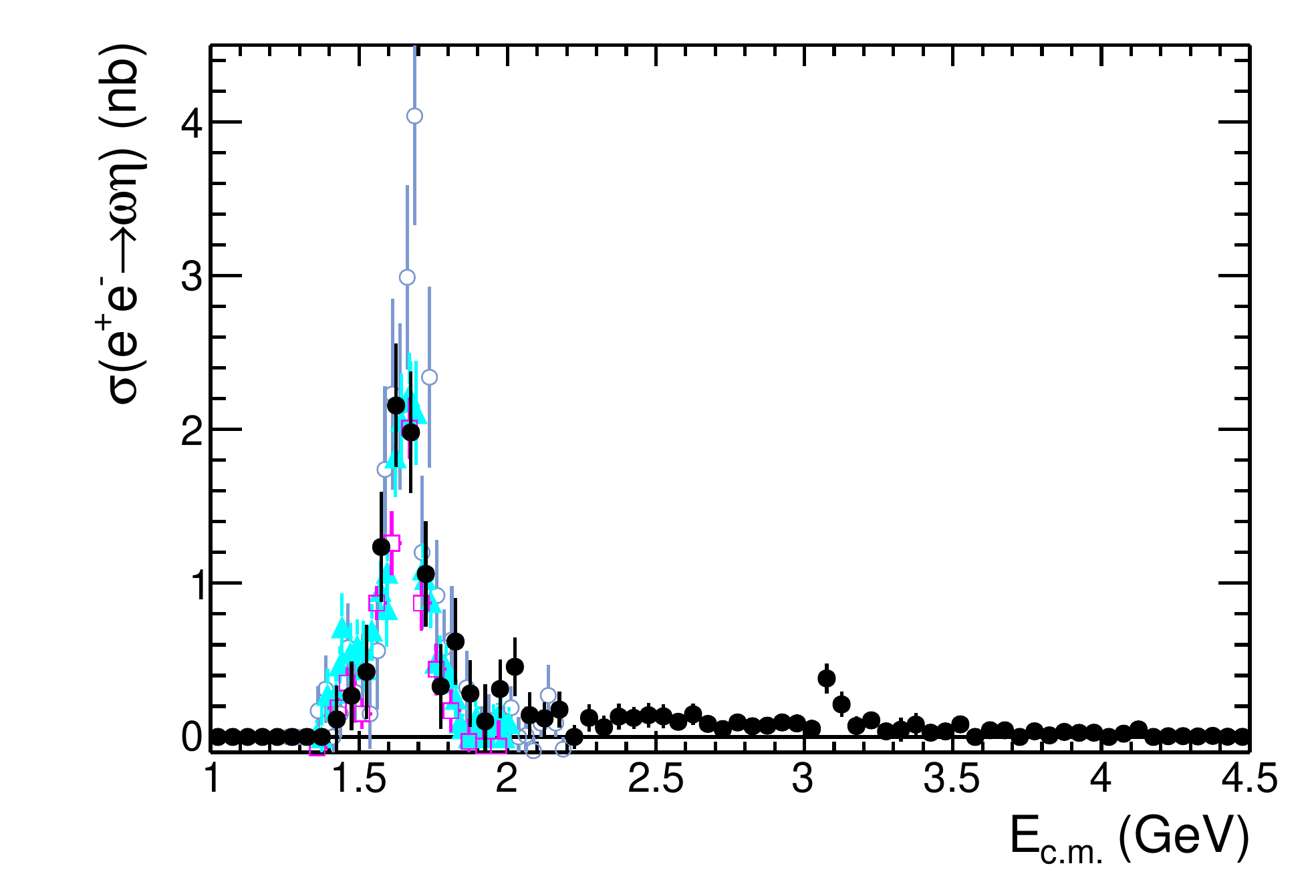}
 \put(-135,72){
  \includegraphics[width=0.47\linewidth]{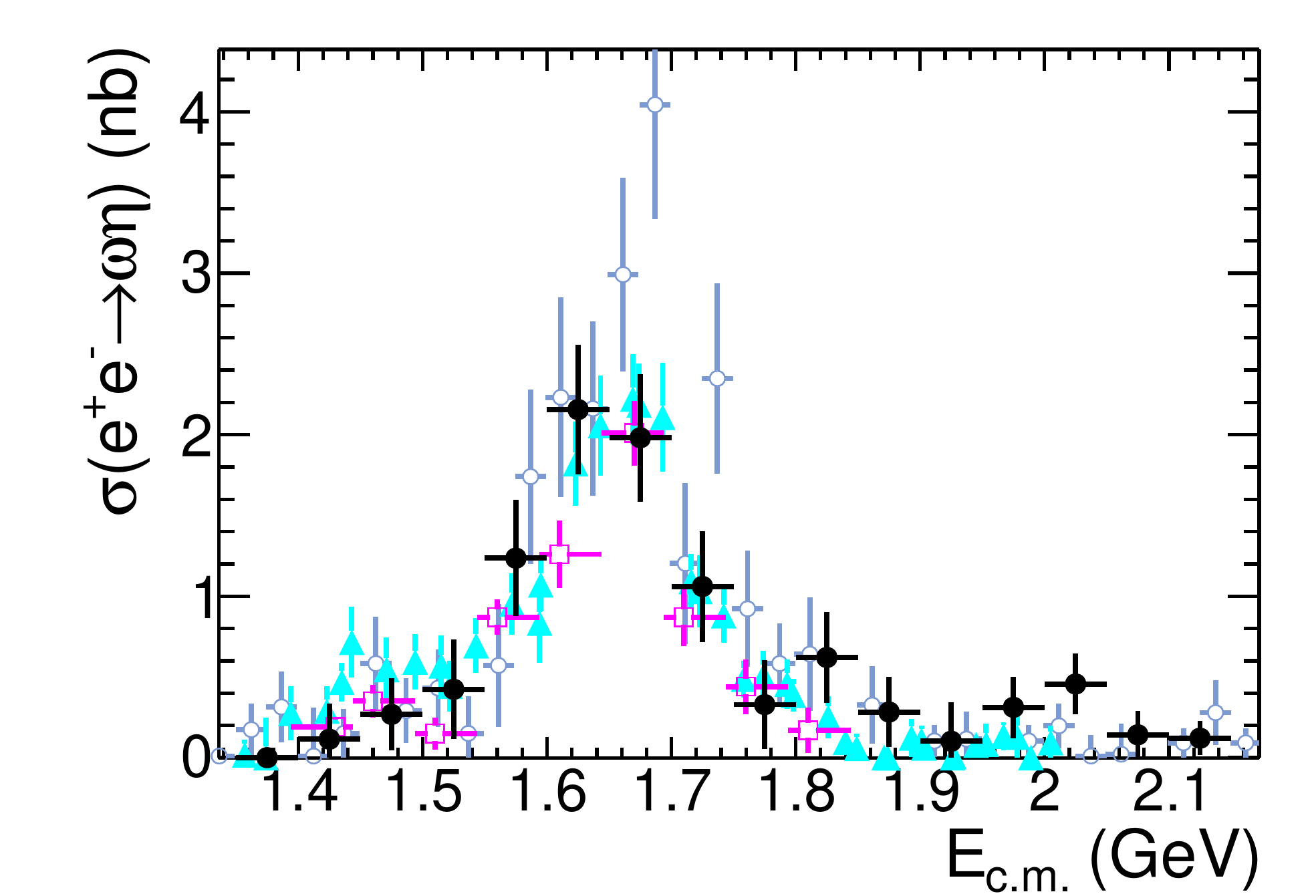}}
  \vspace{-0.5cm}
  \caption{
Comparison of the current results (dots) for
the $\epem\to\eta\omega$ cross section with those
from \babar~  with $\eta\to\pipi\piz$, shown by open circles~\cite{isr6pi},
from the SND experiment with $\eta\to\gamma\gamma$,  shown by squares~\cite{sndeta3pi}, and
with those from the CMD-3 experiment, also based on
$\eta\to\gamma\gamma$, shown by triangles~\cite{cmdeta3pi}. 
The insert shows an expanded view of the resonant region.
}
\label{xs_omegaeta}
\end{center}
\end{figure}

\subsection{\bf\boldmath The $\eta\omega$ intermediate state}
\label{Sec:etaomega}
To determine the contribution of the $\eta\omega$
intermediate state to the $\eta\pipi\piz$ events,
we select the $3\piz$ combination with the invariant
mass closest to the nominal $\eta$ mass, $m_\text{min}(3\piz)$,  and search for the
$\omega$ signal in the remaining
$\pipi\piz$ combination. Figure~\ref{3pi0omega}(a) shows the
$m_\text{min}(3\piz)$ distribution;
Fig.~\ref{3pi0omega}(b) is the distribution for the corresponding
$\pipi\piz$ invariant mass in the event.  An additional requirement $m_\text{min}(3\piz)
<$0.7~\gevcc  is applied.
Prominent $\omega$ and $\phi$ peaks are seen. The latter arises from the $\epem\to\eta\phi,
\phi\to\pipi\piz$ reaction.

 We fit the events of Fig.~\ref{3pi0omega}(b)
using a double-Gaussian function to describe the
signal from the $\omega$ and $\phi$ peaks,  and a polynomial to
describe the background.
We obtain $351\pm43$ and $100\pm32$  $\eta\omega$ and $\eta\phi$ events, respectively.
The number of $\eta\omega$ and $\eta\phi$ events as a function of the six-pion invariant mass
is determined by performing an analogous
fit to the events in each 0.05~\gevcc interval of $m(\pipi4\piz)$.  
The resulting distributions are shown in Fig.~\ref{nevetaomega}.

Using Eq.~(\ref{xseq}), we determine the cross section for the
$\epem\to\eta\omega$ process.
The results, accounting for the
$\eta$ branching fractions, are reported in Fig.~\ref{xs_omegaeta} and listed in
Table~\ref{ometa_table}.
Systematic uncertainties in this measurement are the same as those listed in Table~\ref{error_tab}.
Figure~\ref{xs_omegaeta} shows our measurement
in comparison to the \babar~ result~\cite{isr6pi} (open circles),
the SND result~\cite{sndeta3pi} (squares), and 
the CMD-3 result~\cite{cmdeta3pi} (triangles).
The insert shows an expanded view of the resonant region, where
signal from the $\omega(1650)$ dominates.
These previous results are based on different
$\eta$ decay modes ($\eta\to\pipi\piz$ for \babar,
$\eta\to\gamma\gamma$ for SND and CMD-3) from that considered here.
The  results from the different experiments are seen to agree within the 
uncertainties.
Including the results of the present study,
we have thus now measured the  $\epem\to\eta\omega$ cross
section in two different $\eta$ decay modes.

The observed contribution from the $\epem\to\eta\phi$ reaction is
small and we do not calculate its cross section.

\begin{figure}[tbh]
\begin{center}
\includegraphics[width=0.85\linewidth]{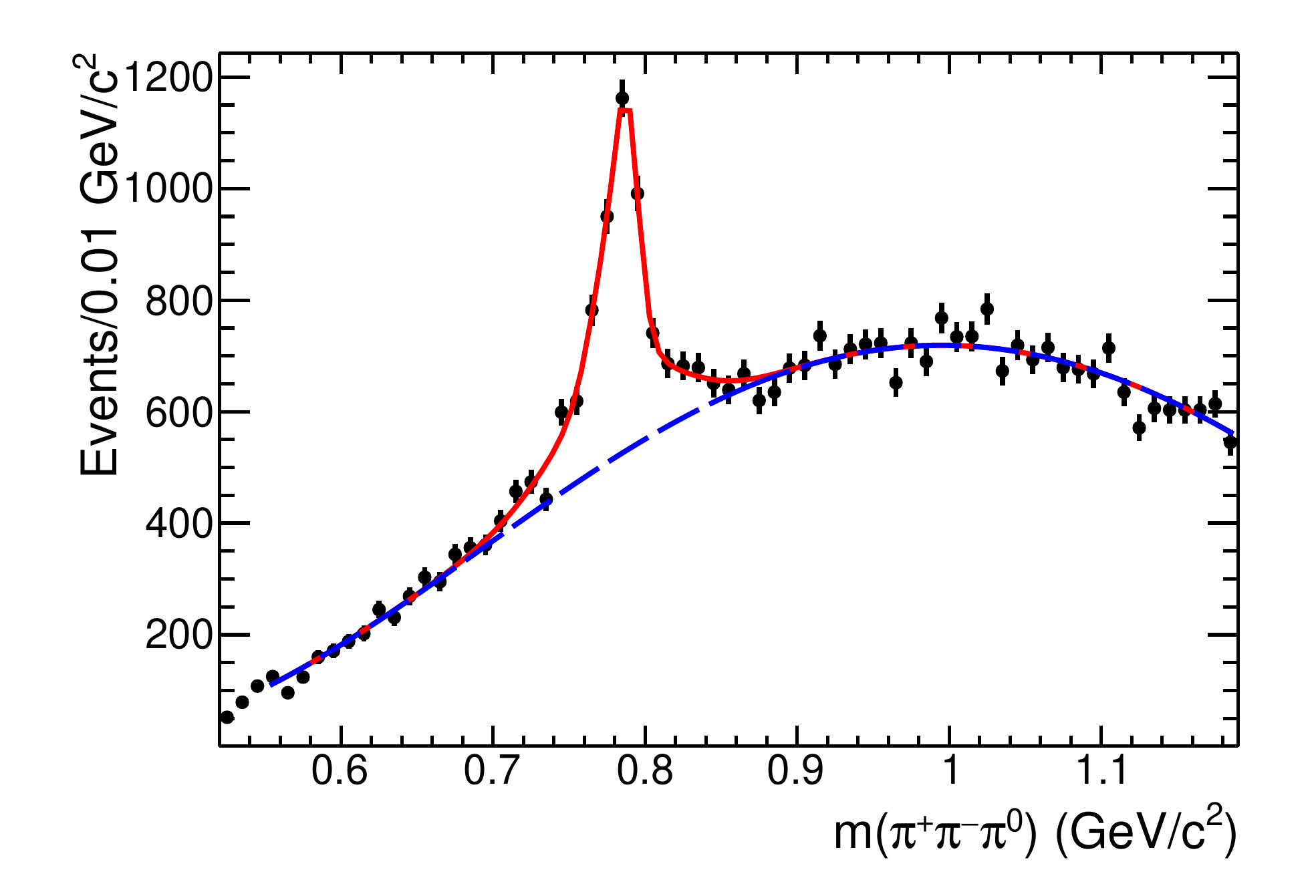}
\put(-50,100){\makebox(0,0)[lb]{\bf(a)}}\\
\vspace{-0.2cm}
\includegraphics[width=0.85\linewidth]{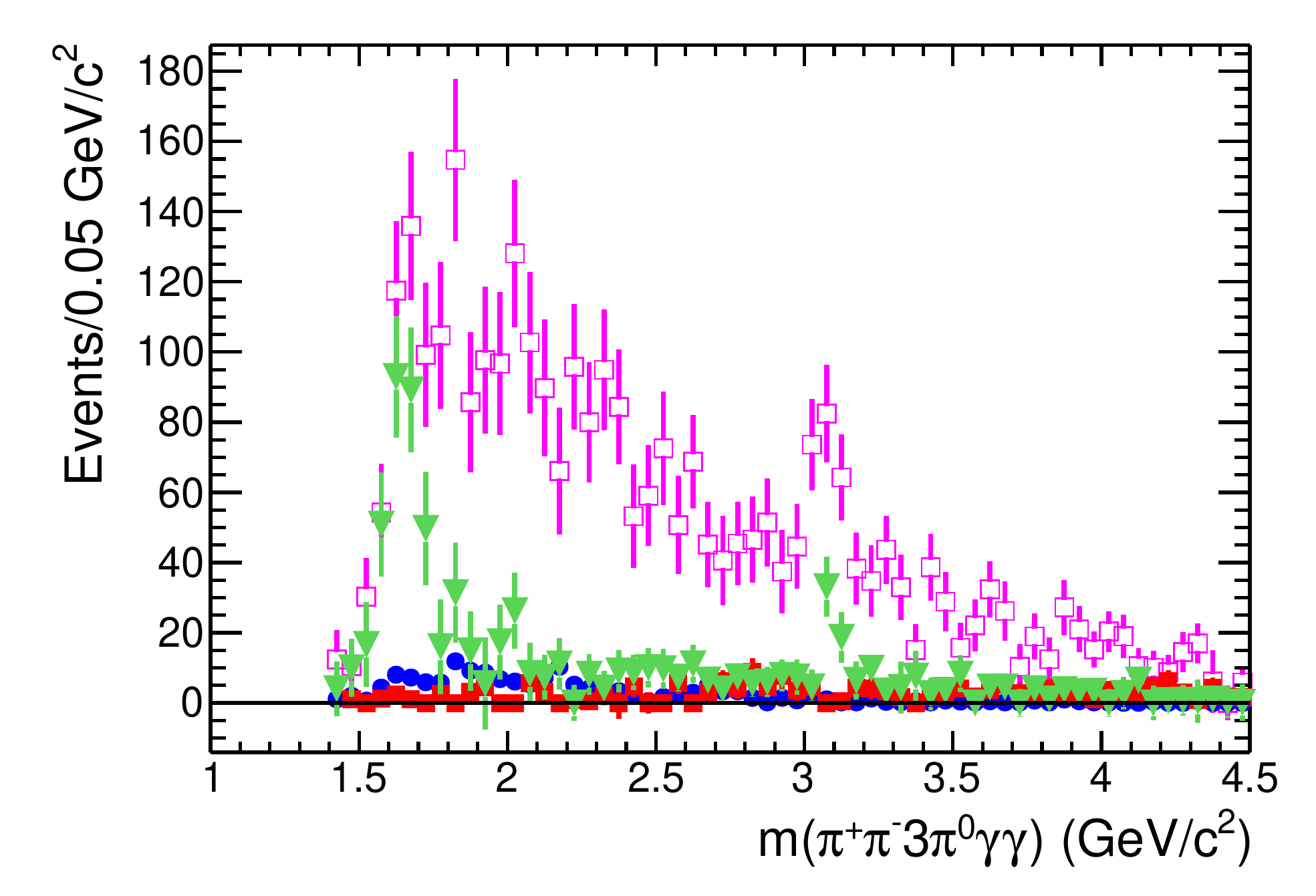}
\put(-50,100){\makebox(0,0)[lb]{\bf(b)}}
\vspace{-0.5cm}
\caption{(a) The $\pipi\piz$ invariant mass for data. The solid curve
  shows the fit function for signal (based on a fit to MC simulation) plus the
  polynomial for
 the combinatorial background (dashed curve). 
(b) The mass distribution of the $\pipi4\piz$ events in the 
$\omega$ peak (open squares) and the estimated contribution for
$\omega\eta$ (triangles), $\omega2\piz$ (circles), and $uds$ (filld squares).
}
\label{nevomega2pi0}
\end{center}
\end{figure}
\begin{figure}[tbh]
  \begin{center}
    \vspace{-0.2cm}
  \includegraphics[width=0.9\linewidth]{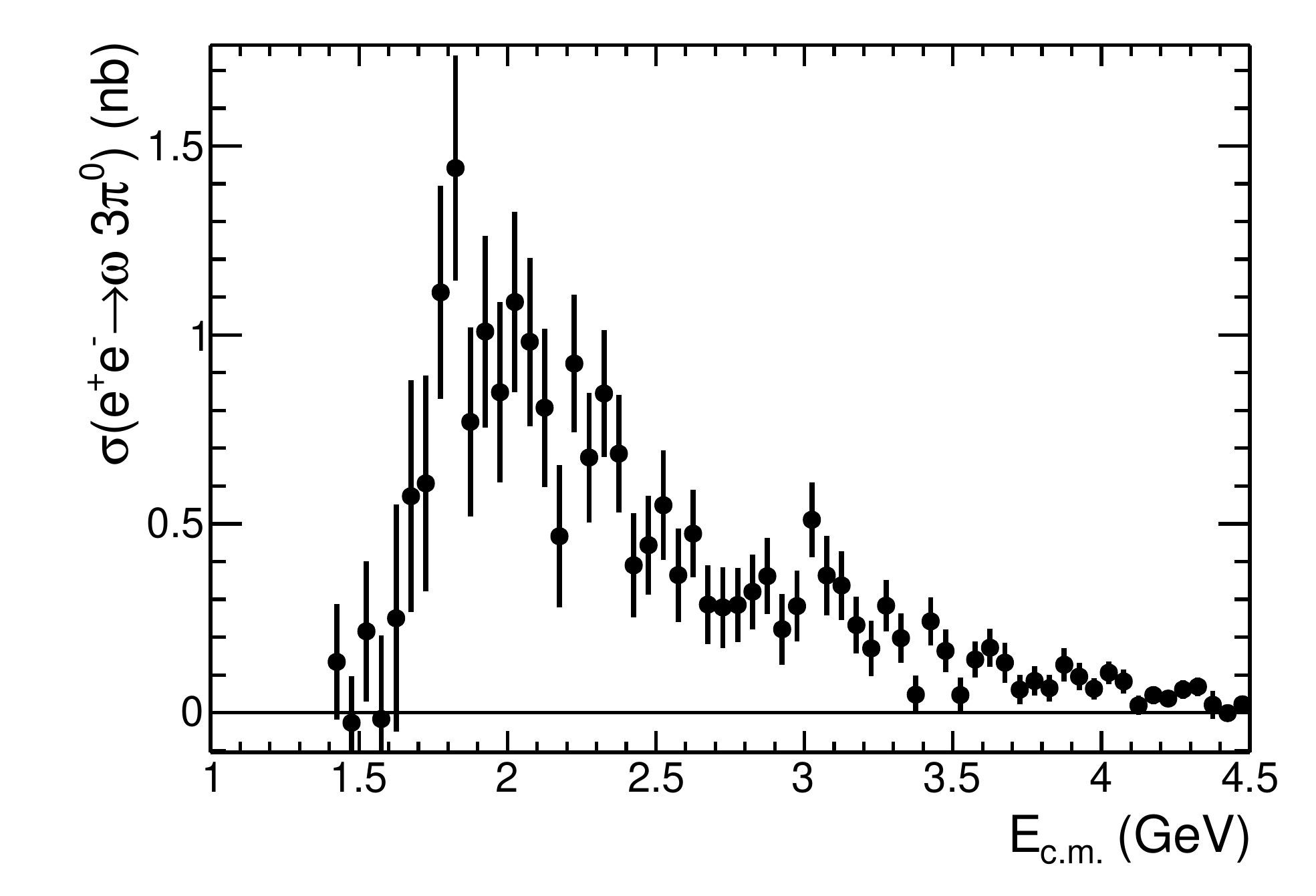}
\vspace{-0.5cm}
\caption{The energy dependent $\epem\to\omega3\piz$ cross
  section  in the $\pipi4\piz$ mode. 
}
\label{xs_2pi0omega}
\end{center}
\end{figure}

\input{xs3pi0omega_table}

\subsection{\bf\boldmath The $\omega3\piz$ intermediate state}
\label{Sec:omegapipi}
%
To determine the contribution of the $\omega3\piz$
 intermediate state, we fit the events of Fig.~\ref{3pivs7pi}(a)
 using a BW function to model the signal and a polynomial to model the background. 
The BW function is convolved with a Gaussian distribution that accounts for the detector
resolution.
The result of the fit is shown in
Fig.~\ref{nevomega2pi0}(a).
We obtain $2808\pm180$ $\omega3\piz$ events. 
The number of  $\omega3\piz$ events as a function of the six-pion invariant mass
is determined by performing an analogous
fit of events in Fig.~\ref{3pivs7pi}(c) in each 0.05~\gevcc interval of $m(\pipi4\piz)$.  
The resulting distribution is shown  in
Fig.~\ref{nevomega2pi0}(b).

For the $\epem\to\omega3\piz$ channel, there can be a peaking background from
$\epem\to\omega2\piz$ when the fourth $\piz$ is formed from background photons.
A simulation of this reaction
with proper normalization leads to the peaking-background
estimation, which is found to be small as shown in
Fig.~\ref{nevomega2pi0}(b).
There is also a small peaking background from the
generic $uds$ reaction.
Finally, we need to remove events with correlated $\omega$ and
$\eta$ production in the $\omega\eta$ final state, described in
Sec.~\ref{Sec:etaomega} and shown in Fig.~\ref{nevomega2pi0}(b).
These contributions are subtracted from the $\omega3\piz$
signal candidate distribution.

The  $\epem\to\omega3\piz$ cross section, not associated with $\eta\omega$ and corrected
for the $\omega\to\pipi\piz$ branching fraction,
is shown in Fig.~\ref{xs_2pi0omega} and summarized in Table~\ref{omega3pi0_table}.
The uncertainties are statistical only.
The systematic uncertainties are about 12\%.
No previous measurement exists for this process.
The cross section exhibits a rise at threshold,
a decrease at large \Ecm with a signal from $J/\psi$, and a possible resonance activity
 around 1.7-2.0~GeV.
\begin{figure}[tbh]
\begin{center}
\includegraphics[width=0.42\linewidth]{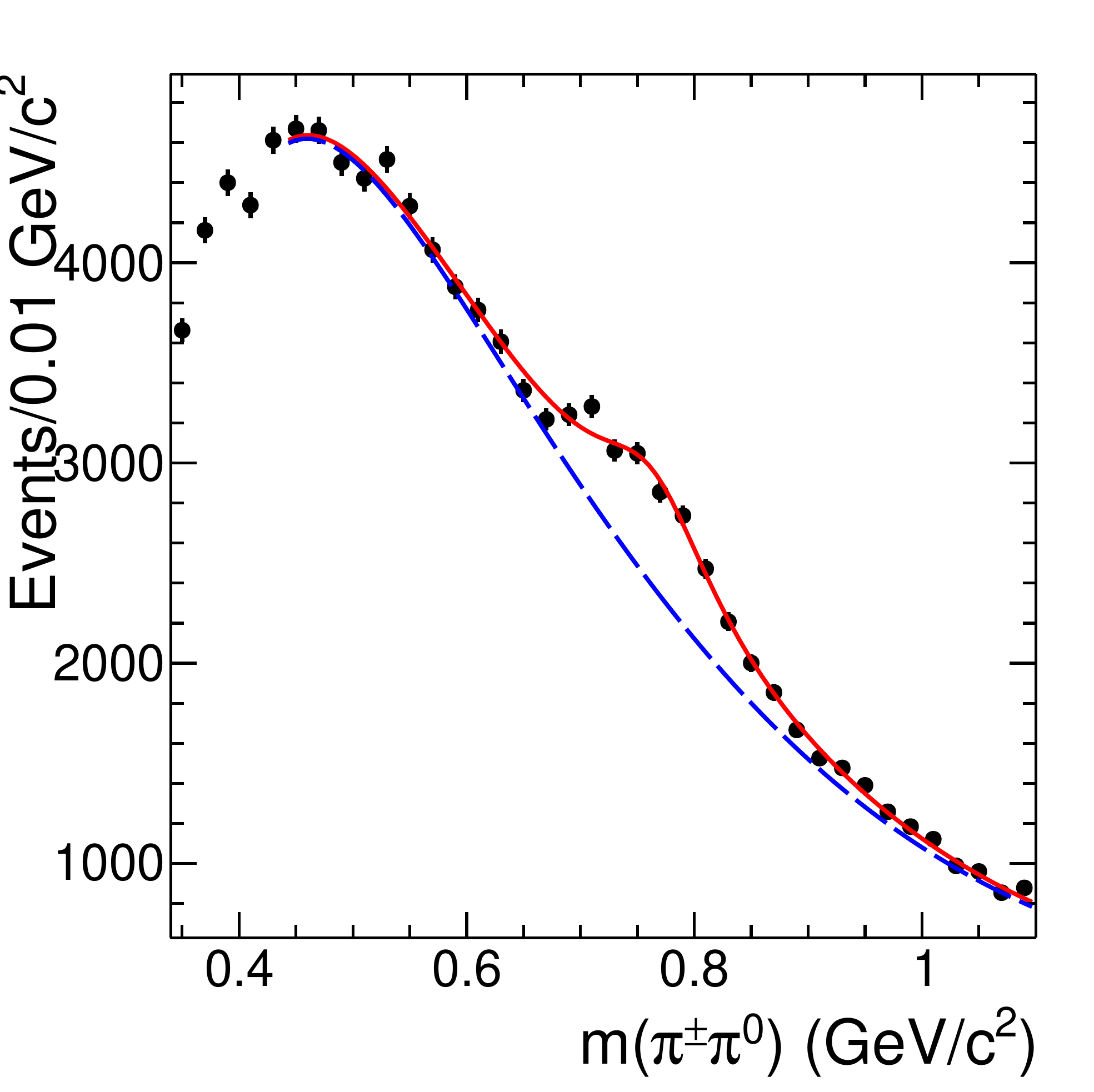}
\put(-50,80){\makebox(0,0)[lb]{\bf(a)}}
\includegraphics[width=0.6\linewidth]{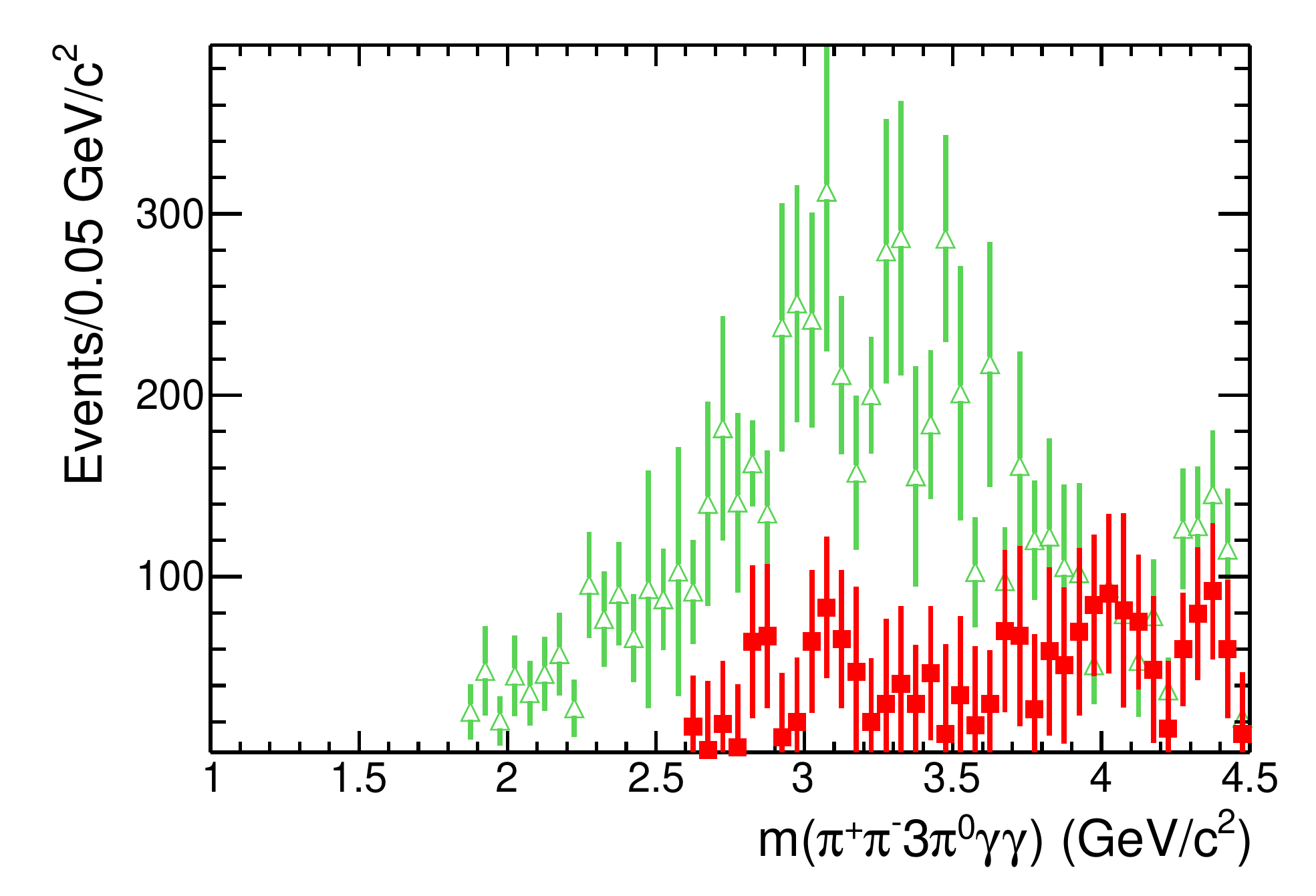}
\put(-30,80){\makebox(0,0)[lb]{\bf(b)}}
\vspace{-0.3cm}
\caption{(a) The $\pi^{\pm}\piz$ invariant mass for data. 
The dashed curve shows the fit to the combinatorial background. The solid curve 
is the sum of the background curve and the BW function for the
$\rho^{\pm}$.
(b) The result of the $\rho$ fit in bins of 0.05~\gevcc
in the $\pipi3\piz\gamma\gamma$ mass. The squares show the 
contribution from $uds$ background. 
}
\label{pipi0slices}
\end{center}
\end{figure}
\begin{figure}[h]
\begin{center}
\includegraphics[width=0.9\linewidth]{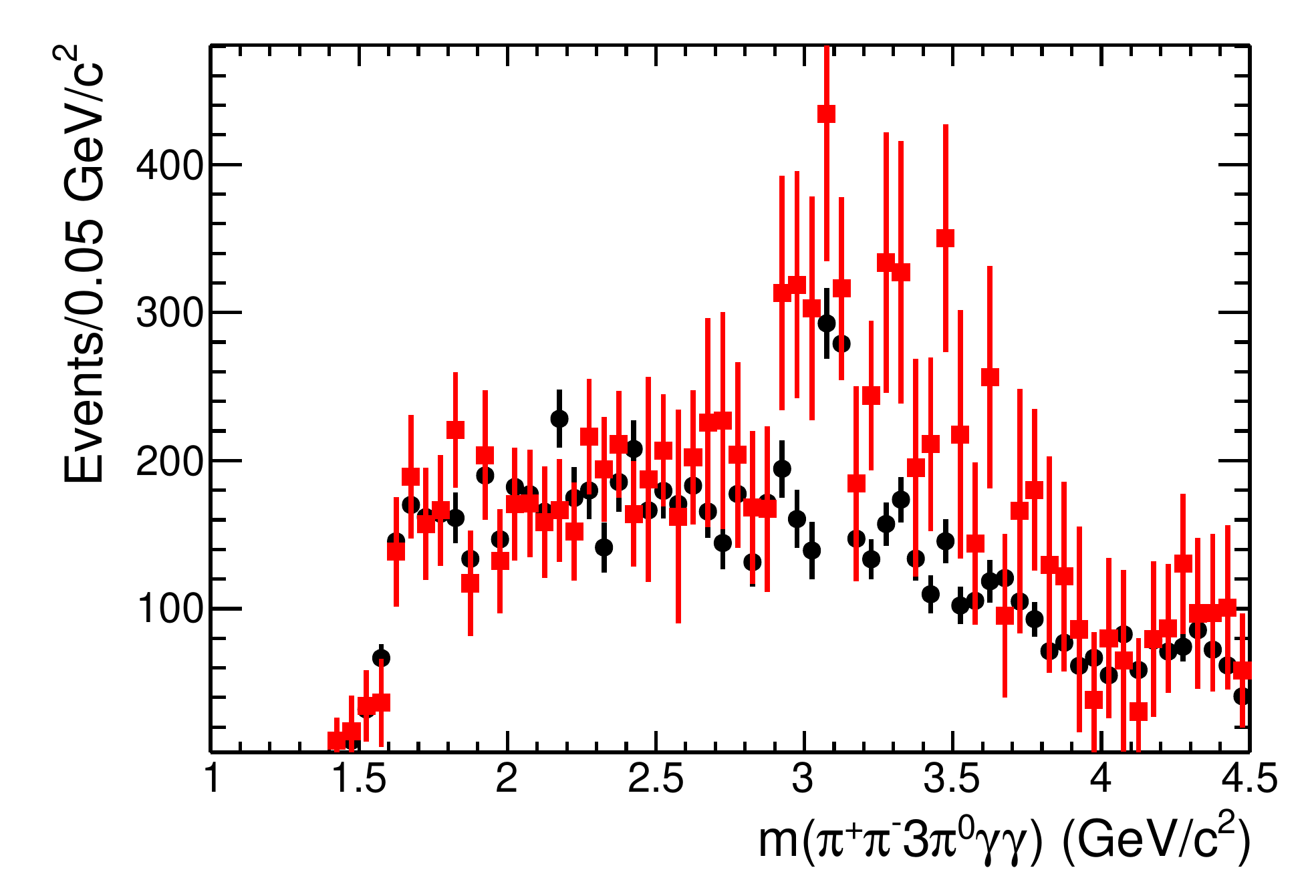}
\vspace{-0.5cm}
\caption{
 The circles show the number of events determined from the $\piz$
fit. The squares show the 
sum of the number of events with an $\eta$, $\omega$, $\rho$, or $uds$ 
contribution.
}
\label{sumallev}
\end{center}
\end{figure}

\subsection{\bf\boldmath The $\rho(770)^{\pm}\pi^{\mp}3\piz$ intermediate state}
\label{sec:rhoselectpi0}
A similar approach is followed to study events
with a $\rho^{\pm}$ meson in the intermediate state.
Because the $\rho$ meson is broad, a BW
function is used to describe the signal shape.
There are eight $\rho^\pm$ candidates per event, leading to
a large combinatoric background.
To extract the contribution of the $\rho^{\pm}\pi^{\mp}3\piz$ intermediate state we
fit the events in Fig.~\ref{pipi0vs7pi}(a) with a BW function
to describe the signal and a polynomial to describe the background.
The parameters of the $\rho$ resonance are taken from Ref.~\cite{PDG}.
The result of the fit  is shown in Fig.~\ref{pipi0slices}(a). 
We obtain $5965\pm667$ combinations with $\rho^{\pm}$ signals. The
distribution of these events vs the six-pion invariant mass is shown
by the triangle symbols in Fig.~\ref{pipi0slices}(b), while a
similar fit for the $uds$ simulation is shown by squares.
The $uds$ background dominates at higher energies.

We expect
more than one $\rho^{\pm}$ per event,
namely that there is a significant production
of $\epem\to\rho^+\rho^-2\piz$.
Because of the large combinatoric background, we do not perform a
study of correlated $\rho^+\rho^-$ production.

A similar study of the $\rho^{0}4\pi^{0}$ final state yields $407\pm45$
events in total, but obtained mass dependence is not reliable for this contribution.  

\subsection{\bf\boldmath The sum of intermediate states}
\label{sec:sum}

The circle symbols in Fig.~\ref{sumallev}  show the
total number of $\pipi4\piz$ events,
already shown in Fig.~\ref{nev_2pi4pi0_data}.
We perform a sum of 
the number of $\eta\pipi\piz$, $\omega3\piz$, $\eta\omega$,  $uds$
and $\rho^{\pm}\pi^{\mp}3\piz$ intermediate state
candidates, found as described in the previous sections,
and we show this sum by the square symbols in Fig.~\ref{sumallev}.
This summed curve
is seen to be in agreement with the total number of
$\pipi4\piz$ events except in the region above 2.5 GeV, where the sum is
dominated by the $\rho^{\pm}$ signal extraction. The observed overcount 
indicates a possible contribution from correlated 
 $\rho^+\rho^-$ production.

\input{xs2pi3pi0eta_table}

\section{\bf\boldmath The $\pipi3\piz\eta$ final state}
\subsection{Determination of  the number of events }

\begin{figure}[tbh]
\begin{center}
\includegraphics[width=0.9\linewidth]{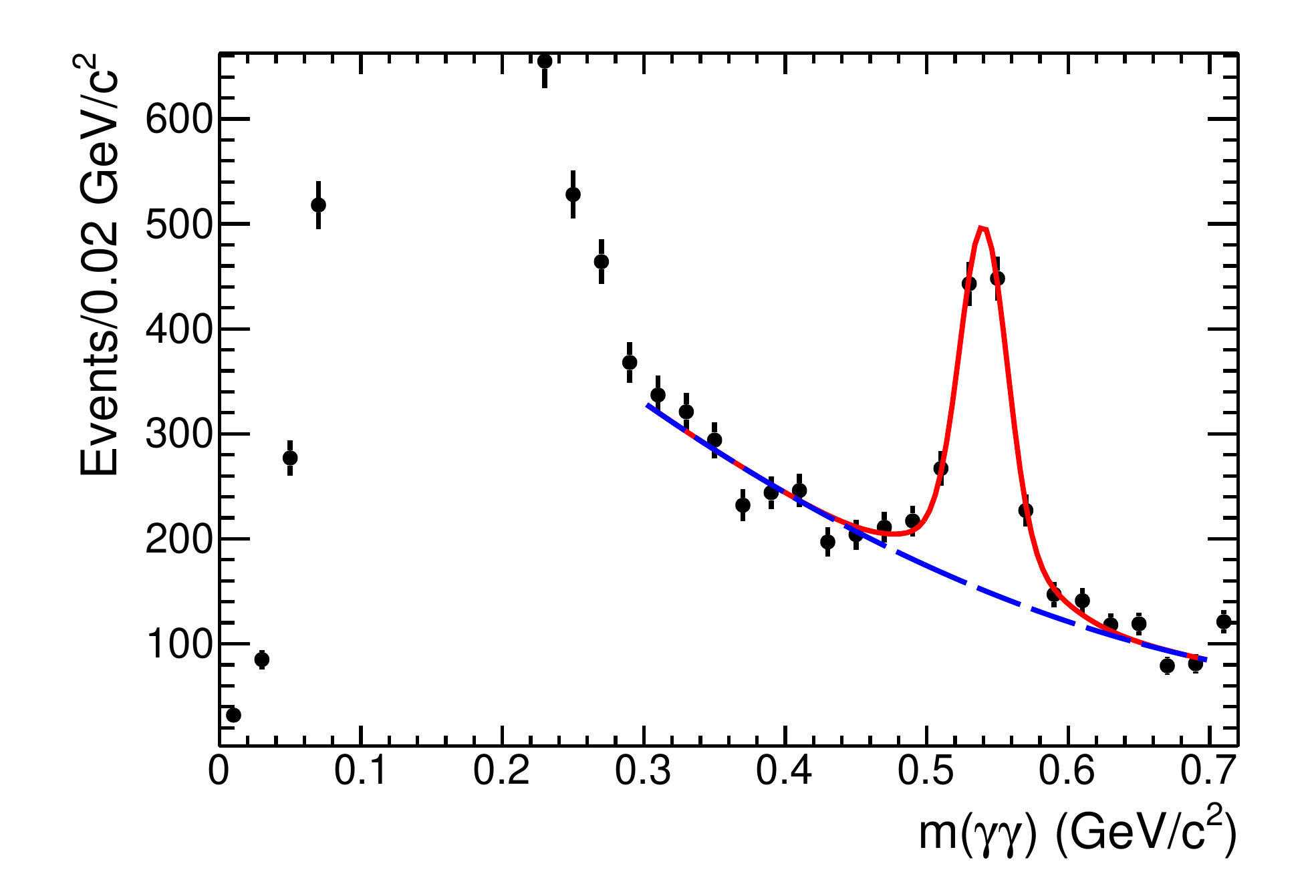}
\put(-50,120){\makebox(0,0)[lb]{\bf(a)}}\\
\includegraphics[width=0.9\linewidth]{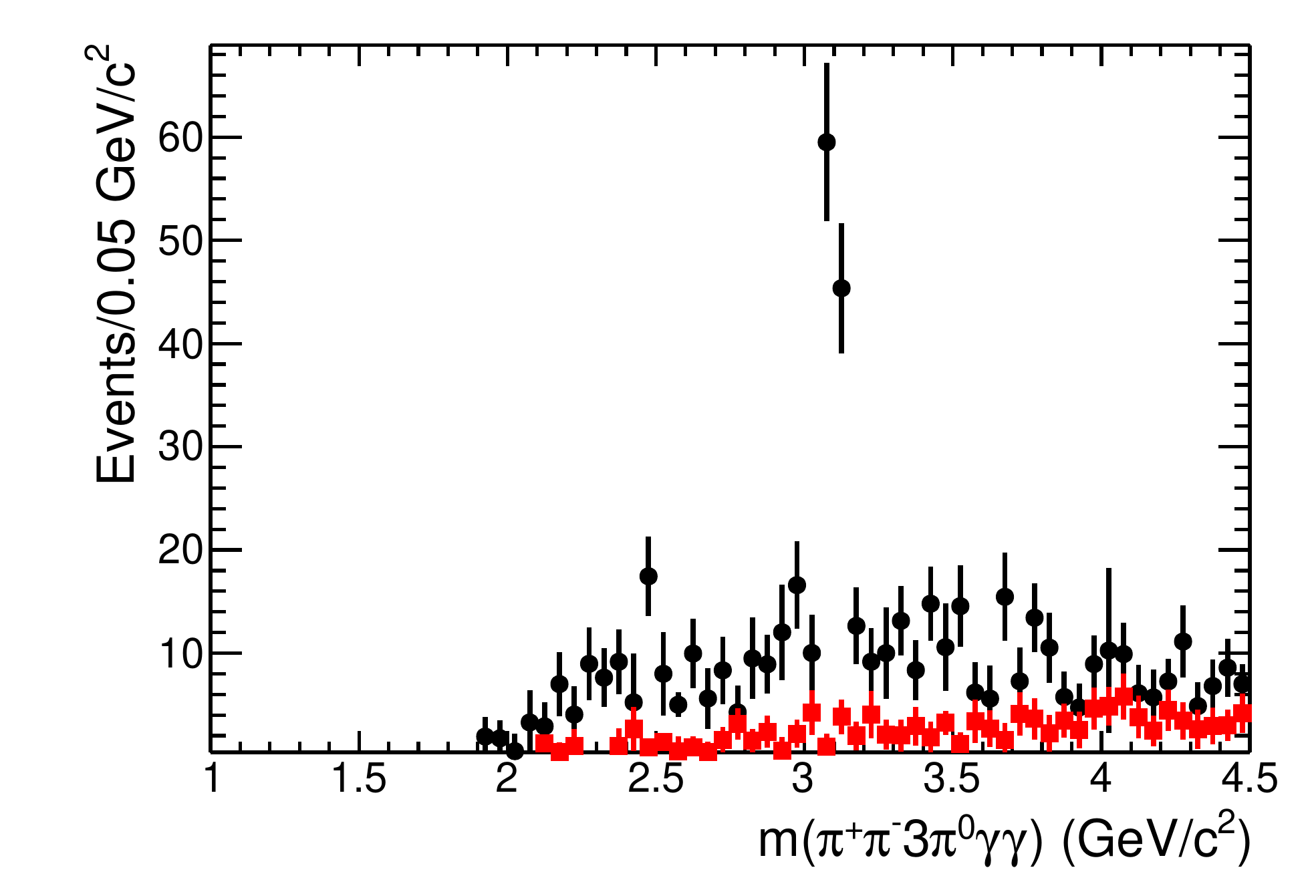}
\put(-50,120){\makebox(0,0)[lb]{\bf(b)}}
\vspace{-0.5cm}
\caption{(a) The expanded view of Fig.~\ref{2pi2pi0_bkg}(b).
  The solid curve
  shows the sum of background and the two-Gaussian fit function used to obtain the number of events
  with an $\eta$.
(b) The invariant mass distribution for the
$\pipi3\piz\eta$ events obtained from the $\eta$ signal fit. The
contribution of the $uds$ background events is shown by the squares.
}
\label{meta_data_fit}
\end{center}
\end{figure}

An analogous approach to that described above for
$\epem\to\pipi4\piz$ events is used to study
$\epem\to\pipi3\piz\eta$ events.
We fit the $\eta$ signal in the fourth-photon-pair
invariant mass distribution (cf., Fig.~\ref{2pi2pi0_bkg}(b)) to
the sum of
two Gaussians with a common mean, while the relatively smooth 
background is described by a second-order polynomial function, as shown
in Fig.~\ref{meta_data_fit}(a).  We obtain $870\pm52$ events.
Figure~\ref{meta_data_fit}(b) shows the mass distribution of these events.
\subsection{Peaking background}\label{sec:udsbkg2}

The major background producing an $\eta$ peak is the
non-ISR background, in particular $\epem\to\pipi4\piz\eta$
in which one of the neutral pions decays asymmetrically, producing a 
photon interpreted as ISR.  The $\eta$ peak from the $uds$ simulation
is visible in Fig.~\ref{udsbkg}.
  We fit the $\eta$
peak in the $uds$ simulation in intervals of 0.05~\gevcc in
$m(\pipi3\piz\gamma\gamma)$. 

To normalize the $uds$ simulation,
we form the di-photon invariant mass distribution of
 the ISR candidate with all the remaining photons in the event.
Comparing the number of events in the $\piz$ peaks
in data and $uds$ simulation, we assign
a scale factor of $1.5\pm0.2$ to the simulation.
The results are shown by the squares in Fig.~\ref{meta_data_fit}(b).
We subtract these events from the data distribution.

\begin{figure}[tbh]
\begin{center}
\includegraphics[width=0.95\linewidth]{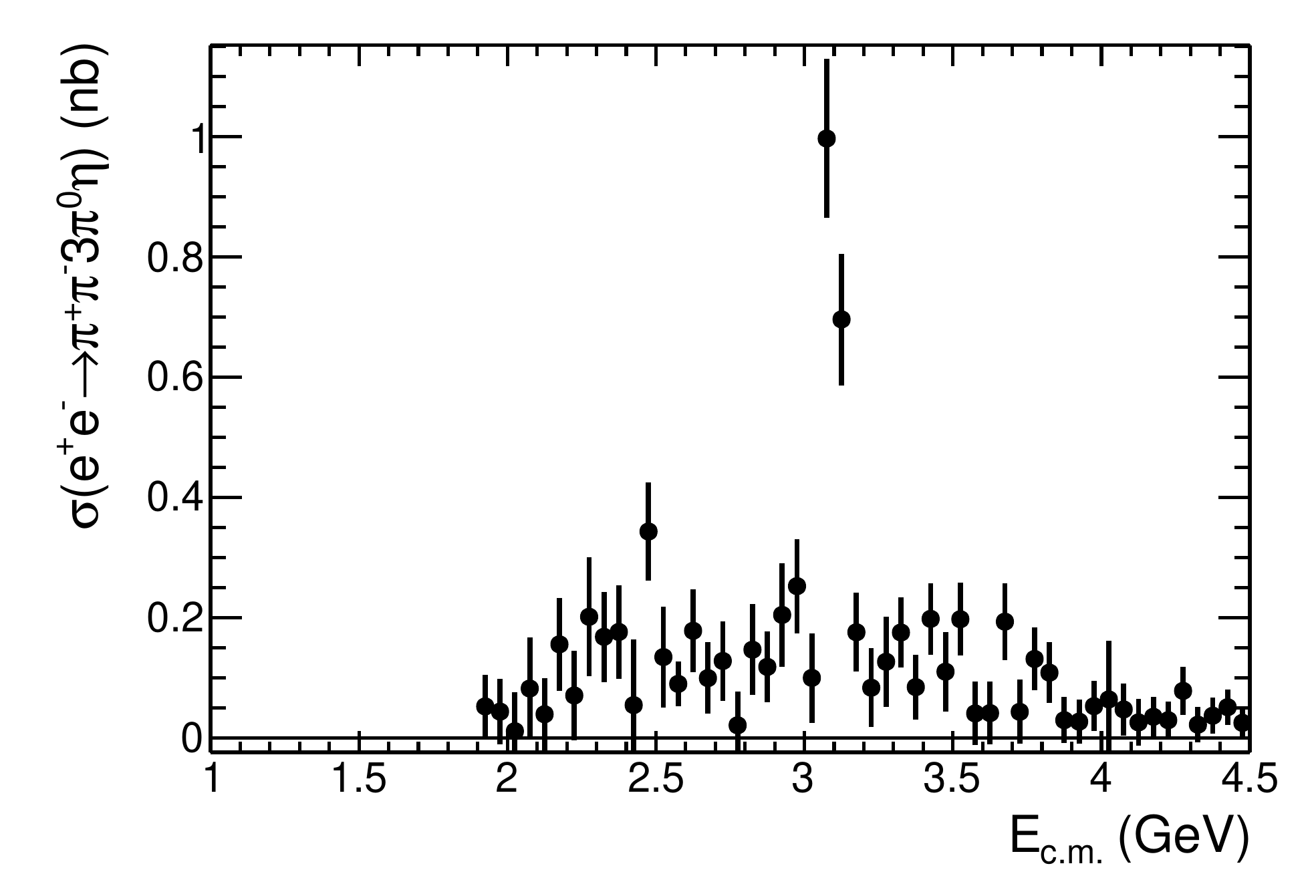}
\vspace{-0.5cm}
\caption{
The energy-dependent  cross section
for   $\epem\to\pipi3\piz\eta$.
The uncertainties are statistical only.
}
\label{2pi3pi0eta_ee_babar}
\end{center}
\end{figure} 
\subsection{\boldmath Cross section for $\epem\to\pipi3\piz\eta$}
\label{2pi2pi0eta}
The cross section for $\epem\to\pipi3\piz\eta$ is determined
using Eq.~(\ref{xseq}).  We assume the same detection efficiency for the photons from the $\piz\to\gamma\gamma$ and $\eta\to\gamma\gamma$ decays. 
The results are shown in
Fig.~\ref{2pi3pi0eta_ee_babar} and
listed in Table~\ref{5pieta_table}. These are the first results for
this process. 
The systematic uncertainties and corrections
are the same as those presented in Table~\ref{error_tab} except
that the uncertainty in the detection efficiency increases to 15\%.

The cross section is sizable above 2 GeV. Since this is above the energy range
         where the final states are summed for the ($g_\mu-2$) value calculation,
         we do not attempt to study the intermediate channels.

\section{\bf\boldmath The $J/\psi$ region}
\subsection{\bf\boldmath The $\pipi4\piz$  final state}
\begin{figure}[tbh]
\includegraphics[width=0.50\linewidth]{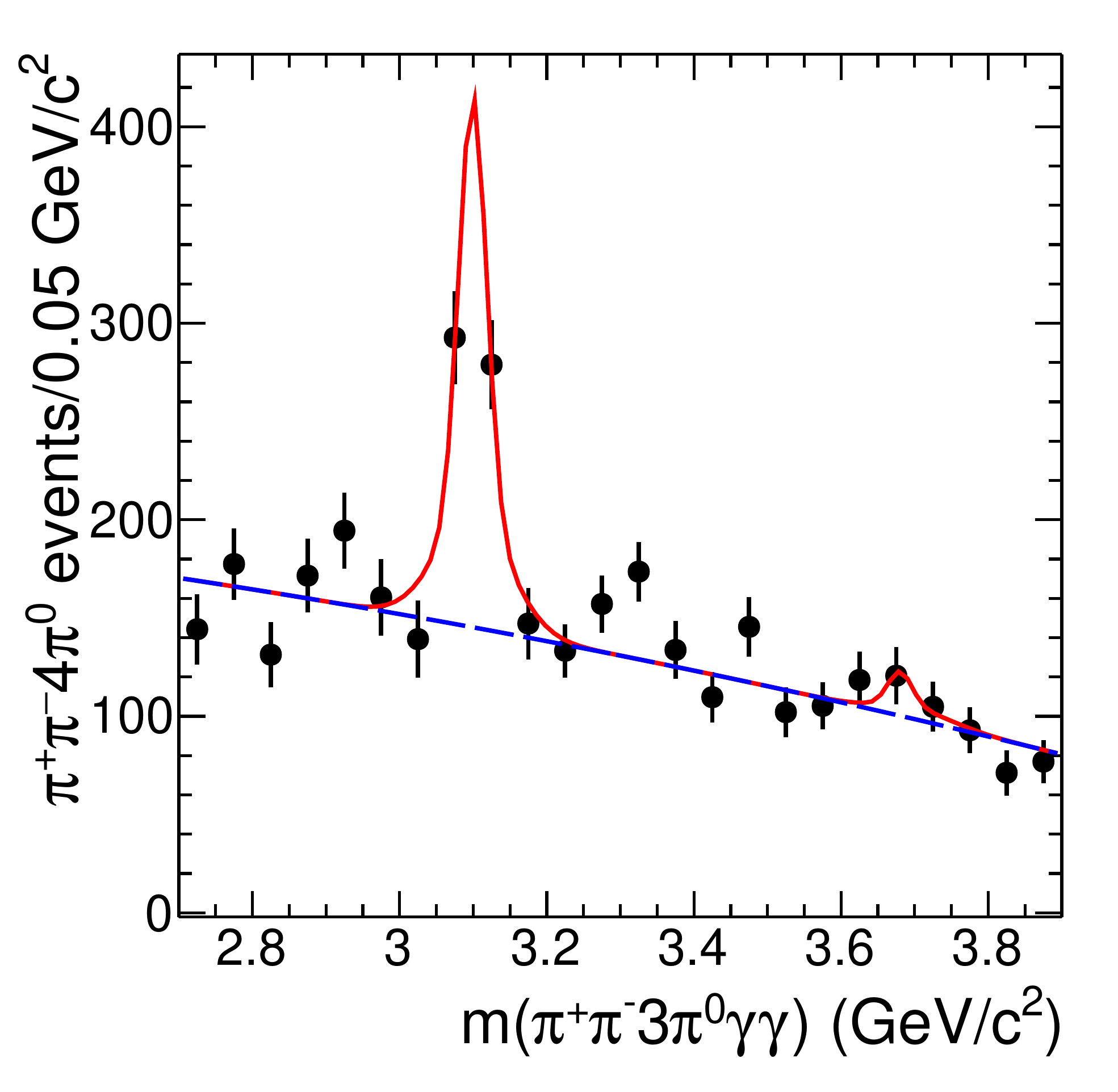}
\put(-40,90){\makebox(0,0)[lb]{\bf(a)}}
\includegraphics[width=0.50\linewidth]{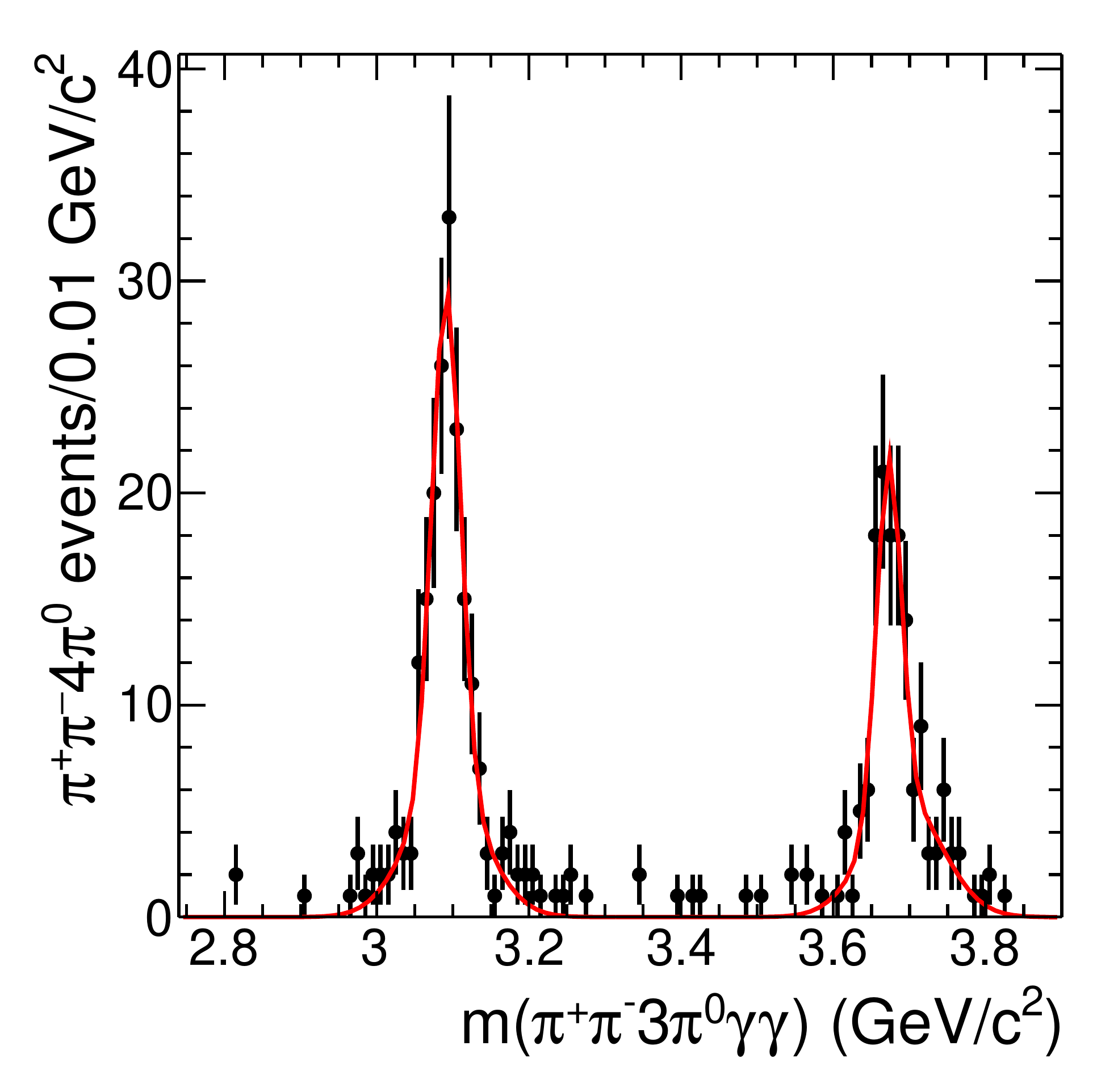}
\put(-40,90){\makebox(0,0)[lb]{\bf(b)}}
\vspace{-0.4cm}
\caption{(a)
The $\pipi4\piz$ mass distribution for ISR-produced
$\epem\to\pipi4\piz$ events in the $J/\psi$--$\psi(2S)$
        region.  
(b) The MC-simulated signals. The $J/\psi$--$\psi(2S)$ signals ratio is arbitrary. The curves show the fit functions
described in the text.
}
\label{jpsi}
\end{figure}

Figure~\ref{jpsi}(a) shows an expanded view of the $J/\psi$ mass region 
from Fig.~\ref{nev_2pi4pi0_data} for the six-pion data sample. 
Signals from  $J/\psi\to\pipi4\piz$ and $\psi(2S)\to 
\pipi4\piz$  are  seen. 
The non-resonant background distribution is  well described by the second-order polynomial function
in this region.

\begin{figure*}[tbh]
\includegraphics[width=0.25\linewidth]{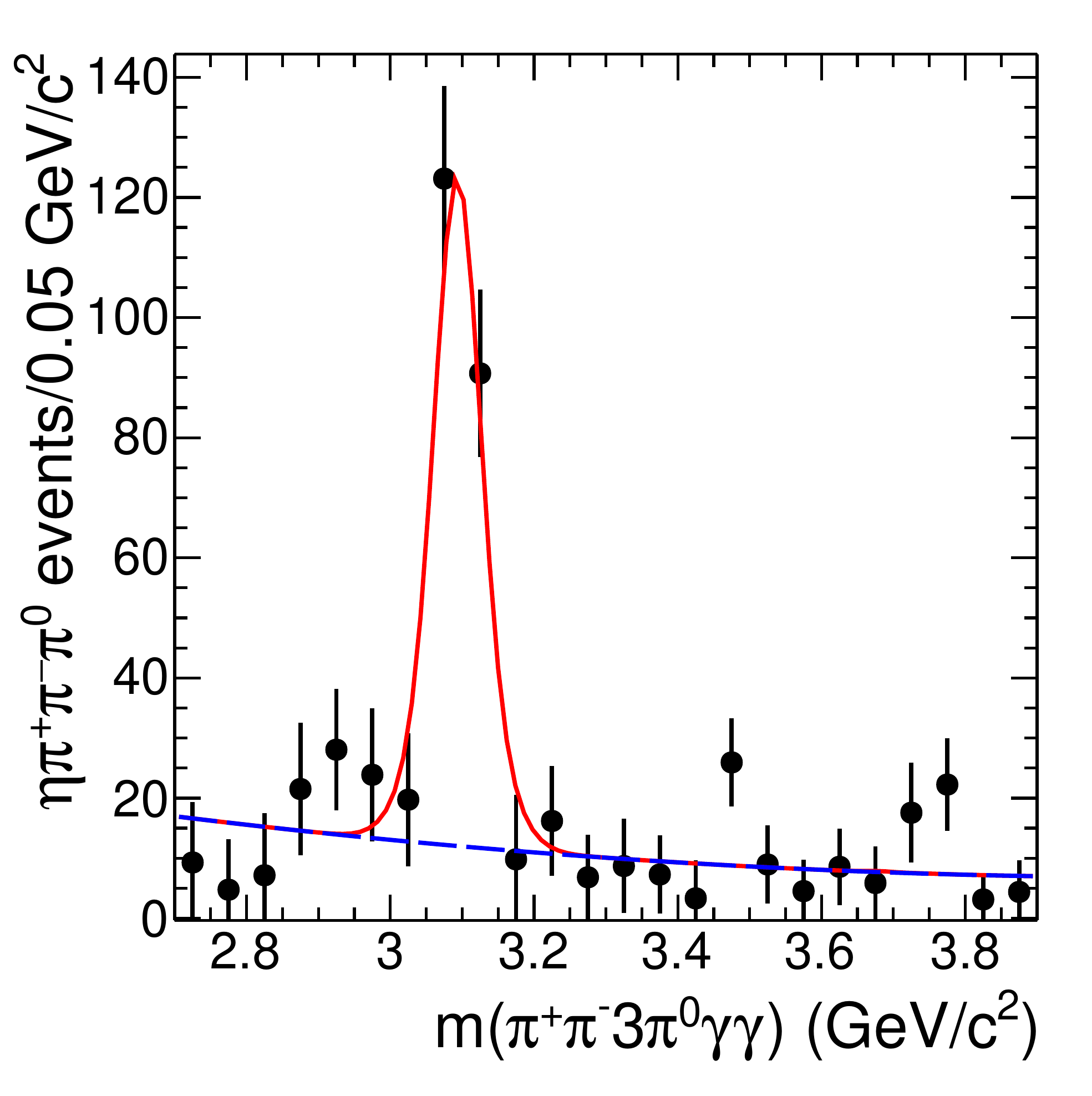}
\put(-40,90){\makebox(0,0)[lb]{\bf(a)}}
\includegraphics[width=0.25\linewidth]{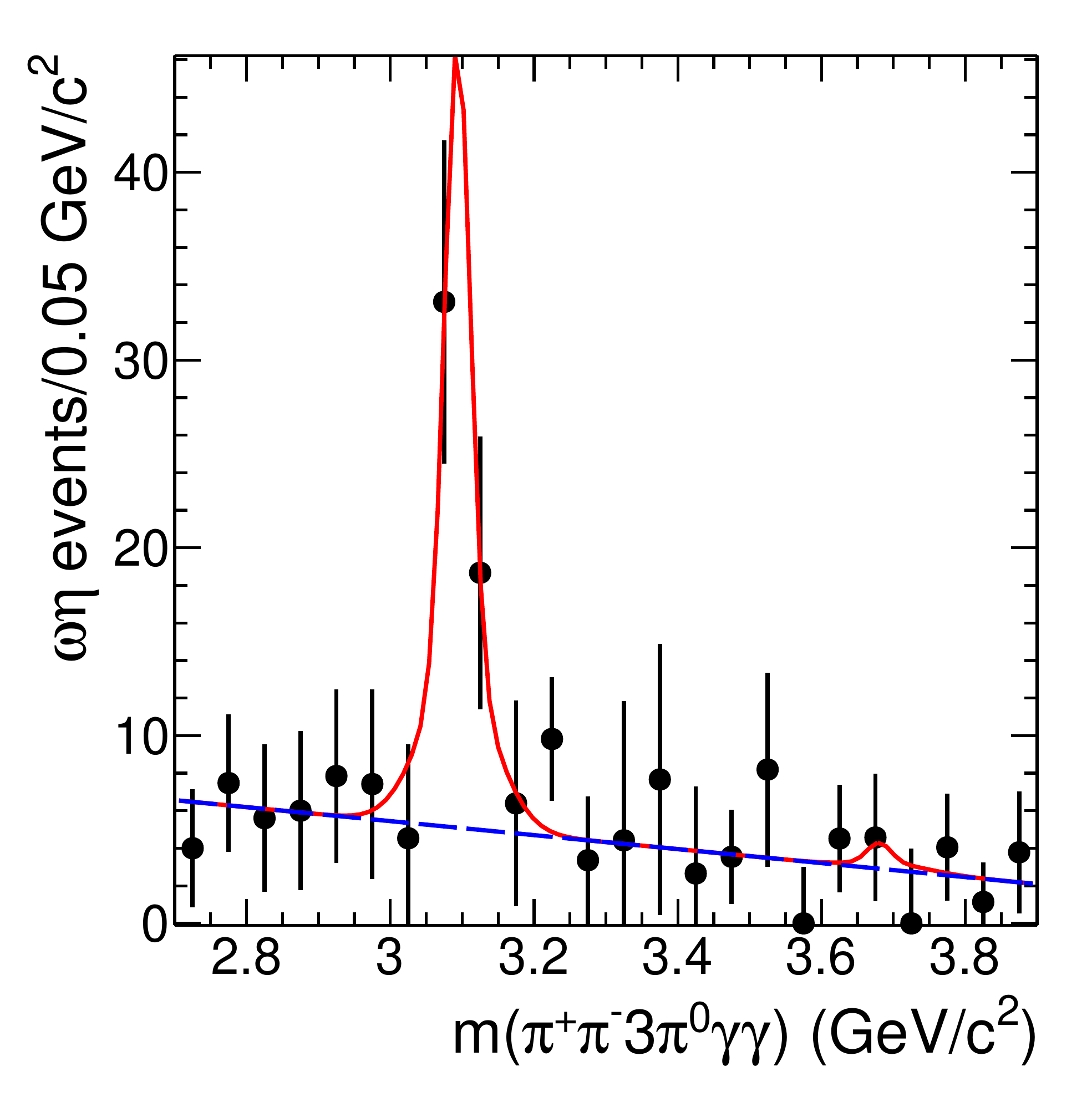}
\put(-40,90){\makebox(0,0)[lb]{\bf(b)}}
\includegraphics[width=0.25\linewidth]{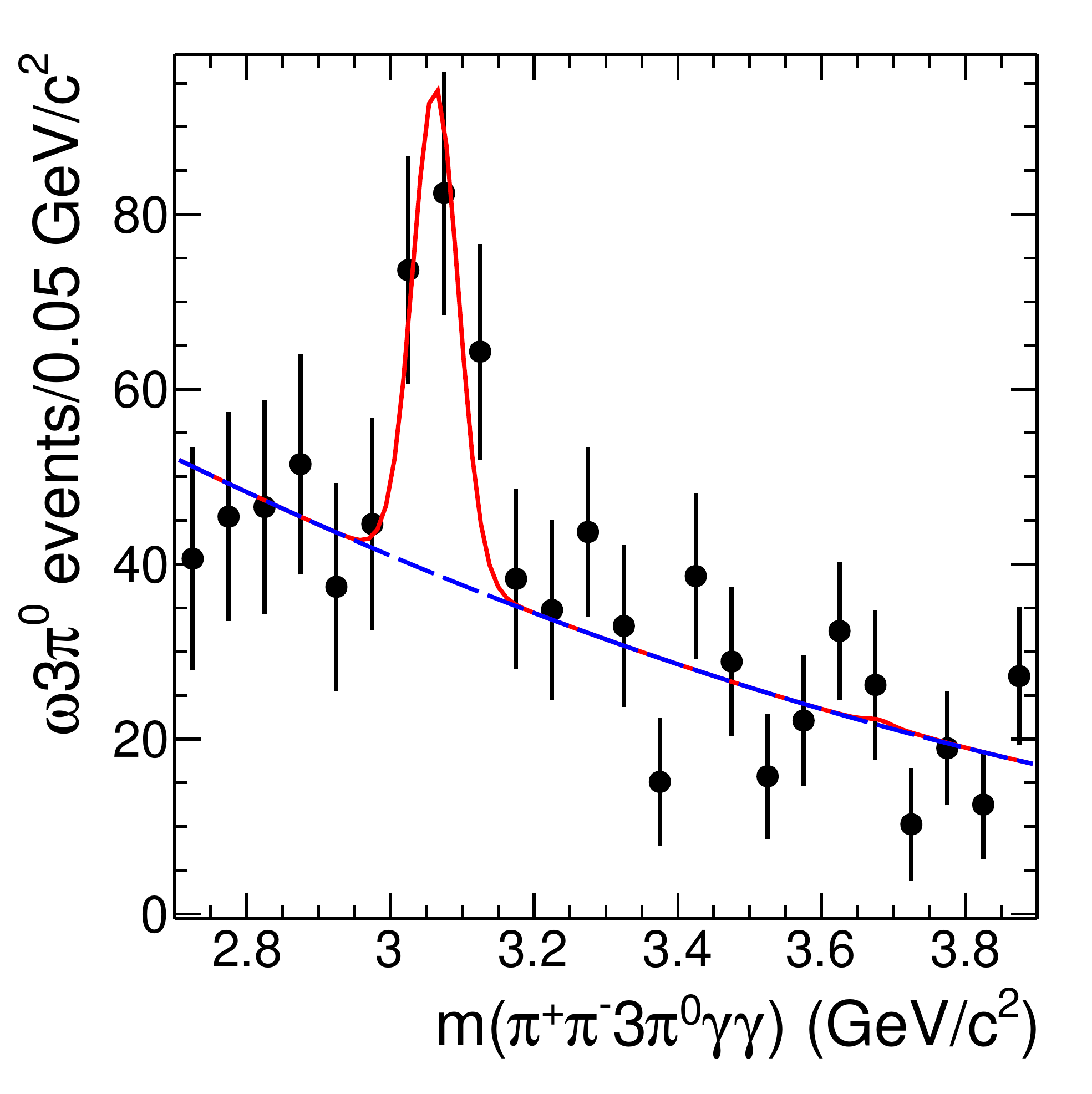}
\put(-40,90){\makebox(0,0)[lb]{\bf(c)}}
\includegraphics[width=0.25\linewidth]{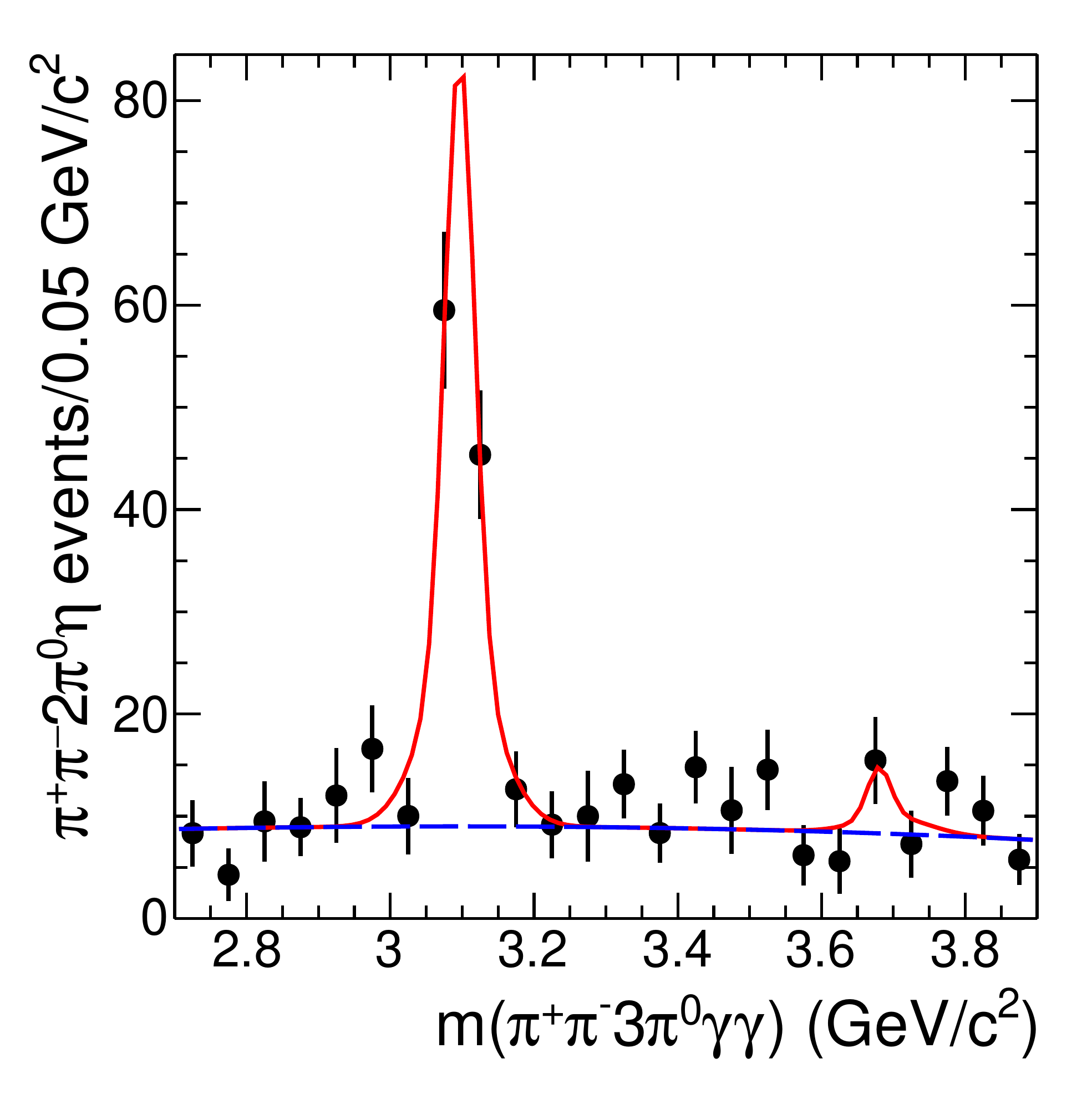}
\put(-40,90){\makebox(0,0)[lb]{\bf(d)}}
\vspace{-0.3cm}
\caption{
The $J/\psi$ region for the  $\pipi4\piz$ events for the selection
of (a)  $\eta\pipi\piz$, (b) $\eta\omega$,  (c)  $\omega3\piz$, and
(d)  $\pipi3\piz\eta$ intermediate states. 
The curves show the fit
functions described in the text.
}
\label{jpsistates}
\end{figure*}    

The observed peak shapes are not purely Gaussian because of radiation
effects and resolution, 
as seen in the simulated signal distributions
shown in Fig.~\ref{jpsi}(b).  The sum of two Gaussians  is used in the fit.
We obtain $340\pm42$  $J/\psi$ events and $28\pm19$
$\psi(2S)$ events.
Using the results for the number of events, the detection
efficiency, and the ISR luminosity,
we determine the product:
\begin{eqnarray}
  B_{J/\psi\to 6\pi}\cdot\Gamma^{J/\psi}_{ee}
  = \frac{N(J/\psi\to\pipi4\piz)\cdot m_{J/\psi}^2}%
           {6\pi^2\cdot d{\cal L}/dE\cdot\epsilon^{\rm MC}\cdot\epsilon^\text{corr}\cdot C} \label{jpsieq}\\
  = (35.8\pm4.4\pm 5.4)~\ev\ ,\nonumber
\end{eqnarray}
where $\Gamma^{J/\psi}_{ee}$ is
the electronic width, $d{\cal L}/dE =
  180~\invnb/\mev$ is the ISR luminosity 
at the $J/\psi$ mass $m_{J/\psi}$, $\epsilon^{\rm MC} = 0.018\pm0.002$ is the detection
efficiency from simulation with the corrections $\epsilon^\text{corr}=0.85$, discussed in 
Sec.~\ref{sec:Systematics},
and  $C = 3.894\times 10^{11}~\nb\mev^2$ is a
conversion constant ~\cite{PDG}. We estimate the systematic uncertainty for this
region to be 15\%.  
The subscript ``$6\pi$'' for the branching fraction refers
to the $\pipi4\piz$ final state exclusively.

Using $\Gamma^{J/\psi}_{ee} =5.55\pm0.14~\kev$ ~\cite{PDG}, we obtain
$B_{J/\psi\to 6\pi} = (6.5\pm 0.8\pm 1.0)\times 10^{-3}$; no
other measurements for this channel exist. 

Using Eq.(\ref{jpsieq}) and the result $d{\cal L}/dE =
  228~\invnb/\mev$  at the $\psi(2S)$ mass, we obtain:
\begin{eqnarray*}
  B_{\psi(2S)\to 6\pi}\cdot\Gamma^{\psi(2S)}_{ee}
  &=& (3.3\pm2.3\pm0.5)~\ev\ .
\end{eqnarray*}
With $\Gamma^{\psi(2S)}_{ee} =2.34\pm0.06~\kev$ ~\cite{PDG}  we
find $B_{\psi(2S)\to 6\pi} = (1.4\pm 1.0\pm 0.2)\times
10^{-3}$. For this channel also, no previous result exists.

\subsubsection{\bf\boldmath The $\eta\pipi\piz$,  $\eta\omega$ intermediate states}

Figure~\ref{jpsistates}(a) shows 
an expanded view of Fig.~\ref{neveta2pi} with the $\pipi4\piz$ mass
distribution for events obtained by a fit to the $3\piz$
mass distribution to select events with an $\eta$.
The two-Gaussian fit, implemented as described above, yields
$200\pm16$ and $<20$ events at 90\% C.L.  for
the $J/\psi$ and $\psi(2S)$, respectively. 
Using Eq.(\ref{jpsieq}) we obtain:
\begin{eqnarray*}
  B_{J/\psi\to\eta\pipi\piz}\cdot B_{\eta\to3\piz}\cdot\Gamma^{J/\psi}_{ee}
  &=&\\ (21.1\pm1.7\pm3.2)\ev\ ,\\
  B_{\psi(2S)\to\eta\pipi\piz}\cdot B_{\eta\to3\piz}\cdot\Gamma^{\psi(2S)}_{ee}
  & < &  3~\ev\ .
\end{eqnarray*} 
Using $B_{\eta\to3\piz} = 0.3268$ and the value of $\Gamma_{ee}$ from
Ref.~\cite{PDG}, we
obtain $B_{J/\psi\to\eta\pipi\piz} = (11.9\pm 0.9\pm 2.3)\times 10^{-3}$
and $B_{\psi(2S)\to\eta\pipi\piz} < 3.5\times
10^{-3}$ at 90\% C.L.
There are no other measurements of these decays.

Similarly, the expanded view of Fig.~\ref{nevetaomega} is shown in
Fig.~\ref{jpsistates}(b) for the  subsample of $\pipi4\piz$  events
with $\eta\to3\piz$  and an additional signal from
$\omega\to\pipi\piz$.  The fit yields $47\pm20$ events
corresponding to
\begin{eqnarray*}
 B_{J/\psi\to\eta\omega}\cdot B_{\eta\to3\piz}\cdot B_{\omega\to\pipi\piz}\cdot\Gamma^{J/\psi}_{ee}
  &=&\\ (4.9\pm2.1\pm0.7)\ev\ ,
\end{eqnarray*} 
which yields$B_{J/\psi\to\eta\omega} = (3.0\pm 1.3\pm
0.5)\times 10^{-3}$, compatible with the current
world average result  $B_{J/\psi\to\eta\omega} =
(1.74\pm0.20)\times 10^{-3}$~\cite{PDG}.

We can set only an upper limit for the $\psi(2S)\to\eta\omega$
decay: we observe $<20$ events corresponding to
$B_{\psi(2S)\to\eta\omega}< 14\times 10^{-4}$ at 90\% C.L.,  consistent
with the world average value $<1.1\times 10^{-5}$~\cite{PDG}.

\subsubsection{\bf\boldmath The $\omega3\piz$ intermediate state}

The expanded view of Fig.~\ref{nevomega2pi0}(b) is shown in
Fig.~\ref{jpsistates}(c).   The fit yields $89\pm22$ for the $J/\psi\to\omega3\piz$ events
corresponding to
\begin{eqnarray*}
  B_{J/\psi\to\omega3\piz)}\cdot B_{\omega\to\pipi\piz}\cdot\Gamma^{J/\psi}_{ee}
  &=&\\ (9.4\pm2.3\pm1.5)\ev\ ,\\
B_{J/\psi\to\omega3\piz} = (1.9\pm 0.5\pm0.3)\times 10^{-3}.
\end{eqnarray*}
We use $B_{\omega\to\pipi\piz} = 0.892$ from Ref.~\cite{PDG}.
No other measurements are available for this decay mode.

We can set only an upper limit for the $\psi(2S)\to\omega3\piz$
decay: we observe $<14$ events corresponding to
$B_{\psi(2S)\to\omega3\piz}< 8\times 10^{-4}$ at 90\% C.L. which is
the only measured limit for this decay.
\begin{table*}[tbh]
\caption{
  Summary of the $J/\psi$ and $\psi(2S)$ branching fractions.
  }
\label{jpsitab}
\begin{tabular}{r@{$\cdot$}l  r@{.}l@{$\pm$}l@{$\pm$}l 
                              r@{.}l@{$\pm$}l@{$\pm$}l
  r@{.}l@{$\pm$}l }
  \hline
\multicolumn{2}{c}{Measured} & \multicolumn{4}{c}{Measured}    &  
\multicolumn{7}{c}{$J/\psi$ or $\psi(2S)$ Branching Fraction  (10$^{-3}$)}\\
\multicolumn{2}{c}{Quantity} & \multicolumn{4}{c}{Value (\ev)} &
\multicolumn{4}{c}{Derived, this work}    & 
\multicolumn{3}{c}{PDG~\cite{PDG}} \\
\hline
$\Gamma^{J/\psi}_{ee}$  &  $B_{J/\psi \to \pipi\ppz\ppz}$  &
  35& 8 & 4.4 & 5.4  &   ~~~6&5 & 0.8 & 1.0  &   \multicolumn{3}{c}{no entry}  \\

$\Gamma^{J/\psi}_{ee}$  &  $B_{J/\psi  \to \eta\pipi\piz}
                           \cdot B_{\eta   \to \ppz\piz}$  &
  21&1 & 1.7 & 3.2  &   11&9 & 0.9 & 2.3  & \multicolumn{3}{c}{no entry}  \\
  
$\Gamma^{J/\psi}_{ee}$  &  $B_{J/\psi  \to \omega\eta}  
                        \cdot B_{\omega\to \pipi\piz} \cdot B_{\eta   \to \ppz\piz}$  &
   4&9& 2.1 & 0.7 &   3&0 & 1.3 & 0.5  & ~~~~1&74 & 0.20\\

$\Gamma^{J/\psi}_{ee}$  &  $B_{J/\psi  \to \omega\ppz\piz}
                           \cdot B_{\omega    \to \pipi\piz}$  &
  9&4 & 2.3 & 1.5  &   1&9 & 0.5 & 0.3  &    \multicolumn{3}{c}{no entry}   \\

  $\Gamma^{J/\psi}_{ee}$  &  $B_{J/\psi  \to \pipi\ppz\piz\eta}
                            \cdot B_{\eta   \to \gamma\gamma}$ &
   10&6& 1.6& 1.6 &   4&9 & 0.8 & 0.8  &  \multicolumn{3}{c}{no entry}\\

$\Gamma^{\psi(2S)}_{ee}$  &  $B_{\psi(2S) \to\pipi\ppz\ppz} $  &
   3&3& 2.3& 0.5 &   1&4  & 1.0  & 0.2   &  \multicolumn{3}{c}{no entry} \\

 $\Gamma^{\psi(2S)}_{ee}$  &  $B_{\psi(2S) \to\eta\pipi\piz} \cdot B_{\eta   \to \ppz\piz}$  &
   ~~~~$<$3 & \multicolumn{3}{l}{0 at 90\% C.L.} & ~~~~ $<$3 &
                                                     \multicolumn{3}{l}{5
                                                     at 90\% C.L.} &      \multicolumn{3}{c}{no entry}\\ 

$\Gamma^{\psi(2S)}_{ee}$  &  $B_{\psi(2S) \to \omega\eta} \cdot B_{\omega  \to \pipi\piz}\cdot B_{\eta   \to \ppz\piz}$  &
   ~~~~$<$1 & \multicolumn{3}{l}{1 at 90\% C.L.} & ~~~~ $<$1 &
                                                     \multicolumn{3}{l}{4
                                                     at 90\% C.L.} &   ~~~~ $<$0 &  \multicolumn{2}{l}{11
                                                     at 90\% C.L.}  \\

$\Gamma^{\psi(2S)}_{ee}$  &  $B_{\psi(2S) \to\omega\ppz\piz} \cdot B_{\omega  \to \pipi\piz}$  &
   ~~~~$<$1 & \multicolumn{3}{l}{6 at 90\% C.L.} & ~~~~ $<$0 &
                                                     \multicolumn{3}{l}{8
                                                     at 90\% C.L.} &      \multicolumn{3}{c}{no entry}\\ 

 $\Gamma^{\psi(2S)}_{ee}$  &  $B_{\psi(2S) \to\pipi\ppz\piz\eta} \cdot B_{\eta   \to \gamma\gamma}$  &
   ~~~~$<$1 & \multicolumn{3}{l}{9 at 90\% C.L.} & ~~~~ $<$2 &
                                                     \multicolumn{3}{l}{0
                                                     at 90\% C.L.} &      \multicolumn{3}{c}{no entry}\\

\hline
\end{tabular}
\end{table*}

\subsection{\bf\boldmath The $\pipi3\piz\eta$  final state}

The expanded view of Fig.~\ref{meta_data_fit}(b) is shown in
Fig.~\ref{jpsistates}(d).   The fit yields $101\pm16$ for the $J/\psi\to\pipi3\piz\eta$ events
corresponding to
\begin{eqnarray*}
  B_{J/\psi\to\pipi3\piz\eta}\cdot B_{\eta\to\gamma\gamma}\cdot\Gamma^{J/\psi}_{ee}
  &=&\\ (10.6\pm1.6\pm1.6)\ev\ ,\\
B_{J/\psi\to\pipi3\piz\eta} = (4.9\pm 0.8\pm0.8)\times 10^{-3}.
\end{eqnarray*}
We set an upper limit for the $\psi(2S)\to\pipi3\piz\eta$
decay: we observe $<16$ events at 90\% C.L. corresponding to
$B_{\psi(2S)\to\pipi3\piz\eta}< 2.0\times 10^{-3}$.
There are no previous results for this final state.

\subsection{Summary of the charmonium region study}
The rates of  $J/\psi$ and $\psi(2S)$ decays to $\pipi4\piz$,
$\pipi3\piz\eta$, and several intermediate  final
states have been measured. 
The measured products and calculated 
branching fractions are summarized in Table~\ref{jpsitab}
together with the available PDG~\cite{PDG} values for comparison.
Most of the measurements are performed for the first time.

\section{Summary}
\label{sec:Summary}
\noindent
The excellent photon-energy and charged-particle momentum 
resolutions, as well as  the particle
identification capabilities of the \babar\ detector, allow  the
reconstruction of the 
$\pipi4\piz$ and $\pipi3\piz\eta$
final states produced at  center-of-mass energies below 4.5~\gev 
via initial-state radiation  in data collected
at the $\Upsilon(4S)$  mass region.

The cross sections  for the $\epem\to\pipi4\piz$
and the $\epem\to\pipi3\piz\eta$ reactions
have been measured for the first time. The accuracies are  12\% and 15\%, respectively.

The selected multi-hadronic final states in the broad range of accessible
energies provide new information on hadron spectroscopy. The  
observed $\epem\to\omega3\piz$, $\epem\to\eta\pipi\piz$, and $\epem\to\eta\omega$  cross
sections provide  additional information for the hadronic contribution
calculation of the muon $g_\mu-2$.  

The initial-state radiation events allow a study of $J/\psi$ and
$\psi(2S)$ production and a measurement of the corresponding products of
the decay branching fractions and $\epem$ width for most of
the studied channels, the majority of them for the first time.


\section{Acknowledgments}
\label{sec:Acknowledgments}

\input acknowledgements

\newpage

\input {biblio_PRD}
\end{document}

%% file: authors_sep2021_frozen.tex
\author{J.~P.~Lees}
\author{V.~Poireau}
\author{V.~Tisserand}
\affiliation{Laboratoire d'Annecy-le-Vieux de Physique des Particules (LAPP), Universit\'e de Savoie, CNRS/IN2P3,  F-74941 Annecy-Le-Vieux, France}
\author{E.~Grauges}
\affiliation{Universitat de Barcelona, Facultat de Fisica, Departament ECM, E-08028 Barcelona, Spain }
\author{A.~Palano}
\affiliation{INFN Sezione di Bari, I-70126 Bari, Italy}
\author{G.~Eigen}
\affiliation{University of Bergen, Institute of Physics, N-5007 Bergen, Norway }
\author{D.~N.~Brown}
\author{Yu.~G.~Kolomensky}
\affiliation{Lawrence Berkeley National Laboratory and University of California, Berkeley, California 94720, USA }
\author{M.~Fritsch}
\author{H.~Koch}
\author{T.~Schroeder}
\affiliation{Ruhr Universit\"at Bochum, Institut f\"ur Experimentalphysik 1, D-44780 Bochum, Germany }
% ...
\collaboration{The \babar\ Collaboration}
\noaffiliation

%% file: xs2pi4pi0_table.tex
\begin{table*}
\caption{Summary of the $\epem\to\pipi\ppz\ppz$ 
cross section measurement. The uncertainties are statistical only.}
\label{2pi4pi0_tab}
\begin{tabular}{c c c c c c c c c c}
  \hline
$E_{\rm c.m.}$ (GeV) & $\sigma$ (nb)  
& $E_{\rm c.m.}$ (GeV) & $\sigma$ (nb) 
& $E_{\rm c.m.}$ (GeV) & $\sigma$ (nb) 
& $E_{\rm c.m.}$ (GeV) & $\sigma$ (nb)  
& $E_{\rm c.m.}$ (GeV) & $\sigma$ (nb)  
\\
  \hline
1.425 & 0.03 $\pm$ 0.05 &2.075 & 1.69 $\pm$ 0.21 &2.725 & 0.94 $\pm$ 0.14 &3.375 & 0.60 $\pm$ 0.10 &4.025 & 0.02 $\pm$ 0.06 \\ 
1.475 & 0.17 $\pm$ 0.06 &2.125 & 1.46 $\pm$ 0.20 &2.775 & 1.12 $\pm$ 0.14 &3.425 & 0.45 $\pm$ 0.09 &4.075 & 0.18 $\pm$ 0.06 \\ 
1.525 & 0.47 $\pm$ 0.08 &2.175 & 1.96 $\pm$ 0.19 &2.825 & 0.78 $\pm$ 0.13 &3.475 & 0.71 $\pm$ 0.10 &4.125 & 0.11 $\pm$ 0.06 \\ 
1.575 & 0.92 $\pm$ 0.13 &2.225 & 1.46 $\pm$ 0.19 &2.875 & 0.99 $\pm$ 0.14 &3.525 & 0.40 $\pm$ 0.08 &4.175 & 0.18 $\pm$ 0.06 \\ 
1.625 & 1.92 $\pm$ 0.18 &2.275 & 1.44 $\pm$ 0.18 &2.925 & 1.16 $\pm$ 0.14 &3.575 & 0.41 $\pm$ 0.08 &4.225 & 0.10 $\pm$ 0.05 \\ 
1.675 & 2.13 $\pm$ 0.21 &2.325 & 1.11 $\pm$ 0.15 &2.975 & 0.92 $\pm$ 0.14 &3.625 & 0.48 $\pm$ 0.09 &4.275 & 0.13 $\pm$ 0.05 \\ 
1.725 & 1.99 $\pm$ 0.20 &2.375 & 1.45 $\pm$ 0.18 &3.025 & 0.68 $\pm$ 0.15 &3.675 & 0.46 $\pm$ 0.09 &4.325 & 0.21 $\pm$ 0.06 \\ 
1.775 & 1.88 $\pm$ 0.20 &2.425 & 1.60 $\pm$ 0.17 &3.075 & 1.75 $\pm$ 0.17 &3.725 & 0.32 $\pm$ 0.10 &4.375 & 0.13 $\pm$ 0.05 \\ 
1.825 & 1.83 $\pm$ 0.20 &2.475 & 1.15 $\pm$ 0.15 &3.125 & 1.61 $\pm$ 0.16 &3.775 & 0.24 $\pm$ 0.08 &4.425 & 0.07 $\pm$ 0.05 \\ 
1.875 & 1.48 $\pm$ 0.18 &2.525 & 1.33 $\pm$ 0.16 &3.175 & 0.75 $\pm$ 0.13 &3.825 & 0.10 $\pm$ 0.09 &4.475 & 0.01 $\pm$ 0.04 \\ 
1.925 & 1.96 $\pm$ 0.21 &2.575 & 1.26 $\pm$ 0.16 &3.225 & 0.72 $\pm$ 0.10 &3.875 & 0.18 $\pm$ 0.07 & &  \\ 
1.975 & 1.49 $\pm$ 0.20 &2.625 & 1.30 $\pm$ 0.15 &3.275 & 0.75 $\pm$ 0.10 &3.925 & 0.13 $\pm$ 0.06 & &  \\ 
2.025 & 1.76 $\pm$ 0.21 &2.675 & 1.07 $\pm$ 0.14 &3.325 & 0.85 $\pm$ 0.11 &3.975 & 0.10 $\pm$ 0.06 & &  \\ 
\hline
\end{tabular}
\end{table*}

%% file: xs3pieta_table.tex
\begin{table*}
\caption{Summary of the $\epem\to\eta\pipi\piz$ 
cross section measurement. The uncertainties are statistical only.}
\label{3pieta_table}
\begin{tabular}{c c c c c c c c c c}
  \hline
$E_{\rm c.m.}$ (GeV) & $\sigma$ (nb)  
& $E_{\rm c.m.}$ (GeV) & $\sigma$ (nb) 
& $E_{\rm c.m.}$ (GeV) & $\sigma$ (nb) 
& $E_{\rm c.m.}$ (GeV) & $\sigma$ (nb)  
& $E_{\rm c.m.}$ (GeV) & $\sigma$ (nb)  
\\
  \hline
1.425 & 0.08 $\pm$ 0.46 &2.075 & 0.96 $\pm$ 0.53 &2.725 & 0.21 $\pm$ 0.23 &3.375 & 0.14 $\pm$ 0.12 &4.025 & 0.07 $\pm$ 0.07 \\ 
1.475 & 0.90 $\pm$ 0.39 &2.125 & 0.71 $\pm$ 0.54 &2.775 & 0.11 $\pm$ 0.19 &3.425 & 0.06 $\pm$ 0.12 &4.075 & 0.07 $\pm$ 0.07 \\ 
1.525 & 0.45 $\pm$ 0.61 &2.175 & 1.33 $\pm$ 0.48 &2.825 & 0.16 $\pm$ 0.23 &3.475 & 0.47 $\pm$ 0.13 &4.125 & 0.18 $\pm$ 0.07 \\ 
1.575 & 1.57 $\pm$ 0.75 &2.225 & 0.11 $\pm$ 0.42 &2.875 & 0.47 $\pm$ 0.24 &3.525 & 0.16 $\pm$ 0.12 &4.175 & 0.07 $\pm$ 0.05 \\ 
1.625 & 4.80 $\pm$ 0.99 &2.275 & 0.77 $\pm$ 0.42 &2.925 & 0.60 $\pm$ 0.22 &3.575 & 0.08 $\pm$ 0.09 &4.225 & 0.12 $\pm$ 0.05 \\ 
1.675 & 5.09 $\pm$ 1.01 &2.325 & 0.39 $\pm$ 0.37 &2.975 & 0.50 $\pm$ 0.23 &3.625 & 0.15 $\pm$ 0.11 &4.275 & 0.09 $\pm$ 0.05 \\ 
1.725 & 4.07 $\pm$ 0.95 &2.375 & 0.88 $\pm$ 0.33 &3.025 & 0.41 $\pm$ 0.23 &3.675 & 0.10 $\pm$ 0.10 &4.325 & 0.00 $\pm$ 0.02 \\ 
1.775 & 2.35 $\pm$ 0.82 &2.425 & 1.03 $\pm$ 0.37 &3.075 & 2.53 $\pm$ 0.32 &3.725 & 0.29 $\pm$ 0.14 &4.375 & 0.02 $\pm$ 0.05 \\ 
1.825 & 3.05 $\pm$ 0.76 &2.475 & 0.26 $\pm$ 0.33 &3.125 & 1.83 $\pm$ 0.28 &3.775 & 0.36 $\pm$ 0.12 &4.425 & 0.04 $\pm$ 0.04 \\ 
1.875 & 0.31 $\pm$ 0.66 &2.525 & 0.65 $\pm$ 0.25 &3.175 & 0.20 $\pm$ 0.21 &3.825 & 0.05 $\pm$ 0.06 &4.475 & 0.06 $\pm$ 0.04 \\ 
1.925 & 2.13 $\pm$ 0.75 &2.575 & 0.08 $\pm$ 0.26 &3.225 & 0.32 $\pm$ 0.18 &3.875 & 0.07 $\pm$ 0.08 & & \\ 
1.975 & 1.04 $\pm$ 0.65 &2.625 & 1.04 $\pm$ 0.31 &3.275 & 0.13 $\pm$ 0.14 &3.925 & 0.00 $\pm$ 0.14 & & \\ 
2.025 & 0.65 $\pm$ 0.60 &2.675 & 0.53 $\pm$ 0.28 &3.325 & 0.17 $\pm$ 0.15 &3.975 & 0.18 $\pm$ 0.08 & &  \\
\hline
\end{tabular}
\end{table*}

%% file: xsomegaeta_table.tex
\begin{table*}
\caption{Summary of the $\epem\to\eta\omega$ 
cross section measurement. The uncertainties are statistical only.}
\label{ometa_table}
\begin{tabular}{c c c c c c c c c c}
  \hline
$E_{\rm c.m.}$ (GeV) & $\sigma$ (nb)  
& $E_{\rm c.m.}$ (GeV) & $\sigma$ (nb) 
& $E_{\rm c.m.}$ (GeV) & $\sigma$ (nb) 
& $E_{\rm c.m.}$ (GeV) & $\sigma$ (nb)  
& $E_{\rm c.m.}$ (GeV) & $\sigma$ (nb)  
\\
  \hline
1.425 & 0.11 $\pm$ 0.22 &2.075 & 0.14 $\pm$ 0.15 &2.725 & 0.05 $\pm$ 0.04 &3.375 & 0.08 $\pm$ 0.08 &4.025 & 0.00 $\pm$ 0.03 \\ 
1.475 & 0.27 $\pm$ 0.22 &2.125 & 0.12 $\pm$ 0.11 &2.775 & 0.09 $\pm$ 0.05 &3.425 & 0.03 $\pm$ 0.05 &4.075 & 0.02 $\pm$ 0.04 \\ 
1.525 & 0.42 $\pm$ 0.31 &2.175 & 0.18 $\pm$ 0.12 &2.825 & 0.07 $\pm$ 0.05 &3.475 & 0.04 $\pm$ 0.03 &4.125 & 0.05 $\pm$ 0.02 \\ 
1.575 & 1.24 $\pm$ 0.36 &2.225 & 0.00 $\pm$ 0.08 &2.875 & 0.07 $\pm$ 0.05 &3.525 & 0.08 $\pm$ 0.05 &4.175 & 0.00 $\pm$ 0.04 \\ 
1.625 & 2.16 $\pm$ 0.40 &2.275 & 0.12 $\pm$ 0.09 &2.925 & 0.09 $\pm$ 0.06 &3.575 & 0.00 $\pm$ 0.03 &4.225 & 0.00 $\pm$ 0.02 \\ 
1.675 & 1.98 $\pm$ 0.40 &2.325 & 0.06 $\pm$ 0.07 &2.975 & 0.09 $\pm$ 0.06 &3.625 & 0.04 $\pm$ 0.03 &4.275 & 0.01 $\pm$ 0.03 \\ 
1.725 & 1.06 $\pm$ 0.34 &2.375 & 0.13 $\pm$ 0.09 &3.025 & 0.05 $\pm$ 0.06 &3.675 & 0.04 $\pm$ 0.03 &4.325 & 0.00 $\pm$ 0.04 \\ 
1.775 & 0.33 $\pm$ 0.28 &2.425 & 0.12 $\pm$ 0.07 &3.075 & 0.38 $\pm$ 0.10 &3.725 & 0.00 $\pm$ 0.04 &4.375 & 0.01 $\pm$ 0.01 \\ 
1.825 & 0.62 $\pm$ 0.28 &2.475 & 0.14 $\pm$ 0.08 &3.125 & 0.21 $\pm$ 0.08 &3.775 & 0.04 $\pm$ 0.03 &4.425 & 0.00 $\pm$ 0.03 \\ 
1.875 & 0.28 $\pm$ 0.22 &2.525 & 0.13 $\pm$ 0.08 &3.175 & 0.07 $\pm$ 0.06 &3.825 & 0.01 $\pm$ 0.02 &4.475 & 0.00 $\pm$ 0.03 \\ 
1.925 & 0.10 $\pm$ 0.24 &2.575 & 0.10 $\pm$ 0.05 &3.225 & 0.11 $\pm$ 0.04 &3.875 & 0.03 $\pm$ 0.03 & &  \\ 
1.975 & 0.31 $\pm$ 0.19 &2.625 & 0.15 $\pm$ 0.07 &3.275 & 0.04 $\pm$ 0.04 &3.925 & 0.03 $\pm$ 0.02 & &  \\ 
2.025 & 0.46 $\pm$ 0.19 &2.675 & 0.08 $\pm$ 0.05 &3.325 & 0.05 $\pm$ 0.08 &3.975 & 0.03 $\pm$ 0.02 & &  \\ 
\hline
\end{tabular}
\end{table*}

%% file: xs3pi0omega_table.tex
\begin{table*}
\caption{Summary of the $\epem\to\omega\ppz\piz$ 
cross section measurement. The uncertainties are statistical only.}
\label{omega3pi0_table}
\begin{tabular}{c c c c c c c c c c}
  \hline
$E_{\rm c.m.}$ (GeV) & $\sigma$ (nb)  
& $E_{\rm c.m.}$ (GeV) & $\sigma$ (nb) 
& $E_{\rm c.m.}$ (GeV) & $\sigma$ (nb) 
& $E_{\rm c.m.}$ (GeV) & $\sigma$ (nb)  
& $E_{\rm c.m.}$ (GeV) & $\sigma$ (nb)  
\\
  \hline
1.425 & 0.13 $\pm$ 0.15 &2.075 & 0.98 $\pm$ 0.22 &2.725 & 0.28 $\pm$ 0.11 &3.375 & 0.05 $\pm$ 0.05 &4.025 & 0.11 $\pm$ 0.03 \\ 
1.475 & -0.03 $\pm$ 0.12 &2.125 & 0.81 $\pm$ 0.21 &2.775 & 0.29 $\pm$ 0.10 &3.425 & 0.24 $\pm$ 0.06 &4.075 & 0.08 $\pm$ 0.03 \\ 
1.525 & 0.22 $\pm$ 0.19 &2.175 & 0.47 $\pm$ 0.19 &2.825 & 0.32 $\pm$ 0.10 &3.475 & 0.16 $\pm$ 0.06 &4.125 & 0.02 $\pm$ 0.03 \\ 
1.575 & -0.02 $\pm$ 0.22 &2.225 & 0.92 $\pm$ 0.18 &2.875 & 0.36 $\pm$ 0.10 &3.525 & 0.05 $\pm$ 0.05 &4.175 & 0.05 $\pm$ 0.02 \\ 
1.625 & 0.25 $\pm$ 0.30 &2.275 & 0.68 $\pm$ 0.17 &2.925 & 0.22 $\pm$ 0.09 &3.575 & 0.14 $\pm$ 0.05 &4.225 & 0.04 $\pm$ 0.02 \\ 
1.675 & 0.57 $\pm$ 0.31 &2.325 & 0.84 $\pm$ 0.17 &2.975 & 0.28 $\pm$ 0.09 &3.625 & 0.17 $\pm$ 0.05 &4.275 & 0.06 $\pm$ 0.03 \\ 
1.725 & 0.61 $\pm$ 0.29 &2.375 & 0.69 $\pm$ 0.16 &3.025 & 0.51 $\pm$ 0.10 &3.675 & 0.13 $\pm$ 0.05 &4.325 & 0.07 $\pm$ 0.02 \\ 
1.775 & 1.11 $\pm$ 0.28 &2.425 & 0.39 $\pm$ 0.14 &3.075 & 0.36 $\pm$ 0.10 &3.725 & 0.06 $\pm$ 0.04 &4.375 & 0.02 $\pm$ 0.04 \\ 
1.825 & 1.44 $\pm$ 0.30 &2.475 & 0.44 $\pm$ 0.13 &3.125 & 0.34 $\pm$ 0.09 &3.775 & 0.08 $\pm$ 0.04 &4.425 & 0.00 $\pm$ 0.02 \\ 
1.875 & 0.77 $\pm$ 0.25 &2.525 & 0.55 $\pm$ 0.15 &3.175 & 0.23 $\pm$ 0.07 &3.825 & 0.06 $\pm$ 0.04 &4.475 & 0.02 $\pm$ 0.01 \\ 
1.925 & 1.01 $\pm$ 0.25 &2.575 & 0.36 $\pm$ 0.12 &3.225 & 0.17 $\pm$ 0.07 &3.875 & 0.13 $\pm$ 0.04 & &  \\ 
1.975 & 0.85 $\pm$ 0.24 &2.625 & 0.47 $\pm$ 0.12 &3.275 & 0.28 $\pm$ 0.07 &3.925 & 0.10 $\pm$ 0.04 & &  \\ 
2.025 & 1.09 $\pm$ 0.24 &2.675 & 0.29 $\pm$ 0.10 &3.325 & 0.20 $\pm$ 0.06 &3.975 & 0.06 $\pm$ 0.03 & &  \\ 
\hline
\end{tabular}
\end{table*}

%% file: xs2pi3pi0eta_table.tex
\begin{table*}
\caption{Summary of the $\epem\to\pipi\ppz\piz\eta$ 
cross section measurement. The uncertainties are statistical only.}
\label{5pieta_table}
\begin{tabular}{c c c c c c c c c c}
  \hline
$E_{\rm c.m.}$ (GeV) & $\sigma$ (nb)  
& $E_{\rm c.m.}$ (GeV) & $\sigma$ (nb) 
& $E_{\rm c.m.}$ (GeV) & $\sigma$ (nb) 
& $E_{\rm c.m.}$ (GeV) & $\sigma$ (nb)  
& $E_{\rm c.m.}$ (GeV) & $\sigma$ (nb)  
\\
  \hline
1.925 & 0.05 $\pm$ 0.05 &2.475 & 0.34 $\pm$ 0.08 &3.025 & 0.10 $\pm$ 0.07 &3.575 & 0.04 $\pm$ 0.05 &4.125 & 0.03 $\pm$ 0.04 \\ 
1.975 & 0.04 $\pm$ 0.05 &2.525 & 0.13 $\pm$ 0.08 &3.075 & 1.00 $\pm$ 0.13 &3.625 & 0.04 $\pm$ 0.05 &4.175 & 0.04 $\pm$ 0.03 \\ 
2.025 & 0.01 $\pm$ 0.06 &2.575 & 0.09 $\pm$ 0.04 &3.125 & 0.70 $\pm$ 0.11 &3.675 & 0.19 $\pm$ 0.06 &4.225 & 0.03 $\pm$ 0.03 \\ 
2.075 & 0.08 $\pm$ 0.08 &2.625 & 0.18 $\pm$ 0.07 &3.175 & 0.18 $\pm$ 0.07 &3.725 & 0.04 $\pm$ 0.05 &4.275 & 0.08 $\pm$ 0.04 \\ 
2.125 & 0.04 $\pm$ 0.06 &2.675 & 0.10 $\pm$ 0.06 &3.225 & 0.08 $\pm$ 0.07 &3.775 & 0.13 $\pm$ 0.05 &4.325 & 0.02 $\pm$ 0.03 \\ 
2.175 & 0.16 $\pm$ 0.08 &2.725 & 0.13 $\pm$ 0.07 &3.275 & 0.13 $\pm$ 0.07 &3.825 & 0.11 $\pm$ 0.05 &4.375 & 0.04 $\pm$ 0.03 \\ 
2.225 & 0.07 $\pm$ 0.07 &2.775 & 0.02 $\pm$ 0.06 &3.325 & 0.18 $\pm$ 0.06 &3.875 & 0.03 $\pm$ 0.04 &4.425 & 0.05 $\pm$ 0.03 \\ 
2.275 & 0.20 $\pm$ 0.10 &2.825 & 0.15 $\pm$ 0.08 &3.375 & 0.08 $\pm$ 0.05 &3.925 & 0.03 $\pm$ 0.04 &4.475 & 0.02 $\pm$ 0.02 \\ 
2.325 & 0.17 $\pm$ 0.08 &2.875 & 0.12 $\pm$ 0.06 &3.425 & 0.20 $\pm$ 0.06 &3.975 & 0.05 $\pm$ 0.04 & &  \\ 
2.375 & 0.18 $\pm$ 0.08 &2.925 & 0.20 $\pm$ 0.09 &3.475 & 0.11 $\pm$ 0.07 &4.025 & 0.06 $\pm$ 0.10 & &  \\ 
2.425 & 0.05 $\pm$ 0.11 &2.975 & 0.25 $\pm$ 0.08 &3.525 & 0.20 $\pm$ 0.06 &4.075 & 0.05 $\pm$ 0.04 & &  \\ 
\hline
\end{tabular}
\end{table*}

%% file: acknowledgements.tex
We are grateful for the 
extraordinary contributions of our \pep2\ colleagues in
achieving the excellent luminosity and machine conditions
that have made this work possible.
The success of this project also relies critically on the 
expertise and dedication of the computing organizations that 
support \babar.
The collaborating institutions wish to thank 
SLAC for its support and the kind hospitality extended to them. 
This work is supported by the
US Department of Energy
and National Science Foundation, the
Natural Sciences and Engineering Research Council (Canada),
the Commissariat \`a l'Energie Atomique and
Institut National de Physique Nucl\'eaire et de Physique des Particules
(France), the
Bundesministerium f\"ur Bildung und Forschung and
Deutsche Forschungsgemeinschaft
(Germany), the
Istituto Nazionale di Fisica Nucleare (Italy),
the Foundation for Fundamental Research on Matter (The Netherlands),
the Research Council of Norway, the
Ministry of Education and Science of the Russian Federation, 
Ministerio de Econom\'{\i}a y Competitividad (Spain), the
Science and Technology Facilities Council (United Kingdom),
and the Binational Science Foundation (U.S.-Israel).
Individuals have received support from 
the Marie-Curie IEF program (European Union) and the A. P. Sloan Foundation (USA). 


%% file: note2739.bbl
\begin{thebibliography}{99}
%
\bibitem{dehz} 
M. ~Davier, A. ~Hoecker, B. ~Malaescu, and Z. ~Zhang, 
Eur.~Phys.~ J. C {\bf 77}, 827 (2017);
Fred Jegerlehner, EPJ~Web~Conf. {\bf 166}, 00022 (2018);
A.~Keshavarzi, D.~Nomura, T.~Teubner,  
Phys.\ Rev. D {\bf97}, 114025 (2018). 
%
\bibitem{theoryg2}T.~Aoyama {\em et al.}, Phys.\  Rep.\  {\bf 887}, 1 (2020).
%
\bibitem{fermilab}B.~Abi {\em et al.}, Phys.\ Rev.\ Lett.\ {\bf 126}, 141801 (2021).
%
\bibitem{baier} V.~N.~Baier and V.~S.~Fadin, Phys.\ Lett. B {\bf 27}, 223 
(1968).
%
\bibitem{arbus} A.~B.~Arbuzov {\em et al.}, J. High Energy Phys. {\bf 9812}, 
009 (1998).
%
\bibitem{kuehn} S.~Binner, J.H.~K\"uhn and K.~Melnikov, 
Phys.\ Lett. B {\bf 459}, 279 (1999).
%
\bibitem{ivanch} M.~Benayoun {\em et al.}, 
Mod.\ Phys.\ Lett. A {\bf 14}, 2605 (1999).
%
\bibitem{Druzhinin1} B. Aubert {\em et al.} (\babar\ Collaboration),
  Phys.\ Rev. D {\bf 69}, 011103 (2004).
%
\bibitem{isr3pi} B.\ Aubert {\em et al.} (\babar\ Collaboration),
Phys.\ Rev. D {\bf 70}, 072004 (2004). 
 %
\bibitem{isr4pi} B.\ Aubert {\em et al.} (\babar\ Collaboration),
Phys.\ Rev. D {\bf 71}, 052001 (2005).
%
\bibitem{isr6pi} B.\ Aubert {\em et al.} (\babar\ Collaboration),
Phys.\ Rev. D {\bf 73}, 052003 (2006).
%
\bibitem{isr5pi} B.\ Aubert {\em et al.} (\babar\ Collaboration),
  Phys.\ Rev. D {\bf 76}, 092005 (2007).
%
\bibitem{isrkkpi} B.\ Aubert {\em et al.} (\babar\ Collaboration),
Phys.\ Rev. D {\bf 77}, 092002 (2008).
%
\bibitem{isr2pi} B.\ Aubert {\em et al.} (\babar\ Collaboration),
Phys.\ Rev.\ Lett. {\bf 103}, 231801 (2009);
J.\ P.\ Lees {\em et al.} (\babar\ Collaboration),
Phys.\ Rev. D {\bf 86}, 032013 (2012).
%
\bibitem{isr2k2pi} B.\ Aubert {\em et al.}  (\babar\ Collaboration),
Phys.\ Rev. D {\bf 86}, 012008 (2012).
%
\bibitem{isr2k} J.\ P.\ Lees {\em et al.} (\babar\ Collaboration),
Phys.\ Rev. D {\bf 88}, 032013 (2013).
%
\bibitem{isr2p} J.P.~Lees et al. (\babar\ Collaboration),
Phys.\ Rev. D {\bf 88}, 072009 (2013).
%
\bibitem{isrkskl} J.~P.~Lees {\em et al.} (\babar\ Collaboration),
Phys.\ Rev. D {\bf 89}, 092002 (2014).
%
\bibitem{isr2pi2pi0} J.~P.~Lees {\em et al.} (\babar\ Collaboration),
  Phys.\ Rev. D {\bf 96}, 092007 (2017).
%
\bibitem{isr2pi3pi0} J.~P.~Lees {\em et al.} (\babar\ Collaboration),
  Phys.\ Rev. D {\bf 98}, 112015 (2018).
%
\bibitem{isretapipi} J.~P.~Lees {\em et al.} (\babar\ Collaboration),
  Phys.\ Rev. D {\bf 97}, 052007 (2018).
%
\bibitem{isr4pi3pi0} J.~P.~Lees {\em et al.} (\babar\ Collaboration),
  Phys.\ Rev. D {\bf 103}, 092001 (2021).
%
\bibitem{lumi} J.~P.~Lees {\it et al.}  (\babar\ Collaboration),
Nucl.\ Instr.\  Meth. A {\bf 726}, 203 (2013).
%
\bibitem{babar} B.\ Aubert {\em et al.}  (\babar\ Collaboration),
Nucl.\ Instr.\  Meth.  A {\bf 479}, 1 (2002);
B.\ Aubert {\it et al.}  (\babar\ Collaboration),
Nucl.\ Instr.\  Meth. A {\bf 729}, 615 (2013).
%
\bibitem{kuehn2}
H.~Czy\.z and J.~H.~K\"uhn, Eur.\ Phys.\ J. C {\bf 18}, 497 (2001).
%
\bibitem{kuraev} A.~B.~Arbuzov {\em et al.},
J.\ High Energy Phys. {\bf 9710}, 001 (1997).
%
\bibitem{strfun} M.~Caffo, H.~Czy\.z, E.~Remiddi, Nuovo Cim.  A {\bf 110},
 515  (1997); Phys.\ Lett. B {\bf 327}, 369 (1994). 
%
\bibitem{PHOTOS} E.~Barberio, B.~van~Eijk and Z.~Was, Comput.\ Phys.\
Commun. {\bf 66}, 115 (1991).
%
\bibitem{PDG} P.~A. ~Zyla {\em et al.} (Particle Data Group),
  Prog. Theor. Exp. Phys. 2020, 083C01 (2020).
%
\bibitem{GEANT4}  S.\ Agostinelli {\em et al.}  (\geant
  Collaboration),
  Nucl.\ Instr.\  Meth. A {\bf 506}, 250 (2003).
%
\bibitem{jetset} T.~Sj\"ostrand, Comput. Phys.\ Commun. {\bf 82}, 74 (1994).
%
\bibitem{koralb} S.~Jadach and Z.~Was, 
Comput.\ Phys.\ Commun. {\bf 85}, 453 (1995).
%
\bibitem{sndeta3pi} M.~N.~Achasov {\em et al.}  (SND Collaboration),
  Phys.\ Rev. D {\bf 99}, 112004 (2019).
%
\bibitem{eta2pisnd} M.~N.~Achasov {\em et al.}  (SND Collaboration),
Phys.\ Rev. D {\bf 91}, 052013 (2015). 
%
%
\bibitem{cmdeta3pi}R.~R.~Ahmetshin  {\em et al.}  (CMD3 Collaboration), 
  Phys.\ Lett. B {\bf 773}, 150 (2017).
\end{thebibliography}
